\documentclass[aps,prx,reprint]{revtex4-1}
\usepackage[english]{babel}
\usepackage{amsmath}
\usepackage{amssymb}
\usepackage{braket}
\usepackage{graphicx}
\usepackage{dsfont}
\usepackage{hyperref}
\usepackage{xcolor}
\usepackage{verbatim}
\hypersetup{
    colorlinks,
    linkcolor={red},
    citecolor={blue},
    urlcolor={blue}
}

\usepackage{mdframed}

\mdfdefinestyle{nolines}{%
  hidealllines=true,%
  nobreak=true, 
  leftmargin=0pt,
  rightmargin=0pt,
  innerleftmargin=0pt,
  innerrightmargin=0pt,
}

\newtheorem{theorem}{Theorem}
\newtheorem{corollary}[theorem]{Corollary}
\newtheorem{lemma}[theorem]{Lemma}
\newtheorem{proposition}[theorem]{Proposition}
\newtheorem{definition}[theorem]{Definition}
\newtheorem{conjecture}[theorem]{Conjecture}

\usepackage[dvipsnames]{xcolor}

\begin{document}
\title{{Barriers to dequantization of near-term molecular electronic structure methods}}

\author{Oskar Leimkuhler}
\email{ol22@berkeley.edu}
\affiliation{Department of Chemistry, University of California, Berkeley}
\affiliation{Berkeley Quantum Information and Computation Center, University of California, Berkeley, CA 94720, USA}
\author{K. Birgitta Whaley}
\email{whaley@berkeley.edu}
\affiliation{Department of Chemistry, University of California, Berkeley}
\email{whaley@berkeley.edu}
\affiliation{Berkeley Quantum Information and Computation Center, University of California, Berkeley, CA 94720, USA}

\date{\today}

\begin{abstract}
We prove classical simulation hardness, under the generalized $\mathsf{P}\neq\mathsf{NP}$ conjecture, for quantum circuit families with applications in near-term {molecular electronic ground and excited state} estimation. The proof exploits a connection to particle number conserving matchgate circuits with fermionic magic state inputs, which are shown to be universal for quantum computation under postselection, and are therefore not classically simulable in the worst case, in either the strong (multiplicative) or weak (sampling) sense. We apply these results to {non-orthogonal quantum eigensolvers} designed for near-term hardware by ruling out {sample-based} dequantization strategies {to additively approximate the off-diagonal matrix elements between orbital-rotated {correlated} reference states {that satisfy} an irreducible fermionic magic property, which combines anticoncentration in any orbital basis with an average-case multiplicative approximation hardness conjecture for the associated matrix polynomials}. {We apply this complexity analysis to} two choices of ansatz that incorporate both static and dynamic correlations to model the electronic eigenstates of molecular systems. {These are} linear combinations of orbital-rotated matrix product states, which are preparable in linear depth, and linear combinations of states prepared by generalized unitary coupled cluster with singles and doubles or unitary cluster Jastrow circuits of polynomial depth, for which computing the expectation values of local fermionic observables up to a constant additive error is $\mathsf{BQP}$-complete. We discuss the implications for achieving practical quantum advantage in resolving the electronic structure of catalytic systems composed from multivalent transition metal atoms using near-term quantum hardware.
\end{abstract}

\maketitle

\section{Introduction}
\label{sec:introduction}

This article addresses a missing link in the literature on near-term quantum computing. On the one hand, polynomial-depth quantum circuit families such as instantaneous quantum polynomial (IQP) circuits \cite{shepherd_temporally_2009,bremner_classical_2010}, random circuits \cite{aaronson_complexity-theoretic_2017,arute_quantum_2019}, and graph states \cite{ghosh_complexity_2023}, have proven quantum speedups in the simulation complexity but lack immediate practical applications. On the other hand, near-term {quantum algorithms to prepare molecular electronic ground and excited states}, one of the most anticipated industrial and scientific applications for quantum computers, have relied on algorithmic quantum advantage arguments \cite{peruzzo_variational_2014,baek_say_2023,leimkuhler2025quantum}, meaning there is no known classical algorithm to perform the same task efficiently. This is in contrast to a complexity theoretic quantum speedup, which would guarantee that no such classical algorithm could exist \textit{in principle}, under reasonable assumptions. {We would then say that the quantum algorithm cannot be \emph{dequantized} by any efficient classical algorithm}. The widely adopted unitary coupled-cluster ansatz with single and double excitations (UCCSD) \cite{mcclean_theory_2016,anand_quantum_2022} has not previously been shown to provide a complexity theoretic speedup under conservation of particle number (PN), a constraint which is needed to accurately describe the electronic structure of closed molecular systems. Furthermore, related ans\"atze require circuits of super-linear depth in the system size \cite{tilly_variational_2022}, limiting their practical implementation in the near term. This raises the question: does there exist an efficient near-term algorithm for {molecular electronic ground and excited states}, with circuit depth that scales linearly in the system size, that also has a provable {exponential separation} in the simulation complexity? 

In this work we {make progress toward answering} this question in two stages. First, we prove our main technical result, which guarantees a worst case exponential separation between the quantum and classical simulation complexity of orbital rotation circuits with four-qubit entangled input states. An orbital rotation is a PN conserving transformation of the fermionic tensor product space, which is mapped under the Jordan-Wigner transformation to a family of efficiently simulable quantum circuits known as matchgates \cite{valiant_quantum_2002,terhal_classical_2002,jozsa_matchgates_2008}. These orbital rotation circuits can be factorized into linear depth \cite{kivlichan_quantum_2018,jiang_quantum_2018}, suggestive of this being a lower bound for performing useful tasks in electronic structure. It has been shown that generic matchgate circuits with four-qubit magic state inputs are universal under postselection and are therefore classically hard to simulate unless the polynomial hierarchy collapses \cite{hebenstreit_all_2019,hebenstreit_computational_2020}. In this work, we extend this analysis to the subset of matchgates which conserve PN, thus implementing a quantum chemical orbital rotation with a fixed number of electrons. We also prove $\mathsf{BQP}$-completeness of closed simulation up to a constant additive error for a more general class of quantum chemistry circuits, namely generalized UCCSD circuits consisting of single and double excitation gates with polynomial depth in the system size.

In the second part, we apply our proof of simulation hardness to {non-orthogonal quantum eigensolvers} for {molecular electronic ground and excited state} estimation \cite{huggins_non-orthogonal_2020,baek_say_2023,leimkuhler2025quantum,marti-dafcik_spin_2025}, which perform a classical diagonalization of the Hamiltonian (energy) operator in a subspace of low-depth {correlated} reference states that are expressed in different orbital bases, providing a route to accurate computation of electronic energies for molecular systems possessing both static and dynamic electron correlations. These {quantum multi-reference} methods specifically target an important class of systems containing many significant molecules involved in catalysis and bond-breaking, for which there are no {reliable} classical algorithms, and which are generally regarded as either presenting a major computational challenge or even classically intractable. {Under reasonable assumptions,} our result rules out efficient computation of both the off-diagonal Hamiltonian and overlap matrix elements up to a {constant or inverse-polynomial} additive error approximation by mid-circuit $\ell^2$-norm sampling \cite{tang_quantum-inspired_2019}, {for {correlated} reference states that provide \emph{irreducible fermionic magic}, which we define by anticoncentration of the probability distribution in any orbital basis together with average-case hardness of {multiplicative} strong simulation} {under Haar-random orbital rotations.}

{We then apply this complexity analysis to two classes of reference state that have been proposed for practical implementations.}
The first of these are {low-rank-factorized UCCD \cite{motta_low_2021} or unitary cluster Jastrow (UCJ) \cite{matsuzawa_jastrow-type_2020} circuits}, which in linear combination provide the ansatz for the non-orthogonal quantum eigensolver (NOQE) \cite{baek_say_2023}. 
In this case, we prove that computing the expectation values of local fermionic observables is $\mathsf{BQP}$-complete {given a number of Jastrow layers that scales polynomially in the system size, and we conjecture that a constant number of Jastrow layers is sufficient to prepare states that provide irreducible fermionic magic}. 
{The second class of {correlated} reference states we consider are} matrix product states (MPS) in rotated orbital bases, which are prepared by quantum circuits of linear depth in the system size \cite{leimkuhler2025quantum}. 
A linear combination of these states provides the ansatz for our recently developed tensor network quantum eigensolver (TNQE) \cite{leimkuhler2025quantum}, which extends the density matrix renormalization group (DMRG) \cite{white_density_1992} optimization routine beyond the typical constraint of one-dimensional area law entanglement, and has achieved significantly more accurate and resource-efficient ground state energy estimates than a single-reference UCCSD benchmark in preliminary tests on small molecules. {We demonstrate that an MPS with polynomial bond dimension in the system size can prepare fermionic magic states and can also prepare states that perfectly anticoncentrate in the Fock space, motivating the conjecture that an MPS with polynomial bond dimension can also provide irreducible fermionic magic.}

We present these results as evidence {in support of the hypothesis} that super-polynomial quantum advantage in quantum chemistry is theoretically achievable using methods suitable for near-term hardware, noting that both of these ansatz classes are designed to compactly represent energy eigenstates of systems characterized by a combination of static and dynamic electron correlation.

\section{Overview of main results}
\label{sec:main_result}
{A schematic illustrating the connections between all lemmas, theorems, and corollaries presented in this work is provided in Appendix \ref{app:summary_of_results}}

Our main technical result is to prove that orbital rotation circuits, which we denote by $\hat{G}$, are universal under postselection when provided with four-qubit fermionic magic state inputs (Theorem \ref{thm:main_result}). From this follow two simulation hardness results: $\mathsf{GapP}$-hardness of closed simulation up to a constant multiplicative factor (Corollary \ref{cor:strong_sim_formal}), as well as worst case hardness of classical sampling (Corollary \ref{cor:weak_sim_formal}), under the widely held complexity theoretic assumption that the polynomial hierarchy does not collapse \cite{hangleiter_computational_2023,fujii_commuting_2017,bremner_classical_2010}. This improves on previous results demonstrating $\#\mathsf{P}$-hardness of closed simulation of PN conserving matchgate circuits (also known as \emph{passive fermionic linear optics}) with magic state inputs, up to exact precision \cite{ivanov_computational_2017}, or up to an exponentially small additive error \cite{oszmaniec_fermion_2022}. A worst-to-average case reduction for these weaker results had also been proven {in Ref. \citealp{oszmaniec_fermion_2022}}, as well as an anticoncentration property in the output probabilities. By conjecturing that the average case hardness of closed simulation should hold up to a multiplicative factor, Ref. \citealp{oszmaniec_fermion_2022} then argued that classical sampling up to a constant additive error is intractable due to a reduction based on Stockmeyer's theorem \cite{stockmeyer_complexity_1983,aaronson_computational_2011,hangleiter_computational_2023}. Our Corollary \ref{cor:strong_sim_formal} now strengthens this argument by proving the multiplicative hardness conjecture for a worst case instance. 

Corollaries \ref{cor:strong_sim_formal} and \ref{cor:weak_sim_formal} can be used to demonstrate worst case quantum speedups in the simulation complexity of any quantum ansatz that is sufficiently expressive to encode any instance of $\hat{G}$ with four-qubit magic state inputs. This has broad implications for the complexity of quantum chemical simulations, given the importance of orbital rotations in electronic structure methods. Furthermore, the family of generalized UCCSD circuits consisting of one- and two-body fermionic excitation gates can encode arbitrary logical quantum computations with constant overhead in the numbers of qubits and non-matchgates (Corollary \ref{cor:universality_uccsd}). This implies that, for polynomial-depth circuits from this family, computing the output probabilities up to a constant additive error is $\mathsf{BQP}$-complete (Corollary \ref{cor:bqp_complete_formal}). Another consequence of these proofs is that nearest-neighbor Givens and phase rotations plus controlled-$Z$ gates constitute a universal gate set for quantum chemistry (Corollary \ref{cor:universal_gateset}). Definitions for all of the terms in Theorem \ref{thm:main_result} and its corollaries are given in Section \ref{sec:definitions}, and their proofs are given in Section \ref{sec:universality}. 

In Section \ref{sec:implications}, we show how these results apply to the off-diagonal matrix element computations in {non-orthogonal quantum eigensolvers}, which diagonalize within a subspace of correlated low-depth reference states expressed in different orbital bases. Corollary \ref{cor:bqp_complete_formal} now implies that computing an additive approximation to the expected energy using the multi-reference NOQE ansatz, with polynomial-depth generalized {UCCSD or UCJ} reference states, is $\mathsf{BQP}$-hard (see Section \ref{sec:bqp_qexpt}). Furthermore, Theorem \ref{thm:main_result} and its corollaries now imply that computing the off-diagonal matrix elements between non-universal {correlated} reference states that {provide \emph{irreducible fermionic magic} under orbital rotations} are robust against dequantization by tensor network contraction, matchgate simulation, or any sampling-based approach (see Section \ref{sec:sampling_robustness}). {We formalize this by Definition \ref{def:itbc}, in terms of two properties that must coincide: anticoncentration over a constant fraction of Slater determinants in any orbital basis, and average-case $\mathsf{\# P}$-hardness of approximating the associated matrix polynomials up to a constant multiplicative factor, for a fraction $1-o(1)$ of orbital rotations sampled uniformly according to the Haar measure over $\text{U}(n)$, where $n$ is the number of orbitals (i.e., fermionic modes). We refer to these matrix polynomials, arising from the overlaps between irreducible fermionic magic states and orbital-rotated Slater determinants, as \emph{{irreducible} fermionic magic polynomials} (IFMPs). These have mixed determinant-permanent character, resembling `pseudo-permanents', for which it has long been conjectured that multiplicative factor approximations are classically intractable for average-case instances \cite{aaronson_complexity-theoretic_2017,aaronson_computational_2011,hangleiter_computational_2023,movassagh2018efficientunitarypathsquantum}.}

{Computing the off-diagonal overlap matrix elements by mid-circuit $\ell^2$-norm sampling is ruled out by two immediate implications of Definition \ref{def:itbc}: average-case hardness of additive-approximate sampling under the generalized $\mathsf{P}\neq\mathsf{NP}$ conjecture (Proposition \ref{prop:ifmp_sampling}), and average-case $\mathsf{\#P}$-hardness of closed simulation up to an exponentially small additive error (Proposition \ref{prop:ifmp_hardness}). Two examples of {correlated} reference state families that are conjectured to provide instances that satisfy Definition \ref{def:itbc} are given in Section \ref{sec:applications}, including UCJ circuits with a constant number of Jastrow layers $L$ independent of $n$ (Conjecture \ref{conj:jastrow_ifm}), and matrix product states with bond dimension $\chi=\text{poly}(n)$ (Conjecture \ref{conj:mps_ifm}), both of which can be prepared by circuits of depth $O(n)$.} 
By contrast, for corresponding \emph{single-reference} ans\"atze, both of these circuit families allow for efficient classical algorithms to compute the expectation values, {which is} proven for the single-reference UCJ ansatz with a single Jastrow layer {($L=1$)} in Appendix \ref{app:jastrow_dequantized} {(although classical tractability may not hold for $L>1$)}. 

Discussion of these examples is preceded by an explanation of static and dynamic correlation in Section \ref{sec:correlation}, which situates our results in relation to molecular electronic structure concepts. 
Our {complexity analysis} now suggests a distinction between tractable classical calculations for weakly correlated electronic systems described by single-reference states, as pointed out in Ref.~\cite{lee_evaluating_2023}, and the classical intractability of the more challenging strongly correlated systems, for which there are no scalable classical methods allowing accurate description with both static and the essential dynamic correlations. 
\section{Preliminaries}
\label{sec:definitions}

This article draws on concepts from chemistry, physics, and computer science. Here we lay out the terms and definitions that we will use in the rest of the paper for the benefit of readers from each research field. An accessible summary of the main complexity classes referred to in this work ($\mathsf{P}$, $\mathsf{NP}$, $\mathsf{PH}$, $\mathsf{PP}$, $\mathsf{BPP}$, $\mathsf{BQP}$, $\#\mathsf{P}$, and $\mathsf{GapP}$) is provided in Appendix \ref{app:complexity}, and an overview of basic electronic structure terminology is provided in Appendix \ref{app:el_struct_term}.

\subsection{Quantum circuit simulation}
\label{sec:simulation}

Let $\ket{x}$, $\ket{y}$ denote arbitrary computational basis states of a register of $n$ qubits. Let $\hat{U}$ denote a family of polynomial size quantum circuits, by which we mean that $\hat{U}$ could be any quantum circuit of $\text{poly}(n)$ gates on $n$ qubits satisfying some particular rules of construction. For example, $\hat{U}$ might have a restriction on the allowed types of quantum gates, or on the connectivity of the qubits. Let $\ket{\psi}$ denote the corresponding family of quantum states, defined by $\ket{\psi}=\hat{U}\ket{x}$. \textit{Closed simulation} refers to the computation of an output probability
\begin{align}
P = |\braket{y|\psi}|^2 = |\braket{y|\hat{U}|x}|^2.
\end{align}
While exact closed simulation implies that arbitrarily many digits of precision may be obtained, typically one of two notions of approximate closed simulation are considered. Closed simulation up to a \textit{multiplicative factor} refers to the computation of approximate probabilities $\tilde{P}$ up to some constant factor $c\geq1$ such that
\begin{align}
\frac{1}{c}P \leq \tilde{P} \leq cP,
\label{eq:mult_error}
\end{align}
while closed simulation up to an \textit{additive error} approximation refers to computing $\tilde{P}$ satisfying
\begin{align}
|\tilde{P}-P| \leq \epsilon,
\label{eq:add_error}
\end{align}
where {one often chooses either $\epsilon=O(1)$ or $\epsilon=1/\text{poly}(n)$}. Because $P\leq 1$, a multiplicative approximation is typically much harder to compute than an additive one (for example, Eq. \ref{eq:mult_error} would be exact whenever $P=0$). While the term \textit{strong simulation} is often used interchangeably with closed simulation, we will use it to refer specifically to multiplicative approximation (Eq. \ref{eq:mult_error}). We will say that a circuit family is efficiently strongly simulable if there exists a classical algorithm to compute $P$ up to a multiplicative factor, for all $\hat{U}$ of $\text{poly}(n)$ gates on $n$ qubits, with $\text{poly}(n)$ time and memory requirements. 

Note that a quantum computer does not directly enable strong simulation of $\hat{U}$. Instead, a quantum register prepared in the state $\ket{\psi}$ allows computational basis vectors $\ket{y}$ to be sampled from the probability distribution,
\begin{align}
\ket{y} \sim P(y), \quad P(y) \equiv |\braket{y|\psi}|^2,
\label{eq:weak_sim}
\end{align}
by collapsing the superposition under measurement (we will sometimes use $\ket{y}\sim\ket{\psi}$ as short-hand for Eq. \ref{eq:weak_sim}).
 This is known as \textit{weak simulation}, and a quantum circuit family is said to be efficiently weakly simulable if there exists a randomized classical algorithm that samples in the computational basis with probabilities according to Eq. \ref{eq:weak_sim}, with $\text{poly}(n)$ time and memory requirements. While an ideal quantum computer would in principle enable efficient sampling from the exact probability distribution, any realistic quantum computer will be subject to gate and measurement errors. The notions of multiplicative or additive error can be applied to sampling from an approximate distribution, $\ket{y}\sim\tilde{P}(y)$, satisfying Eq. \ref{eq:mult_error} or \ref{eq:add_error}. {It is also common to define additive approximate sampling in terms of the total variation distance between the distributions, in which case $\max_y|P(y)-\tilde{P}(y)|\leq\epsilon$.} Unless otherwise stated, we shall assume a multiplicative approximation to the probability distribution when discussing weak simulation.

While an additive error approximation of the output probabilities is generally a more realistic standard, it is typically much easier to prove simulation hardness up to a multiplicative factor, which does not imply the former. For example, suppose that $|\braket{y|\hat{U}|x}|^2$ is not strongly simulable according to Eq. \ref{eq:mult_error}, but can be factorized into a pair of unitaries $\hat{U}=\hat{U}_a\hat{U}_b$ which are efficiently simulable. It was demonstrated in Ref. \citealp{tang_quantum-inspired_2019} that if one can efficiently sample computational basis states $\ket{z}\sim\hat{U}_b\ket{x}$, and efficiently compute the overlaps $\braket{y|\hat{U}_a|z}$ and $\braket{z|\hat{U}_b|x}$ {up to an exponentially small additive error}, then one can efficiently obtain a {constant additive error} approximation to $\braket{y|\hat{U}|x}$ {(see Appendix \ref{app:sampling_dequantization})}. We will refer to this dequantization scheme as \textit{mid-circuit} $\ell^2$-\textit{norm sampling}. 

\subsection{Universality and postselection}
\label{sec:postselection}

We will say that a quantum circuit family $\hat{U}$ is \textit{universal} for quantum computation if any logical quantum circuit $\hat{W}$ defined on $\nu$ qubits, with $\mu=\text{poly}(\nu)$ two-qubit gates, can be encoded within an instance of $\hat{U}$ defined on a larger register of $n=\text{poly}(\nu,\mu)$ qubits using $\text{poly}(n)$ gates. The output probabilities of $\hat{U}$ then cannot be efficiently approximated up to an additive error by a probabilistic classical algorithm unless $\mathsf{BQP}=\mathsf{BPP}$ \cite{arad_quantum_2010} (see Appendix \ref{app:complexity}). Furthermore, exactly computing the output probabilities of $\hat{U}$ is $\mathsf{GapP}$-hard \cite{hangleiter_computational_2023}. While $\#\mathsf{P}$ and $\mathsf{GapP}$ are equivalent under polynomial-time reductions ($\mathsf{P^{\# P}=P^{GapP}}$), the approximation of a $\mathsf{\#P}$ sum up to a multiplicative factor is enabled by Stockmeyer's algorithm in $\mathsf{BPP}^\mathsf{NP}\subseteq\mathsf{\Sigma_3}$ \cite{stockmeyer_complexity_1983}, whereas approximating a $\mathsf{GapP}$ sum up to a multiplicative factor is also a $\mathsf{GapP}$-hard problem \cite{hangleiter_computational_2023}. This is the basis for quantum speedups in the weak simulation of universal quantum circuits, as an efficient classical sampler would enable a multiplicative approximation of the output probabilities by Stockmeyer's algorithm in $\mathsf{\Sigma_3}$ \cite{aaronson_computational_2011,hangleiter_computational_2023}. By invoking Toda's theorem \cite{toda_pp_1991}, which states that
\begin{align}
\mathsf{PH}\subseteq \mathsf{P^{\# P}}=\mathsf{P^{PP}},
\label{eq:toda_main}
\end{align}
we would then have $\mathsf{PH}\subseteq\mathsf{P^{GapP}}\subseteq\mathsf{\Sigma_3}$, so the polynomial hierarchy would collapse to the third level.

If $\hat{U}$ contains additional restrictions such that it is non-universal, then certain simulation hardness results can still be shown provided it is universal under \textit{postselection}. This refers to deterministically selecting the outcome of a measurement, the same as projecting the quantum state of the register onto the desired measurement outcome, which may have an arbitrarily small nonzero amplitude, and renormalizing. We will say that a restricted quantum circuit family $\hat{U}$ is universal under postselection if an arbitrary logical quantum computation $\hat{W}$ defined on $\nu$ qubits, with $\mu=\text{poly}(\nu)$ gates, may be encoded within a larger register of $n=\text{poly}(\nu,\mu)$ qubits using an instance of $\hat{U}$ with $\text{poly}(n)$ gates, by postselecting on the measurement outcomes of some subset of the qubits. It follows that any computation that could be encoded by postselecting on $\hat{W}$ could also be encoded by postselecting on $\hat{U}$ with polynomial overhead. This is sometimes written as $\mathsf{post}\text{-}\hat{U}=\mathsf{postBQP}$.

This postselected complexity class is characterized by
\begin{align}
\mathsf{postBQP}=\mathsf{PP},
\label{eq:aaronson_proof}
\end{align}
which was proven by Aaronson \cite{aaronson_quantum_2005}. Together with Toda's theorem (Eq. \ref{eq:toda_main}), this implies that 
$\mathsf{PH} \subseteq \mathsf{P}^{\mathsf{PP}} = \mathsf{P}^{\mathsf{postBQP}}$, so an efficiently computable solution to any problem in $\mathsf{postBQP}$ would imply the same for any problem in $\mathsf{PH}$. In Ref. \citealp{fujii_commuting_2017} it was shown that, given $\mathsf{post}\text{-}\hat{U}=\mathsf{postBQP}$, an efficient strong simulation of $\hat{U}$ by a deterministic classical algorithm would mean $\mathsf{postBQP}$ is efficiently computable in polynomial time, thus the polynomial hierarchy collapses completely ($\mathsf{P}=\mathsf{PH}$, implying $\mathsf{P}=\mathsf{NP}$). If this deterministic classical algorithm were to be replaced by a randomized one, then at the very least $\mathsf{PH}\subseteq\mathsf{BPP}\subseteq \mathsf{\Sigma_2}$. Equivalently, the argument in Ref. \citealp{fujii_commuting_2017} implies that the strong simulation of $\hat{U}$ up to a small multiplicative factor is $\mathsf{GapP}$-hard. Furthermore, it was previously shown in Ref. \citealp{bremner_classical_2010} that an efficient weak simulation of $\hat{U}$ would imply $\mathsf{postBQP}=\mathsf{postBPP}$. It is known that $\mathsf{P^{postBPP}}$ is contained in the third level of the polynomial hierarchy \cite{han_threshold_1997}, so if $\hat{U}$ were weakly simulable by an efficient classical algorithm it would follow that $\mathsf{PH}\subseteq\mathsf{P}^\mathsf{postBPP}\subseteq \mathsf{\Sigma_3}$, i.e., the polynomial hierarchy would collapse to the third level. We summarize these results in the following lemmas, the proofs of which are contained in Refs. \citealp{fujii_commuting_2017} and \citealp{bremner_classical_2010} respectively.

\begin{lemma}
\emph{(Fujii and Morimae \cite{fujii_commuting_2017})}
If $\hat{U}$ is universal under postselection, then the strong simulation of $\hat{U}$ up to a multiplicative factor $1\leq c<\sqrt{2}$ is $\mathsf{GapP}$-hard. If $\hat{U}$ is efficiently strongly simulable by a deterministic classical algorithm then $\mathsf{P}=\mathsf{PH}$, or if by a randomized classical algorithm then $\mathsf{PH}\subseteq\mathsf{BPP}\subseteq\mathsf{\Sigma_2}$.
\label{lem:strong_sim}
\end{lemma}

\begin{lemma}
\emph{(Bremner, Jozsa, and Shepherd \cite{bremner_classical_2010})}
If $\hat{U}$ is universal under postselection, and is efficiently weakly simulable with $1\leq c<\sqrt{2}$ by a randomized classical algorithm, then $\mathsf{PH}\subseteq\mathsf{P^{postBPP}}\subseteq\mathsf{\Sigma_3}$.
\label{lem:weak_sim}
\end{lemma}

While these results were first derived in the context of IQP circuits \cite{shepherd_temporally_2009,bremner_classical_2010,fujii_commuting_2017}, they hold in general for any quantum circuit family that is universal under postselection. An alternative proof of the hardness of weak simulation (the equivalent of Lemma \ref{lem:weak_sim}) was presented in Ref. \citealp{aaronson_computational_2011} in the context of boson sampling. This argument, based on Stockmeyer's algorithm \cite{stockmeyer_complexity_1983}, follows directly from Lemma \ref{lem:strong_sim}, and is similar to that which applies to sampling from universal circuits. For a comprehensive review of these arguments see Ref. \citealp{hangleiter_computational_2023}.

\subsection{Orbital rotations}
\label{sec:orbitals}

Electronic structure in second quantization is formalized within a Fock space defined over a finite set of $n$ orthonormal single-particle basis functions, known as \textit{molecular orbitals} or MOs \cite{helgaker_second_2000} (see Appendix \ref{app:el_struct_term}). The many-body basis states of the Fock space are the computational basis vectors $\ket{x}=\ket{x_1\cdots x_n}$, where $x_p\in\{0,1\}$ denotes the electron occupancy number of MO $p$. Each computational basis vector corresponds to an antisymmetrized separable wavefunction known as a Slater determinant, constructed so as to incur a phase of $-1$ under particle exchange. Operators in the Fock space are spanned by products of creation and annihilation operators which satisfy the fermionic anticommutation relations,
\begin{align}
\{\hat{a}_p,\hat{a}_q\}_+ = 0, \quad \{\hat{a}^\dag_p,\hat{a}_q\}_+ = \delta_{pq},
\label{eq:ferm_anticomm}
\end{align}
where $\{\cdot,\cdot\}_+$ denotes the anticommutator. Conservation of particle (electron) number, $\eta$, is necessary to accurately describe the physical states of an isolated molecular system. Under this restriction, any physically allowed quantum state must reside within the block of $n\choose\eta$ Slater determinants which span the PN conserving subspace. 
The size of this subspace grows exponentially when $n$ and $\eta$ are increased in proportion. In addition, any physical observable or transformation operator must have an expansion in terms of equal combinations of annihilation and creation operators. For example, the electronic structure Hamiltonian is given by
\begin{align}
\hat{H} = \sum_{pq}^nh_{pq}\,\hat{a}_p^\dag \hat{a}_q + \sum_{pqrs}^nh_{pqrs}\,\hat{a}_p^\dag \hat{a}_q^\dag \hat{a}_r \hat{a}_s. 
\label{eq:el_ham}
\end{align}
The mathematical object of central importance in our study is an \textit{orbital rotation}, which is a unitary transformation of the MOs, characterized by an $n\times n$ unitary coefficient matrix $\mathbf{G}$ (with elements $g_{pq}$). We will use the symbol $\hat{G}$ for the corresponding Fock space operator, which can be understood in terms of its action on the fermionic creation and annihilation operators,
\begin{align}
\hat{G}\hat{a}_p\hat{G}^\dag=\sum_{q=1}^ng_{pq}\hat{a}_q.
\label{eq:orbrot_def}
\end{align}
This represents a subset of the more general class of Bogoliubov transformations, which transform between any pair of fermionic Gaussian states \cite{surace_fermionic_2022}. Under conserved PN, Eq. \ref{eq:orbrot_def} provides an analogous transformation between any pair of Slater determinants in rotated orbitals, forming a subgroup of unitary operators in the Fock space (any product of rotations $\hat{G}_a\hat{G}_b$ is also an orbital rotation, and every $\hat{G}$ has a unique inverse rotation $\hat{G}^\dag$).

A Slater determinant describes the wavefunction of uncorrelated electrons, so a time evolution operator that does not include two-body or higher interactions must transform to another Slater determinant. Thus the orbital rotation in Eq. \ref{eq:orbrot_def} is related to unitary evolution under a one-body (quadratic) Hamiltonian describing non-interacting fermions. In second quantization this correspondence manifests in the Thouless theorem \cite{thouless_stability_1960}, which may be stated as
\begin{align}
\hat{G} = \exp\left({\sum_{pq}^n\tilde{g}_{pq}(\hat{a}_p^\dag\hat{a}_q-\hat{a}_q^\dag\hat{a}_p)}\right),
\label{eq:thouless_thm}
\end{align}
where $\tilde{g}_{pq}$ are matrix elements of $\tilde{\textbf{G}} = \ln(\textbf{G})$. The overlap between two $\eta$-particle Slater determinants $\ket{x}$ and $\ket{y}$ expressed in rotated orbitals is classically efficiently computable, and is given by
\begin{align}
\braket{x|\hat{G}|y} = \det(\bar{\textbf{G}}_{xy}),
\label{eq:sd_ovlp}
\end{align}
where $\bar{\textbf{G}}_{xy}$ is the $\eta\times\eta$ submatrix obtained by selecting the rows and columns of $\textbf{G}$ according to the occupied modes in $\ket{x}$ and $\ket{y}$ respectively \cite{terhal_classical_2002}.

\subsection{Matchgates}
\label{sec:matchgates}

The Jordan-Wigner (JW) transformation \cite{jordan_uber_1928} provides a direct mapping between the Fock space vectors and the computational basis states of a qubit register. Under this mapping the fermion operators are expressed in terms of the Pauli $X$, $Y$, and $Z$ gates as
\begin{align}
\hat{a}_p \mapsto (X_p-iY_p)Z_{p-1}\cdots Z_1,
\label{eq:jordan_wigner}
\end{align}
where $X_p$ denotes an $X$ gate applied on qubit $p$, etc., and we have dropped the hat notation for one- and two-qubit gates. The $Z$ gates on qubits $1,\ldots,p-1$, known as JW strings, are necessary to preserve the fermionic anticommutation relations (Eqs. \ref{eq:ferm_anticomm}). This construction defines an ordering of the qubits along one dimension from $p=1,\ldots,n$. We can then define \textit{nearest-neighbor} gates as those that act only between pairs of qubits $p$ and $p+1$.

Under the JW transformation, non-interacting fermion evolution is mapped onto a restricted quantum circuit family known as \textit{matchgates}, first proposed by Valiant \cite{valiant_quantum_2002}, before the connection to fermionic systems was established \cite{knill_fermionic_2001,terhal_classical_2002}. Matchgate circuits consist exclusively of nearest-neighbor two-qubit gates that are elements of the matchgate set $\{\mathcal{G}\}$, defined in terms of a pair of $2\times2$ unitary matrices $\textbf{u}$,  $\textbf{v}$ as
\begin{align}
\mathcal{G} = \begin{pmatrix} u_{11} & 0 & 0 & u_{12} \\
0 & v_{11} & v_{12} & 0 \\
0 & v_{21} & v_{22} & 0 \\
u_{21} & 0 & 0 & u_{22}
\end{pmatrix}, \quad \det(\textbf{u}) = \det(\textbf{v}).
\label{eq:matchgate}
\end{align}
Under PN conservation we impose the additional restriction that $\textbf{u}$ is diagonal, i.e., $u_{12}=u_{21}=0$. The restriction to nearest-neighbor gates is essential: while matchgate circuits are strongly simulable by a classical algorithm in $\mathsf{P}$, the inclusion of non-nearest neighbor gates, equivalent to the inclusion of the SWAP gate into the gate set, enables universal quantum computation \cite{jozsa_matchgates_2008} ($\mathsf{GapP}$-hard). In order to satisfy the matchgate definition (Eqs. \ref{eq:matchgate}) one must instead use the fermionic SWAP (FSWAP) gate $F\in\{\mathcal{G}\}$, which flips the sign of the $\ket{11}$ state, simulating fermionic antisymmetry under particle exchange \cite{verstraete_quantum_2009},
\begin{align}
F = \begin{pmatrix}
1 & 0 & 0 & 0 \\
0 & 0 & 1 & 0 \\
0 & 1 & 0 & 0 \\
0 & 0 & 0 & -1
\end{pmatrix}.
\label{eq:fswap_gate}
\end{align}

The FSWAP gate is an example of a PN conserving matchgate. Others include the nearest-neighbor Givens rotation gate $G(\theta)$ and the single-qubit phase gate $R(\varphi)$, which are expressed in terms of fermion operators as
\begin{align}
G_p(\theta) &= \exp\big(\theta(\hat{a}_p^\dag\hat{a}_{p+1}-\text{h.c.})\big), \label{eq:givens_fermop}\\[3pt]
R_p(\varphi) &= \exp\big({i\varphi\,\hat{a}_p^\dag\hat{a}_p}\big),
\label{eq:phase_fermop}
\end{align}
respectively. Substitution of the JW transformation (Eq. \ref{eq:jordan_wigner}) into Eqs. \ref{eq:givens_fermop} and \ref{eq:phase_fermop} then yields the two-qubit matchgate representations of $G(\theta)$ and $R(\varphi)\otimes\mathds{1}$:
\begin{align}
G(\theta) &= \begin{pmatrix} 
1 & 0 & 0 & 0 \\
0 & c & -s & 0 \\
0 & s & c & 0 \\
0 & 0 & 0 & 1
\end{pmatrix} \qquad \begin{matrix}
c = \cos \theta \\
s = \sin \theta,
\end{matrix} \label{eq:givens_matchgate}\\[3pt]
R(\varphi)\otimes\mathds{1} &= \begin{pmatrix} 
1 & 0 & 0 & 0 \\
0 & 1 & 0 & 0 \\
0 & 0 & \alpha & 0 \\
0 & 0 & 0 & \alpha
\end{pmatrix} \qquad \alpha = e^{i\varphi}.
\label{eq:phase_matchgate}
\end{align}
Note that these gates are localized on qubits $p$ and $p+1$, due to cancellation of the JW strings on the remaining qubits (this would no longer be the case if the transformation in Eq. \ref{eq:givens_fermop} were applied to non-neighboring qubits). 

The strong simulation of PN conserving matchgate circuits can be understood by their equivalence to orbital rotations through Eq. \ref{eq:thouless_thm}, which are efficiently classically simulable by Eq. \ref{eq:sd_ovlp}, as demonstrated in Ref. \citealp{terhal_classical_2002}. It was later shown in Ref. \citealp{kivlichan_quantum_2018} that a Givens rotation gate on qubits $p$ and $p+1$ (as in Eq. \ref{eq:givens_fermop}) is equivalent to pre-multiplying the coefficient matrix $\mathbf{G}$ by an $n\times n$ Givens rotation matrix that mixes the $p$'th and $(p+1)$'th rows and columns. In this manner, a sequence of $n \choose 2$ Givens rotation gates can be used to perform a QR factorization of $\textbf{G}$, after which it is in diagonal form. The net effect of this analysis is that any orbital rotation may be factorized into a circuit of $O(n^2)$ matchgates,
\begin{align}
\hat{G} = \prod_{q=1}^n R_q(\varphi_q)\prod_{k=1}^{n\choose 2}G_{p_k}(\theta_k),
\label{eq:givens_factorized}
\end{align}
where $(p_k, \theta_k)$ are pairs of register indices and rotation angles that perform the QR factorization, and the final layer of phase gates accounts for the diagonal entries. Eq. \ref{eq:givens_factorized} assumes $\textbf{G}$ is real-valued, but this can be extended to complex-valued $\textbf{G}$ using additional phase gates. These operations can be performed in parallel with linear circuit depth in $n$. We summarize this duality between PN conserving matchgate circuits and orbital rotations, and their factorization into linear depth, in the following lemma, the proof of which is contained in Refs. \citealp{terhal_classical_2002} and \citealp{kivlichan_quantum_2018}.

\begin{mdframed}[style=nolines]
\begin{lemma}
\emph{(From Terhal and DiVincenzo \cite{terhal_classical_2002} and Kivlichan \textit{et al.} \cite{kivlichan_quantum_2018})}
Under the JW transformation, any PN conserving matchgate circuit on $n$ qubits implements an orbital rotation $\hat{G}$, characterized by an $n\times n$ coefficient matrix $\emph{\textbf{G}}$, which may in turn be factorized into a PN conserving matchgate circuit of depth $O(n)$ that implements the same orbital rotation.
\label{thm:orbital_matchgates}
\end{lemma}
\end{mdframed}

\section{Proofs of hardness}
\label{sec:universality}
In this section we prove our main technical results (Theorem \ref{thm:main_result} and its corollaries). The proof of Theorem \ref{thm:main_result} is inspired by Refs. \citealp{jozsa_matchgates_2008,hebenstreit_all_2019} but extended with a new construction in terms of PN conserving matchgates. {Note that these constructions are purely designed to prove our main complexity results, and are not intended for practical quantum algorithm implementations. However, we point out that Corollary \ref{cor:universal_gateset} proves that Givens and phase rotations plus controlled-$Z$ gates are a universal gate set for quantum chemistry, while Corollary \ref{cor:universality_uccsd} proves that polynomial-depth sequences of one- and two-body fermionic excitation gates can encode universal logical computations under particle number conservation, which may be of independent interest for quantum algorithms designers. We also point out that although Theorem \ref{thm:main_result} relies on postselecting on exponentially unlikely measurement outcomes, the corresponding worst case computational hardness results (Corollaries \ref{cor:strong_sim_formal} and \ref{cor:weak_sim_formal}) apply to quantum circuits from the same family \emph{without} postselection, due to Lemmas \ref{lem:strong_sim} and \ref{lem:weak_sim}, which state that \emph{any} nonzero success probability under postselection is sufficient to prove these simulation hardness results.}
\subsection{Proof of Theorem \ref{thm:main_result}}
We begin by proving two minor results, the first of which (Lemma \ref{lem:dualrail_encoding}) establishes that a universal logical quantum circuit can be encoded in a particle conserving subspace, using PN conserving matchgates plus the controlled-$Z$ (C$Z$) gate, with constant overhead in the number of non-matchgates. The second minor result (Lemma \ref{lem:cz_ps}) states that the C$Z$ gate can in turn be implemented under postselection by a PN conserving matchgate circuit with a four-qubit, two-particle fermionic magic state input.

\begin{figure}
\centering
\includegraphics[scale=0.66]{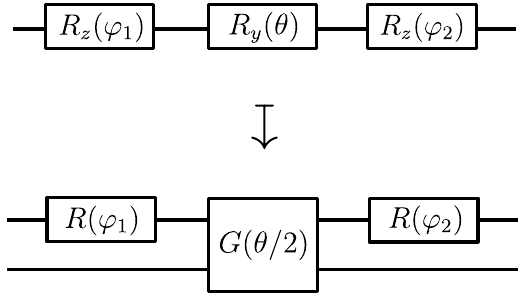}
\caption{A logical single-qubit gate is decomposed into $R_y$ and $R_z$ rotations via Euler angles $(\varphi_1,\theta,\varphi_2)$ up to a global phase (top). Each logical rotation is mapped to a PN conserving matchgate in the dual-rail encoding (bottom). $G(\theta)$ denotes a Givens rotation gate (Eq. \ref{eq:givens_matchgate}) and $R(\varphi)=e^{-i\varphi/2}R_z(\varphi)$ is the generic phase rotation gate (Eq. \ref{eq:phase_matchgate}).}
\label{fig:singlequbit}
\end{figure}

\begin{lemma}
Any logical quantum circuit of $\mu$ two-qubit gates on $\nu$ qubits, with arbitrary connectivity, can be implemented under a dual-rail encoding on $2\nu$ qubits, using only {$\text{poly}(\mu, \nu)$} nearest-neighbor PN conserving matchgates and at most $3\mu$ nearest-neighbor C$Z$ gates.
\label{lem:dualrail_encoding}
\end{lemma}
\emph{Proof.} Because the computational basis states $\ket{0}$ and $\ket{1}$ encode different occupancy numbers under the Jordan-Wigner transformation, and because a particle number conserving unitary is block-diagonal in the different particle number subspaces, an arbitrary quantum computation must be logically encoded within one of the fixed PN subspaces. This can be achieved using a \textit{dual-rail} representation \cite{chuang_simple_1995,arrazola_universal_2022}, wherein the logical $\ket{0}$ and $\ket{1}$ states are respectively encoded by the $\ket{01}$ and $\ket{10}$ states of a physical qubit pair. In this manner any $\nu$-qubit logical computational basis state can be encoded by a Slater determinant of $\nu$ fermions in $2\nu$ orbitals. Any logical single-qubit gate can then be mapped directly to particle number conserving matchgates via its decomposition into Euler angles \cite{nielsen_quantum_2010}, as illustrated in Figure \ref{fig:singlequbit}, corresponding to a complex-valued orbital rotation on the dual-rail qubit.

\begin{figure}
\centering
\includegraphics[scale=0.66]{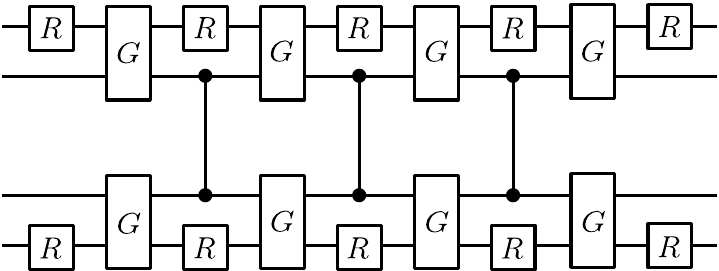}
\caption{A logical two-qubit gate is implemented in the dual-rail encoding via Givens rotations and phase rotation gates applied locally on each dual-rail qubit, and controlled-$Z$ gates applied between the dual-rail qubits (see Appendix \ref{app:dualrail}).}
\label{fig:twoqubit}
\end{figure}

To implement an arbitrary logical computation one gate is required that is not in the PN conserving matchgate set \cite{oszmaniec_universal_2017}. The nearest-neighbor controlled-$Z$ (C$Z$) gate is sufficient, since logical single-qubit rotations together with the logical C$Z$ gate form a universal gate set. In the dual-rail encoding, the logical C$Z$ operation between neighboring dual-rail qubits must flip the sign of the $\ket{10}\ket{10}$ state while leaving the orthogonal subspace unchanged. This is achieved with two physical gates: a $Z$ gate on the third qubit applies a $-1$ phase to the $\ket{10}\ket{10}$ and $\ket{01}\ket{10}$ states, then a C$Z$ gate between the middle two qubits flips the phase of the $\ket{01}\ket{10}$ state back to $+1$. With this extension of the gate set, an arbitrary logical two-qubit gate can be implemented using the construction in Figure \ref{fig:twoqubit}. This follows from the decomposition of a logical two-qubit gate into single-qubit rotations and three CNOT gates \cite{vidal_universal_2004}, which are equivalent to C$Z$ gates under Hadamard conjugation (see Appendix \ref{app:dualrail}). Logical gates between arbitrarily separated dual-rail qubits may be implemented using nearest-neighbor FSWAP networks (because each dual-rail qubit is always occupied by a single particle, these FSWAP networks incur only a global phase of $\pm1$ due to fermionic anticommutation). Therefore a logical circuit of $\mu$ two-qubit gates, with arbitrary connectivity, can be implemented in the dual-rail encoding using only nearest-neighbor PN conserving matchgates and at most $3\mu$ nearest-neighbor C$Z$ gates.
 $\Box$

\begin{lemma}
the C$Z$ gate can be implemented under postselection using only PN conserving matchgates and a four-qubit, two-particle fermionic magic state.
\label{lem:cz_ps}
\end{lemma}
\emph{Proof.} 
The C$Z$ gate between two qubits in computational basis state $\ket{xy}$, where $x,y\in\{0,1\}$, maps
\begin{align}
\ket{xy} \mapsto (-1)^{x\cdot y} \ket{xy}.
\end{align}
This operation cannot be implemented with matchgates alone, but can be implemented under postselection with access to a four qubit, two particle magic state
\begin{align}
\ket{M'} = \frac{1}{2}\left(\ket{1001} + \ket{1010} + \ket{0101} - \ket{0110}\right),
\end{align}
where we have used $\ket{M'}$ to distinguish this from the magic state in Eq. \ref{eq:magic_state}. By postselection we mean that the outcome of a measurement can be enforced, equivalent to projecting onto the desired measurement outcome and re-normalizing the projected quantum state. 

Consider the projectors $\ket{B_\pm}\!\bra{B_\pm}$ which project onto the two-qubit Bell states
\begin{align}
\ket{B_\pm}=\frac{1}{\sqrt{2}}(\ket{01}\pm\ket{10}).
\label{eq:bell_states}
\end{align}
Then it is easily shown that
\begin{align}
\big(\ket{B_+}\!\bra{B_+}\otimes\mathds{1}\otimes\ket{B_+}\!\bra{B_+}\big)\ket{x}\ket{M'}\ket{y} \nonumber \\
= \frac{(-1)^{x\cdot y}}{2}\ket{B_+}\ket{xy}\ket{B_+}.
\end{align}
The projection onto the $\ket{B_+}$ state can be achieved by rotating into the $\ket{B_+}$, $\ket{B_-}$ basis via a Givens rotation with angle $\theta=\pi/4$ and measuring the $\ket{10}$ state.

\begin{figure}
\centering
\includegraphics[scale=0.67]{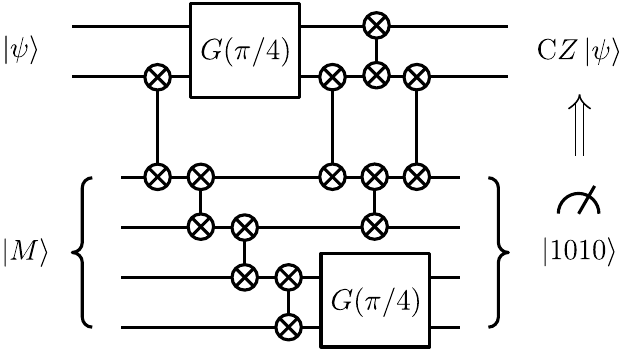}
\caption{A gadget to implement the controlled-$Z$ gate under postselection using only Givens rotation gates $G(\pi/4)$ and FSWAP gates (linked crossed circles, Eq. \ref{eq:fswap_gate}), and a magic state $\ket{M}$ (Eq. \ref{eq:magic_state}). The C$Z$ gate is implemented on the top two qubits provided that the bottom four qubits are measured in the $\ket{1010}$ state, which occurs with probability $1/4$.}
\label{fig:cz_gadget}
\end{figure}

To implement a logical quantum circuit with multiple layers of two-qubit gates under postselection, we require a protocol to swap the magic state in-between the target qubits as needed and then to swap the measurement qubits out to the end of the circuit so that successive logical operations can be applied. This can be achieved using fermionic SWAP (FSWAP) gates, which are particle-number conserving matchgates. When two registers are interchanged via an FSWAP network, a $-1$ phase is incurred whenever the ordering of two particles is interchanged. Since the magic state $\ket{M'}$ has an even number of particles it can be freely moved to any position in the circuit via an FSWAP network without incurring any resultant phase flips. After the application of the Givens rotations, and prior to measurement, rearranging the top and middle qubit pairs incurs an additional phase flip whenever the top qubits are in the $\ket{B_+}$ state and the middle qubits are in the states $\ket{xy}=\ket{10}$ or $\ket{01}$. These phase flips can be incorporated into the magic state which yields
\begin{align}
\ket{M}=\frac{1}{2}(\ket{1001}-\ket{0110}-\ket{0101}-\ket{1010}).
\label{eq:magic_state}
\end{align}
Putting these steps together results in the gadget in Figure \ref{fig:cz_gadget} to implement the C$Z$ gate under postselection, where the bottom four-qubit register can be freely moved around the circuit using FSWAP gates both before and after applying the gadget (and prior to measurement). $\Box$

Note that the construction in Figure \ref{fig:cz_gadget} is similar to the adaptive measurement gadget in Refs. \citealp{hebenstreit_all_2019,hebenstreit_computational_2020} for the SWAP gate using PN non-conserving matchgates. However, in our construction the C$Z$ is not implemented deterministically; the correct operation is only realized upon obtaining a specific measurement outcome, in this case the $\ket{1010}$ state, which occurs with probability $1/4$. Note also that any two-particle state which is equivalent to $\ket{M}$ under PN conserving matchgate operations may be substituted, along with the additional matchgates. The simplest example is the state
\begin{align}
\ket{M''}=\frac{1}{\sqrt{2}}(\ket{1100}+\ket{0011}).
\label{eq:simple_magic_state}
\end{align}
The important feature of the magic state is that it cannot be prepared by any sequence of PN conserving matchgate operations acting on a bitstring state. By the Thouless theorem (Eq. \ref{eq:thouless_thm}), this means that the magic state cannot be prepared using only single-particle fermionic excitation operators. The state in Eq. \ref{eq:simple_magic_state} can instead be prepared by a fermionic two-particle (double) excitation operator. 

We can now prove the main result:
%
%
\begin{theorem}
The family of orbital rotation circuits $\hat{G}$ with four-qubit fermionic magic state inputs is universal under postselection.
\label{thm:main_result}
\end{theorem}
\textit{Proof.} By Lemma \ref{thm:orbital_matchgates}, $\hat{G}$ is described by a PN conserving matchgate circuit. To prove the theorem, we will show that the output state of any quantum computation of $\mu$ two-qubit gates on $\nu$ qubits can be encoded on a register of $n=2\nu+12\mu$ qubits prepared in the state $\hat{G}\ket{\Phi}$, where $\hat{G}$ is an orbital rotation circuit under the JW transformation, $\ket{\Phi}=\ket{01}^{\otimes \nu}\otimes\ket{M}^{\otimes3\mu}$, and $\ket{M}$ is a four-qubit fermionic magic state (Eq. \ref{eq:magic_state}), by postselecting on the measurement outcome $\ket{1010}^{\otimes 3\mu}$ of the last $12\mu$ qubits.
This now follows from Lemmas \ref{lem:dualrail_encoding} and \ref{lem:cz_ps}. Since the magic state has even particle number it can be moved around the register via FSWAP networks without incurring any net fermionic phase flips. The magic states can therefore be initialized on the last $12\mu$ qubits of the quantum register and transported to the correct positions as needed during the computation using FSWAP networks. The measurement qubits may then be returned to the end of the quantum register by FSWAP networks after each use. Then by measuring the last $12\mu$ qubits and postselecting on the outcome $\ket{1010}^{\otimes 3\mu}$, which occurs with probability $1/4^{3\mu}$, the desired logical output state is realized on the first $2\nu$ qubits. $\square$

Note that while $\hat{G}$ is a complex-valued orbital rotation in the above statements, the same results can be shown to hold for real-valued orbital rotations with a constant factor more qubits (see Appendix \ref{app:real_valued}). 

\subsection{Corollaries of Theorem \ref{thm:main_result}}

Theorem \ref{thm:main_result}, together with Lemmas \ref{lem:strong_sim} and \ref{lem:weak_sim}, implies the following:
\begin{corollary}
\emph{(From Theorem \ref{thm:main_result} and Lemma \ref{lem:strong_sim})}
Strong simulation of $\hat{G}\ket{\Phi}$ up to a constant multiplicative factor $1\leq c<\sqrt{2}$ is $\mathsf{GapP}$-hard.
\label{cor:strong_sim_formal}
\end{corollary}
\begin{corollary}
\emph{(From Theorem \ref{thm:main_result} and Lemma \ref{lem:weak_sim})}
If $\hat{G}\ket{\Phi}$ is efficiently weakly simulable by a randomized classical algorithm, up to a multiplicative factor $1\leq c<\sqrt{2}$, then the polynomial hierarchy collapses to the third level.
\label{cor:weak_sim_formal}
\end{corollary}
We make two remarks about Corollaries \ref{cor:strong_sim_formal} and \ref{cor:weak_sim_formal}. First, they are worst case results, meaning that there is at least some instance of $\hat{G}$ for which they hold. Ideally, we should like to extend these results to the \textit{average case}, by which we mean the majority of instances of $\hat{G}$ as characterized by a coefficient matrix, $\textbf{G}$, sampled uniformly according to the Haar measure over the unitary group $\text{U}(n)$. We refer to this as a \textit{Haar-random} instance of $\hat{G}$. Second, as discussed in Section \ref{sec:simulation}, a multiplicative approximation of the output probabilities (Eq. \ref{eq:mult_error}) does not correspond to the level of accuracy typically enabled by a quantum computer subject to physical device errors. A more robust version of these results would be expressed in terms of a constant additive approximation to the output probabilities.

Progress toward both the corresponding average case result and weak simulation up to an additive approximation of the output probabilities can be made using the results of Ref. \citealp{oszmaniec_fermion_2022}, which we now briefly summarize. Let $\ket{\Phi'}=\ket{M}^{\otimes m}$, so that $n=4m$ and $\eta=2m$. Then the following theorems hold.

\begin{theorem}
\emph{(Oszmaniec \textit{et al.} \cite{oszmaniec_fermion_2022}; anticoncentration)} For Haar-random $\hat{G}$, for any computational basis state $\ket{x}$ and any $\delta\in[0,1]$, we have that $|\braket{x|\hat{G}|\Phi'}|^2>\delta/{n \choose \eta}$ with probability greater than $(1-\delta^2)/5.7$. 
\label{thm:anticoncentration}
\end{theorem}

\begin{theorem}
\emph{(Oszmaniec \textit{et al.} \cite{oszmaniec_fermion_2022}; worst-to-average case reduction)} For Haar-random $\hat{G}$, closed simulation of $\hat{G}\ket{\Phi'}$ to exact precision, or to an exponentially small additive error $\epsilon = \exp[-O(m^6)]$, is $\#\mathsf{P}$-hard for a fraction of instances $1-o(m^{-2})$.
\label{thm:wost_to_avg}
\end{theorem}

These theorems were proven in Ref. \citealp{oszmaniec_fermion_2022}, wherein Theorem \ref{thm:wost_to_avg} was used as evidence to support the conjecture that average case $\#\mathsf{P}$-hardness applies not only for an exponentially small additive error $\epsilon=\exp[-O(m^6)]$, but also for a constant multiplicative factor approximation (as in Corollary \ref{cor:strong_sim_formal}). 
Proof of this conjecture, in combination with Theorem \ref{thm:anticoncentration}, would be sufficient to prove that weak simulation of $\hat{G}\ket{\Phi'}$ up to a constant additive error is classically intractable for the average case unless the polynomial hierarchy collapses to the third level \cite{hangleiter_computational_2023,oszmaniec_fermion_2022,aaronson_computational_2011}. 
We note that while Ref. \citealp{oszmaniec_fermion_2022} conjectured average case $\#\mathsf{P}$-hardness of strong simulation up to a multiplicative factor, it was not proven therein even for the hardest case. 
By demonstrating universality under postselection, we have now provided a proof of hardness for the worst case. Thus our Corollary \ref{cor:strong_sim_formal} in the hardest case, as a consequence of our proof of Theorem \ref{thm:main_result}, lends further support to the conjectured average case hardness of strong simulation of $\hat{G}\ket{\Phi'}$ up to a multiplicative factor, and by extension also to the conjectured average case hardness of weak simulation of $\hat{G}\ket{\Phi'}$ up to a constant additive error.

\subsection{Corollaries of Lemma \ref{lem:dualrail_encoding}}

We now discuss implications of the dual-rail encoding for the universality of gate sets used to prepare {correlated wavefunctions for molecular electronic structure calculations.} Lemma \ref{lem:dualrail_encoding} states that any polynomial-sized logical quantum computation can be encoded using only PN conserving matchgates and C$Z$ gates in the dual-rail representation, but does not explicitly state that any quantum state in the PN conserving subspace can be prepared using this gate set (i.e., any superposition over the set of all bitstring states with Hamming weight $\eta$, where the particles need not be evenly distributed among the dual-rail qubits). However, universality under this definition can be easily demonstrated:
\begin{corollary}
Nearest-neighbor Givens and phase rotations plus C$Z$ gates are a universal gate set for quantum chemistry.
\label{cor:universal_gateset}
\end{corollary}
\emph{Proof.} Nearest-neighbor Givens and phase rotations plus C$Z$ gates are sufficient to implement the controlled single excitation gate of Ref. \cite{arrazola_universal_2022}, which can be used to construct any state in the PN conserving subspace. To see this, consider that the logical controlled-$U$ gate maps directly to a controlled single excitation gate in the dual-rail encoding. Since any logical two-qubit gate may be implemented as in Figure \ref{fig:twoqubit}, it follows that the controlled single excitation gate is equivalent to a circuit of Givens and phase rotations plus C$Z$ gates. This circuit can be explicitly derived from the decomposition of the controlled-$U$ gate into two CNOT gates and three single-qubit rotations on the target qubit (see Figure 4.6 in Ref. \cite{nielsen_quantum_2010}). By expanding the target qubit into a dual-rail representation, the logical single qubit rotations are replaced by single excitation gates on the dual-rail qubit, and the CNOT gates are replaced by CSWAP gates which may be further decomposed into C$Z$ gates and additional single excitation gates on the dual-rail qubit. Each single excitation gate may then be factorized into a product of phase and Givens rotation gates as in Figure \ref{fig:singlequbit}. To complete the construction, the logical rotation $R_z(\varphi)$ may be implemented in the dual-rail representation without incurring an unwanted phase on the $\ket{11}$ state of the dual-rail qubit by the product of $R(\varphi/2)\otimes R^\dag(\varphi/2)$. $\Box$

Corollary \ref{cor:universal_gateset} may be useful for designing universal PN conserving `brick-wall' circuits for quantum chemistry, particularly for quantum architectures in which the C$Z$ gate is cheaper to implement than the generic SWAP gate. 
{For example, Corollary \ref{cor:universal_gateset} implies that arbitrary-depth brick-wall circuits of phase rotation + Givens + SWAP, or phase rotation + Givens + C$Z$, are sufficient to prepare any particle number conserving quantum state.} 
The dual-rail encoding of arbitrary logical gates also trivially implies universality (without requiring postselection) for the more general family of quantum chemistry circuits consisting of single (matchgate) and double (two-body) fermionic excitation gates, equivalent to generalized UCCSD circuits, which we will denote by $\hat{\mathcal{U}}$:

\begin{corollary}
\emph{(From Lemma \ref{lem:dualrail_encoding})}
Any logical computation of $\mu$ two-qubit gates on $\nu$ qubits may be encoded on a register of $n=2\nu$ qubits by a PN conserving fermionic circuit, $\hat{\mathcal{U}}$, consisting of $\text{poly}(\nu,\mu)$ single and double excitation gates.
\label{cor:universality_uccsd}
\end{corollary}
\textit{Proof.} By Lemma \ref{lem:dualrail_encoding}, any logical computation of $\mu$ two-qubit gates on $\nu$ qubits may be realized in the dual-rail encoding on a register of $n=2\nu$ qubits, using only $\text{poly}(\nu,\mu)$ PN conserving matchgates and $3\mu$ C$Z$ gates. Under the JW transformation, we may write the C$Z$ gate acting between qubits $p$ and $q$ as a double number excitation, 
\begin{align}
\text{C}Z = \text{exp}(i\pi \hat{n}_p\hat{n}_q),
\end{align}
where $\hat{n}_p=\hat{a}^\dag_p\hat{a}_p$. Thus any logical circuit can be encoded by an instance of $\hat{\mathcal{U}}$ with polynomial overhead, therefore $\hat{\mathcal{U}}$ is universal for quantum computation. $\square$

Corollary \ref{cor:universality_uccsd} closes loopholes in existing proofs, which have shown that PN non-conserving fermionic single and double excitation gates are a universal gate set \cite{bravyi_fermionic_2002}, as are non-fermionic and PN non-conserving generalizations of UCCSD \cite{mcclean_theory_2016}. Under the restriction of conserved PN, it has been shown that an infinite product of generalized unitary single and double excitation operators can parameterize any quantum state in the Fock space \cite{evangelista_exact_2019}, but this has not previously been connected with computational hardness in the low depth regime. Now, because any polynomial-size quantum computation can be encoded with only polynomial overhead, Corollary \ref{cor:universality_uccsd} trivially implies $\mathsf{BQP}$-completeness for closed simulation of polynomial-depth generalized UCCSD circuits up to an additive error:

\begin{corollary}
\emph{(From Corollary \ref{cor:universality_uccsd})}
Closed simulation of $\hat{\mathcal{U}}$ up to a constant additive error $\epsilon$, where $\hat{\mathcal{U}}$ consists of $\text{poly}(n)$ fermionic single and double excitation gates on $n$ qubits, is $\mathsf{BQP}$-complete.
\label{cor:bqp_complete_formal}
\end{corollary}

\section{Non-orthogonal quantum eigensolvers}
\label{sec:implications}
{Here we apply our complexity results to non-orthogonal quantum eigensolvers} \cite{leimkuhler2025quantum,baek_say_2023}. These are hybrid quantum-classical methods designed to address situations with both static and dynamic electronic correlations, for which multi-reference descriptions are needed \cite{tew_electron_2007,ganoe_notion_2024,izsak_measuring_2023} (see Section \ref{sec:correlation}). In contrast to known classical analogues, they are strictly variational, and can be made size-consistent in the appropriate limit. {After introducing the general framework of these methods in Section \ref{sec:mult_ref}, we discuss barriers to dequantization for additively approximating the expectation values of these multi-reference quantum ans\"atze. {We consider the following two cases: (1) the correlated reference states are prepared by polynomial-depth circuits from a family which is universal for quantum computation (universal reference states), and (2) the correlated reference states are prepared by a restricted circuit family which may not be universal (non-universal reference states).} We first show that {the expectation estimation} task is $\mathsf{BQP}$-hard for universal reference states, and is in general at least as hard as computing the overlap matrix elements between reference states from the same family (Proposition \ref{prop:expectation_equiv}). We then investigate the computational hardness of overlap estimation between non-universal reference states that provide irreducible fermionic magic, defined by anticoncentration in any orbital basis combined with a multiplicative hardness conjecture for the associated matrix polynomials that arise from the overlaps with orbital-rotated Slater determinants. We present two further average-case simulation hardness results that then apply to any state that satisfies this definition (Propositions \ref{prop:ifmp_sampling} and \ref{prop:ifmp_hardness}). We demonstrate that these are sufficient to rule out additive dequantization of the overlaps by mid-circuit $\ell^2$-norm sampling up to a constant or inverse-polynomial additive error under complexity theoretic assumptions, with high probability over Haar-random orbital rotations.}

\subsection{General framework}
\label{sec:mult_ref}

The idea behind {non-orthogonal quantum eigensolvers} \cite{leimkuhler2025quantum,baek_say_2023,huggins_non-orthogonal_2020,marti-dafcik_spin_2025}, a variant of quantum subspace methods \cite{motta_subspace_2024}, is to construct a wavefunction ansatz from a linear combination of {correlated} reference states in {non-orthogonal} orbital bases,
\begin{align}
\ket{\psi} ={ \sum_{i=1}^M c_i\ket{u_i}} = \sum_{i=1}^Mc_i\hat{G}_i\ket{\phi_i},
\label{eq:multi_ref}
\end{align}
where the $\hat{G}_i$ operators  rotate each reference state $\ket{\phi_i}$ into a common single-particle basis. {In the following we will use the term ``reference state'' to refer to the correlated reference states in Eq. \ref{eq:multi_ref}, which need not be Slater determinants.} Instead of directly preparing $\ket{\psi}$ with a single quantum circuit, a quantum computer is used to evaluate the Hamiltonian and overlap subspace matrices $\textbf{H}$ and $\textbf{S}$, with matrix elements given by
\begin{align}
h_{ij} &= {\braket{u_i|\hat{H}|u_j}} = \braket{\phi_i|\hat{H}_i\hat{G}_{ij}|\phi_j}, \label{eq:ham_el}\\ 
s_{ij} &={\braket{u_i|u_j}} = \braket{\phi_i|\hat{G}_{ij}|\phi_j},
\label{eq:ovlp_el}
\end{align}
where $\hat{G}_{ij} = \hat{G}_i^\dag\hat{G}_j$ and $\hat{H}_i = \hat{G}_i^\dag\hat{H}\hat{G}_i$. The reference states are {also} generally non-orthogonal ($|s_{ij}|\in[0,1)$), so in order to normalize $\ket{\psi}$ the coefficient vector $\vec{c}$ is subject to the constraint that $\vec{c}^{\,\dag}\textbf{S}\vec{c} = 1$. The coefficients to minimize the expected energy are determined by solving the generalized eigenvalue problem
\begin{align}
\textbf{H}\textbf{C} = \textbf{S}\textbf{C}\textbf{E}
\end{align}
on a classical computer, such that the first column of $\textbf{C}$ corresponds to the lowest valued element $E_1$ of the diagonal eigenvalue matrix, \textbf{E}, which approximates the low energy spectrum. Eq. \ref{eq:multi_ref} represents a natural extension of classical multi-reference methods which diagonalize within a basis of non-orthogonal Slater determinants \cite{thom_hartreefock_2009}, where the reference states are now chosen to be compact correlated wavefunctions. The matrix elements in Eqs. \ref{eq:ovlp_el} can be efficiently resolved up to an additive error $\epsilon$ by a Hadamard test circuit \cite{cleve_quantum_1998}, with $O(1/\epsilon^2)$ circuit repetitions for the overlap matrix elements and $O(\lambda^2/\epsilon^2)$ repetitions for the Hamiltonian matrix elements, where $\lambda$ is the sum of absolute values of the coefficients in the JW decomposition of the Hamiltonian. Provided that {each reference state is prepared by a polynomial-depth quantum circuit, and that} the number of reference states, $M$, the norm of the Hamiltonian, $\lambda$, and the condition number of the subspace, $\kappa(\textbf{S})$, all scale as $\text{poly}(n)$, then computing the lowest expected energy in the subspace up to a constant additive error is contained in $\mathsf{BQP}$ (see Appendix \ref{app:cplex_subs_efficient}). {Note that the polynomial scaling of $\lambda$ is consistently observed in practice~\cite{koridon2021orbital}. 
Intuitions for the scaling of $M$ and $\kappa(\textbf{S})$ depend on the choice of reference states and the method of subspace construction, {as discussed in detail in Sections~\ref{sec:noqe} and~\ref{sec:tnqe}, which also give examples of relevant chemical systems to which these methods can be applied.} {Here the TNQE~(Section \ref{sec:tnqe}) is notable in allowing the condition number to be precisely controlled.
{In contrast to these non-orthogonal quantum algorithms, there} is generally no known classical algorithm to efficiently compute {Eqs. \ref{eq:ham_el} and \ref{eq:ovlp_el}} up to an inverse-polymomial additive error, which is the basis for the claims of algorithmic quantum advantage in Refs. \citealp{baek_say_2023} and \citealp{leimkuhler2025quantum}.}} 

Because the orbital rotations $\hat{G}_{ij}$ are non-local operators, computing the \textit{off-diagonal} ($i\neq j$) matrix elements can be highly non-trivial, even when the reference states $\ket{\phi_i}$ are independently classically simulable. Considering the combined state preparation circuit $\hat{U}_i\ket{0}=\hat{G}_i\ket{\phi_i}$, where $\ket{0}$ denotes the all-zero state, the diagonal matrix elements $h_{ii}=\braket{\hat{U}_i^\dag\hat{H}\hat{U}_i}_0$ are expectation values of a local observable under unitary conjugation, which can be interpreted as dynamical evolution of $\hat{H}$ in the Heisenberg picture, and may be exploitable by Pauli propagation methods \cite{angrisani_classically_2024}. The same argument cannot be applied to the off-diagonal matrix elements $h_{ij}=\braket{\hat{U}_i^\dag\hat{H}\hat{U}_j}_0$, however, which do not represent unitary conjugation. Preparing the multi-reference ansatz $(\sum_ic_i\hat{U}_i)\ket{0}$ on a single quantum register as a linear combination of unitaries (LCU) \cite{childs_hamiltonian_2012} would require a much deeper circuit, by at least a factor of $M$, composed of a highly structured sequence of controlled operations. By contrast, the quantum subspace framework allows for the variational flexibility of the ansatz to be increased systematically through the number of reference states, $M$, without increasing the circuit depth. These features of quantum multi-reference methods make them a promising avenue toward achieving practical quantum advantage in near-term {molecular electronic ground and excited state} preparation.

\subsection{Hardness of expectation estimation with universal reference states}
\label{sec:bqp_qexpt}
{In the following it will be helpful to introduce definitions for the computational tasks of additively approximating the overlap between two quantum states and the expectation value of an observable. Let $\textsc{Ovlp}(u_1,u_2,\epsilon)$ denote the task of computing an approximate overlap, $\tilde{s}_{12}$, satisfying
\begin{align}
|\tilde{s}_{12} - \braket{u_1|u_2}| \leq \epsilon,
\label{eq:ovlp_task}
\end{align}
and let $\textsc{Expv}(\hat{H},\psi,\epsilon)$ denote the task of computing an approximate expectation value $\tilde{E}$ satisfying
\begin{align}
|\tilde{E} - \braket{\psi|\hat{H}|\psi}| \leq \epsilon,
\label{eq:expt_task}
\end{align}
where in both cases $\epsilon$ defines a constant additive error threshold. Now} suppose that the reference state $\ket{\phi_i}$ is prepared by any generalized UCCSD circuit of polynomial depth in $n$. We may then merge the orbital rotation into this circuit, and write
\begin{align}
{\ket{u_i}} = \hat{G}_i\ket{\phi_i} = \hat{\mathcal{U}}_i\ket{x_0},
\end{align}
where $\hat{\mathcal{U}}_i$ now denotes an arbitrary polynomial-depth quantum circuit consisting of single (matchgate) and double (two-body) fermionic excitation gates, and $\ket{x_0}$ is a bitstring state with Hamming weight $\eta$. By Corollary \ref{cor:universality_uccsd}, the state preparation circuit $\hat{\mathcal{U}}_i$ can encode arbitrary polynomial-size logical computations. It follows that an efficient classical {algorithm for $\textsc{Ovlp}(u_i,u_j,\epsilon)$} is ruled out by Corollary \ref{cor:bqp_complete_formal} unless $\mathsf{BQP}=\mathsf{BPP}$.

Furthermore, it is easy to show that {$\textsc{Expv}(\hat{H},\psi,\epsilon)$ is also $\mathsf{BQP}$-hard, where $\hat{H}$ is a} local fermionic observable, such as the electronic structure Hamiltonian in Eq. \ref{eq:el_ham}. This even holds {for the expectation values $\braket{u_i|\hat{H}|u_i}$ with respect to each reference state $\ket{u_i}=\hat{\mathcal{U}}_i\ket{x_0}$ considered independently, and thus extends to single-reference methods such as VQE-UCCSD \cite{peruzzo_variational_2014,mcclean_theory_2016}: 
\begin{proposition}
For any local fermionic observable $\hat{H}$ and any reference state $\ket{u_i}=\hat{\mathcal{U}}_i\ket{x_0}$ prepared by a polynomial-depth quantum-universal PN conserving circuit, $\textsc{Expv}(\hat{H},u_i,\epsilon)$ is $\mathsf{BQP}$-complete. 
\end{proposition}
\emph{Proof.}} Let $\hat{U}$ be a polynomial-depth logical quantum circuit on $\nu$ qubits, which solves a $\mathsf{BQP}$-complete decision problem by determining the probability $P(1)$ of the first qubit being measured in the $\ket{1}$ state when the input is the all-zero state. Let $\hat{\mathcal{U}}$ be a PN conserving fermionic circuit on $n=2\nu$ qubits which encodes $\hat{U}$ in the dual-rail representation, let $\ket{x_0}$ be the dual-rail state which encodes the logical all-zero state, and let $\hat{H}$ be \emph{any} local fermionic observable of the form in Eq. \ref{eq:el_ham}. Now define
\begin{align}
\hat{H}' = \hat{H} + \hat{n}_1,
\end{align}
where $\hat{n}
_1=\hat{a}_1^\dag\hat{a}_1$ is a number operator on the first qubit. Then it is easily shown that
\begin{align}
P(1) = \braket{x_0|\hat{\mathcal{U}}^\dag\hat{H}'\hat{\mathcal{U}}|x_0} - \braket{x_0|\hat{\mathcal{U}}^\dag\hat{H}\hat{\mathcal{U}}|x_0},
\end{align}
thus computing the expectation values on the RHS up to any constant additive error $\epsilon$ would allow one to compute $P(1)$ up to a constant additive error bounded by at most $2\epsilon$. $\Box$

Recall that computing the expected energy of the multi-reference ansatz in Eq. \ref{eq:multi_ref} up to any constant additive error is contained in $\mathsf{BQP}$, provided that the number of reference states, $M$, the sum of the absolute Hamiltonian coefficients, $\lambda$, and the condition number of the subspace, $\kappa(\textbf{S})$, all scale as $\text{poly}(n)$ (see Appendix \ref{app:cplex_subs_efficient}). Therefore, under these conditions, approximating the expected energy of the non-orthogonal multi-reference ansatz with universal reference states, to any constant additive precision, is a $\mathsf{BQP}$-complete problem.

\subsection{{Polynomial-time reduction of overlap estimation to expectation estimation}}
\label{sec:sampling_robustness}

Here we turn our attention to more restricted choices of the reference states, $\ket{\phi_i}$, which are not by themselves sufficient for universal quantum computation. There could be many reasons to restrict the choice of reference state, such as to reduce the circuit depth, or to improve trainability (see Sections \ref{sec:tnqe} and \ref{sec:noqe}). 
In this case the $\mathsf{BQP}$-hardness result for {$\textsc{Expv}(\hat{H},\psi,\epsilon)$, where $\hat{H}$ is a local fermionic observable}, may no longer apply. However, it is easily shown that {$\textsc{Expv}(\hat{H},\psi,\epsilon)$ is at least as hard as $\textsc{Ovlp}(u_i,u_j,\epsilon)$. To state this formally:
\begin{proposition}
Let $\ket{\psi}$ be any normalized linear combination of $M=\text{poly}(n)$ reference states $\{\ket{u_i}\}$ of fixed particle number, prepared by any family of quantum circuits that can also prepare computational basis states on any subset of the qubits. Let $\hat{H}$ be any local fermionic observable. Then $\textsc{Ovlp}(u_i,u_j,\epsilon)$ is polynomial-time Turing reducible to $\textsc{Expv}(\hat{H},\psi,\epsilon)$.
\label{prop:expectation_equiv}
\end{proposition}}
{The proof is provided in Appendix \ref{app:offd_matel_from_qexpt}, where it is shown that six queries to an oracle for $\textsc{Expv}(\hat{H},\psi,\epsilon)$ on $n+2$ qubits are sufficient to solve any instance of $\textsc{Ovlp}(u_i,u_j,4\sqrt{2}\epsilon)$ on $n$ qubits. This means that an efficient black-box algorithm to compute expectation values of the form $\braket{\psi|\hat{H}|\psi}$ up to a constant additive error, where $\ket{\psi}$ is defined in Eq. \ref{eq:multi_ref}, would also allow one to efficiently compute the overlaps of the form ${\braket{u_i|u_j}}=\braket{\phi_i|\hat{G}_{ij}|\phi_j}$ up to {this} level of precision. Thus if $\textsc{Ovlp}(u_i,u_j,\epsilon)$ were shown to {be hard for} a certain complexity class, this would immediately imply the same worst case complexity for $\textsc{Expv}(\hat{H},\psi,\epsilon)$}.

\subsection{{Irreducible fermionic magic states}}
\label{sec:ifm_states}

{We now provide intuition for the conjecture that $\textsc{Ovlp}(u_i,u_j,\epsilon)$ is classically intractable for non-universal reference states of the form $\ket{u_i}=\hat{G}_i\ket{\phi_i}$, where $\ket{\phi_i}$ is chosen from a family of reference states that provide \emph{irreducible fermionic magic}, a property that combines anticoncentration in any orbital basis with average-case multiplicative approximation hardness under Haar-random orbital rotations. Before providing a formal definition of this property, we will first motivate why it is needed. In the following we refer to a \emph{microscopic} quantity as one that decays exponentially in $n$, and a \emph{macroscopic} quantity as one that decays at worst inverse-polynomially in $n$. In practice, microscopic overlaps between the reference states are beneficial, ensuring that the overlap matrix $\textbf{S}$ remains well-conditioned. However, it is important that the off-diagonal Hamiltonian matrix elements $h_{ij} = \braket{\phi_i|\hat{H}_i\hat{G}_{ij}|\phi_j} $ are macroscopic, so that the Hamiltonian subspace matrix $\textbf{H}$ is not effectively diagonal in the reference state basis, and so that the Hadamard test quantum circuit can efficiently extract useful information. We will assume that $\hat{H}_i$ can be decomposed into a linear combination of Pauli strings, and that any Pauli string can be merged into the definition of $\ket{\phi_i}$, so these terms can also be considered as overlap matrix elements between normalized quantum states. For the quantum computer to resolve a meaningful signal from the overlap $s_{ij}=\braket{\phi_i|\hat{G}_{ij}|\phi_j}$, this overlap cannot be microscopic in $n$. Otherwise, the overlap matrix element could be dequantized up to an inverse-polynomial additive error by a trivial classical algorithm that guesses $s_{ij}=0$.

To illustrate this point, the tensor product of magic states $\ket{\Phi'}=\ket{M}^{\otimes m}$, where $n=2\eta=4m$, has exactly two particles per magic state, and thus expands into a linear combination of at most ${4 \choose 2}^m$ computational basis states, which is exponentially smaller than the full PN conserving Fock space of ${n \choose \eta}$ Slater determinants. However, Haar-random orbital rotations $\hat{G}$ anticoncentrate the rotated state $\hat{G}\ket{\Phi'}$ over a constant fraction of the entire Fock space with high probability by Theorem \ref{thm:anticoncentration}. This implies that the absolute overlap $|\braket{\Phi'|\hat{G}|\Phi'}|$ decays exponentially to zero in the limit $n\rightarrow \infty$, with high probability over Haar-random orbital rotations. For such tensor products of magic states, this enables dequantization to inverse-polynomial additive error by a trivial classical algorithm that guesses $\braket{\Phi'|\hat{G}|\Phi'}=0$, except for very specific choices of $\hat{G}$ that do not anticoncentrate $\hat{G}\ket{\Phi'}$. 

Consequently, in order for the quantum computer to resolve a meaningful overlap signal under non-trivial orbital rotations, we should choose reference states $\ket{\phi_i}$ and $\ket{\phi_j}$ that also anticoncentrate over a constant fraction of the Fock space. Furthermore, this anticoncentration property should not be reducible to a microscopic fraction of Slater determinants under any rotation of the orbital basis. To formalize this intuition, we define an \emph{irreducible fermionic magic property} as follows:

\begin{definition}\emph{(Irreducible fermionic magic)}
A family of $n$-qubit, $\eta$-particle quantum states $\ket{\phi}$ provides irreducible fermionic magic (IFM) if it satisfies the following properties.
\begin{enumerate}
\item Irreducible anticoncentration: for any orbital rotation $\hat{G}$, and for a constant fraction $f$ of $\eta$-particle Slater determinants $\ket{x}$ in the limit $\eta\propto  n\rightarrow\infty$, we have 
\begin{align}
|\braket{x|\hat{G}|\phi}|^2 > \delta/{n\choose\eta},
\label{eq:heavy_bitstring}
\end{align}
for some positive constant $\delta$.
\item Average-case hardness of strong simulation: for a fraction $1-o(1)$ of Haar-random orbital rotations $\hat{G}$, and for at least half of Slater determinants $\ket{x}$ that satisfy Eq. \ref{eq:heavy_bitstring}, computing $|\braket{x|\hat{G}|\phi}|^2$ up to a constant multiplicative factor $c$ is $\mathsf{\# P}$-hard.
\end{enumerate}
\label{def:itbc}
\end{definition}
The first property implies that the IFM state $\ket{\phi}$ anticoncentrates over the space of Slater determinants, and that there is no orbital rotation $\hat{G}$ which can be applied to $\ket{\phi}$ to re-concentrate the amplitudes within a microscopic portion of the Fock space. Anticoncentration over a constant fraction of Slater determinants is a minimal requirement for the definition, so this would also be satisfied by states that anticoncentrate over all determinants, or over a fraction that tends asymptotically to $1$ in the limit $\eta\propto n\rightarrow\infty$. The second property builds directly on Conjecture 1 in Ref. \cite{oszmaniec_fermion_2022}, which assumed $\mathsf{\# P}$-hardness of strong simulation of $\hat{G}\ket{M}^{\otimes m}$ up to a multiplicative factor, for a constant fraction of Slater determinants and for a fraction $1-o(1)$ of Haar-random orbital rotations $\hat{G}$. The authors proved Theorem \ref{thm:wost_to_avg} as supporting evidence for their conjecture. This is now further supported by Corollary \ref{cor:strong_sim_formal}, which proves $\mathsf{GapP}$-hardness of multiplicative strong simulation for a worst case instance of $\hat{G}$. 

Definition \ref{def:itbc} is useful because, although anticoncentration is built into the reference states, there is no assumption made on the classical simulability of $\ket{\phi}$. For example, $\ket{\phi}$ may be both strongly and weakly simulable by an efficient classical algorithm. By property 2 of Definition \ref{def:itbc}, computational hardness arises purely due to the fermionic magic resource provided by $\ket{\phi}$ when acted on by orbital rotations. 
As evidenced by Corollary \ref{cor:strong_sim_formal}, comparatively little fermionic magic (just a tensor product of magic states) is sufficient to trigger multiplicative approximation hardness. 

To examine the origins of this hardness mechanism, consider the overlap matrix element $\braket{x|\hat{G}|\phi}$, where $\ket{\phi}$ provides irreducible fermionic magic under Definition \ref{def:itbc}. We call this overlap an \emph{irreducible fermionic magic polynomial} (IFMP):
\begin{align}
\braket{x|\hat{G}|\phi} = \sum_y \braket{y|\phi} \det(\bar{\textbf{G}}_{xy}),
\end{align}
where we have inserted a resolution of the identity and substituted the submatrix determinant expression for the overlaps $\braket{x|\hat{G}|y}$ from Eq. \ref{eq:sd_ovlp}. Since the states are particle-number preserving, this is a sum over ${n\choose\eta}$ simple matrix polynomials of degree $\eta$, weighted by the coefficients $\braket{y|\phi}$, which can have a structured or irregular arrangement of arbitrary complex phases. This expansion has a similar character to mixed determinant-permanent matrix polynomials (this could be described as a `pseudo-permanent'). The intuition behind average-case hardness for approximating matrix polynomials originates in worst-to-average case hardness reductions based on polynomial interpolation \cite{hangleiter_computational_2023, movassagh2018efficientunitarypathsquantum, aaronson_computational_2011, oszmaniec_fermion_2022}, of which Theorem \ref{thm:wost_to_avg} is an example. Under this type of reduction, an efficient black-box algorithm to compute a small number of average-case instances would yield an efficient polynomial-time algorithm to compute the worst case instance up to a multiplicative factor, thus if a worst-case instance encodes a $\mathsf{GapP}$-hard problem, then exactly computing the average-case instances must be at least $\mathsf{\# P}$-hard (this follows because $\mathsf{P}^{\mathsf{\# P}}=\mathsf{P}^{\mathsf{GapP}}$). There are currently no known techniques to prove average-case $\mathsf{\# P}$-hardness of matrix permanents up to a constant multiplicative factor, although this extension is widely conjectured to hold \cite{aaronson_computational_2011,aaronson_complexity-theoretic_2017, hangleiter_computational_2023, movassagh2018efficientunitarypathsquantum}.

Assuming that multiplicative strong simulation is $\mathsf{\# P}$-hard for the average case, then together with the irreducible anticoncentration assumption (property 2 in Definition \ref{def:itbc}), this would imply the following. 
\begin{proposition}
Let $\ket{\phi}$ satisfy Definition \ref{def:itbc}, then there is a positive constant $\epsilon$ such that, for a fraction $1-o(1)$ of Haar-random orbital rotations $\hat{G}$, no randomized classical algorithm can efficiently sample $\ket{x}\sim\tilde{P}(x)$, where $\tilde{P}(x)$ is $\epsilon$-close in total variation distance to the exact distribution $P(x)=|\braket{x|\hat{G}|\phi}|^2$, unless the polynomial hierarchy collapses to the third level.
\label{prop:ifmp_sampling}
\end{proposition}
\begin{proposition}
Let $\ket{\phi}$ satisfy Definition \ref{def:itbc}, then there is a positive constant $\epsilon$ such that, for a fraction $1-o(1)$ of Haar-random orbital rotations $\hat{G}$, and for a constant fraction $f'$ of $\eta$-particle Slater determinants $\ket{x}$, computing $\braket{x|\hat{G}|\phi}$ to an exponentially small additive error $O(\epsilon / {n \choose \eta}^{\frac{1}{2}})$ is $\mathsf{\# P}$-hard.
\label{prop:ifmp_hardness}
\end{proposition}

Proposition \ref{prop:ifmp_sampling} follows directly from Definition \ref{def:itbc} under standard reductions based on Stockmeyer's algorithm \cite{hangleiter_computational_2023,oszmaniec_fermion_2022,aaronson_computational_2011}, and Proposition \ref{prop:ifmp_hardness} is proven rigorously in Appendix \ref{app:proof_prop18}. Note there exists an $\epsilon >0$ that satisfies both propositions for any constants $f$, $\delta$, and $c$ in Definition \ref{def:itbc}. This does not imply that a quantum computer can efficiently solve $\mathsf{\# P}$-hard problems, because the overlaps $\braket{x|\hat{G}|\phi}$ will be exponentially small (i.e., microscopic in $n$).  However, the overlap can be exponentially amplified by constructive interference when the computational basis state on the left is replaced by another IFM state, yielding a macroscopic overlap that the quantum computer can efficiently resolve, thus extracting useful information that is now argued to be inaccessible to efficient classical algorithms.

To claim that computing $\textsc{Ovlp}(u_i,u_j,\epsilon)$ by a Hadamard test circuit presents a barrier to dequantization, it is necessary to rule out any classical randomized algorithm which could compute a constant additive approximation to $s_{ij}=\braket{u_i|u_j}=\braket{\phi_i|\hat{G}_{ij}|\phi_j}$ in time $\text{poly}(n,\epsilon^{-1})$ under complexity theoretic assumptions. We can now provide specific cases where dequantization by the mid-circuit $\ell^2$-norm sampling algorithm \cite{tang_quantum-inspired_2019} is ruled out. Let $\ket{\phi_i}$ and $\ket{\phi_j}$ satisfy Definition \ref{def:itbc}, with $n=2\eta$, and we assume that each reference state anticoncentrates over all $n\choose\eta$ Slater determinants under any rotation of the orbital basis. We also suppose that $\hat{G}_{ij}$ is defined by an $n\times n$ coefficient matrix $\textbf{G}$ that is characteristic of the Haar-random ensemble. 

The mid-circuit $\ell^2$-norm sampling algorithm \cite{tang_quantum-inspired_2019} could obtain a constant additive approximation to $s_{ij}$ if it were possible to factorize the orbital rotation circuit into two parts, $\hat{G}_{ij}=\hat{G}_a\hat{G}_b$, such that one could efficiently sample computational basis states from either $\ket{x}\sim\hat{G}_a^\dag\ket{\phi_i}$ or $\ket{x}\sim\hat{G}_b\ket{\phi_j}$, and efficiently compute the partial overlaps $\braket{\phi_i|\hat{G}_a|x}$ and $\braket{x|\hat{G}_b|\phi_j}$ up to exponentially small additive errors of $O(\epsilon/{n\choose \eta}^{\frac{1}{2}})$ (see Appendix~\ref{app:sampling_dequantization}). However, under the stated assumption on $\hat{G}_{ij}$, and because Haar-random unitaries are left and right invariant, we must have that at least one of $\hat{G}_a$ or $\hat{G}_b$ is also characteristic of the Haar-random ensemble. For example, if $\hat{G}_a$ is a specific structured unitary, then $\hat{G}_b$ cannot also be a structured unitary under our assumption on $\hat{G}_{ij}$ and must therefore be a Haar random unitary. It follows that efficiently sampling from $\ket{x}\sim\hat{G}_b\ket{\phi_j}$ by a randomized classical algorithm is ruled out under Proposition \ref{prop:ifmp_sampling}, and assuming the generalized $\mathsf{P}\neq\mathsf{NP}$ conjecture. One could instead opt to sample from $\ket{x}\sim\hat{G}_a^\dag\ket{\phi_i}$, but it is then not possible to efficiently compute the overlaps $\braket{x|\hat{G}_b|\phi_j}$ up to an additive error of $O(\epsilon/{n \choose \eta}^{\frac{1}{2}})$ for a constant fraction of Slater determinants by Proposition \ref{prop:ifmp_hardness}.

We also expect this barrier to dequantization by $\ell^2$-norm sampling to hold for overlap matrix elements between reference states $\ket{\phi_i}$ and $\ket{\phi_j}$ that irreducibly anticoncentrate over only a constant fraction of Slater determinants, provided that these sets of determinants have macroscopic mutual support. For reference states that irreducibly anticoncentrate over all Slater determinants, or over a fraction that tends to 1 asymptotically in the large $n$ limit, this assumption is guaranteed. 

Recall that in order for a Hadamard test quantum circuit to extract a useful signal, it is necessary that the overlap $|s_{ij}|$ is not microscopic in $n$ (preventing a trivial classical dequantization up to inverse-polynomial additive error by guessing $s_{ij}=0$). If $\ket{\phi_i}$ and $\ket{\phi_j}$ were chosen completely at random, and if $\hat{G}_{ij}$ were also sampled Haar-randomly, then the overlap would be expected to be microscopic in $n$ with high probability. However, given the average case hardness assumption in Definition \ref{def:itbc}, we anticipate that the barrier to dequantization we have described above --- formalized by Propositions \ref{prop:ifmp_sampling} and \ref{prop:ifmp_hardness} for Haar-random orbital rotations --- will also apply in cases where the reference states, $\ket{\phi_i}$ and $\ket{\phi_j}$, are selected to yield a macroscopic overlap by engineering constructive interference between their amplitude distributions under specific orbital rotations. Proving this intuition for physically relevant orbital-rotated reference states, which are optimized to represent molecular electronic ground and excited states, is likely to require analytic techniques that go beyond those employed in this work, and is left as an important open problem.}

\section{Application to quantum chemistry}
\label{sec:applications}

Here we connect the non-orthogonal quantum multi-reference ansatz to the quantum chemistry concepts of static and dynamic electron correlation (Section \ref{sec:correlation}). {In line with the terminology introduced in Section \ref{sec:mult_ref}, we use the term ``reference state'' {here} to refer to a correlated wavefunction prepared by a quantum circuit, and not a single Slater determinant, unless otherwise stated.} We analyze two existing implementations for the calculation of the electronic energies of molecular systems, where accurate energies require the incorporation of both types of correlation. These are the TNQE \cite{leimkuhler2025quantum}, for which the reference states are orbital-rotated MPS (Section \ref{sec:tnqe}), and the NOQE \cite{baek_say_2023}, which employs UCCSD or UCJ reference states (Section \ref{sec:noqe}). Finally, we discuss some concrete applications of these implementations to real-world molecular systems that are of significance for the study of catalysis in Section \ref{sec:candidates}.

{These non-orthogonal eigensolvers occupy a strategic position within the landscape of near-term quantum algorithms for molecular electronic structure. While variants of the variational quantum eigensolver (VQE) methods that were originally targeted to the noisy, intermediate-scale quantum hardware regime are being further developed for near-term early fault tolerant hardware~\cite{tang2021qubit,ramoa2025reducing,huang2026quantum}, the near-term quantum algorithmic landscape is currently dominated by quantum subspace diagonalization (QSD) methods that do not rely on expensive gradient-based hybrid quantum-classical optimization of parameterized quantum circuits. This class of quantum algorithms, originating from quantum susbspace expansions around a VQE wavefunction to mitigate decoherence and to characterize excited states~\cite{mcclean2017hybrid}, has since evolved to encompass any algorithm in which a finite set of states are prepared by low-depth quantum circuits, the Hamiltonian and overlap matrix elements between these states are then measured on quantum hardware, and the obtained subspace matrices are diagonalized on a classical computer~\cite{motta2024subspace}.
QSD methods today include the non-orthogonal NOQE and TNQE, as well as quantum filter diagonalization~\cite{parrish2019quantum,cohn2021quantum} and quantum Krylov methods~\cite{cortes2022quantum,stair2020multireference,shen2023real}, in addition to orthogonal subspace methods such as qSENSE~\cite{patel2026quantum}. }

\subsection{Static and dynamic electron correlation}

\label{sec:correlation}

In practice, the lowest energy Slater determinant can be obtained by the mean-field Hartree-Fock procedure (see Appendix \ref{app:el_struct_term}), that may optimize over MOs which are spin-restricted (RHF), which come in pairs with the same spatial distribution and opposite spin values, or spin-unrestricted (UHF), which allow different spatial distributions for the spin-up and spin-down orbitals. The difference in energy between the optimal Slater determinant and the exact wavefunction is referred to as the \emph{electron correlation energy}, reflecting the inadequacy of the mean-field approximation in the presence of two-body Coulomb interactions. 

For example, the overall wavefunction can have a nonzero value at points where the spatial coordinates of electrons with opposite spins overlap. The repulsive electron-electron potential diverges to infinity at these points, thus the second derivative of the wavefunction must also diverge to ensure cancellation of infinities in the many-body Schr\"odinger equation (see Appendix \ref{app:el_struct_term}), inducing a sharp `cusp' in the exact wavefunction. The set of primitive functions from which the MOs are constructed (the \emph{basis set}) typically consists of functions that are smooth at all points away from the nuclear center. It follows that a single Slater determinant cannot account for the divergence of the second derivative at the electron cusp, and therefore produces inaccurate energy estimates whenever Coulomb correlation cannot be neglected (see e.g. Figure 5 in Ref. \citealp{tew_electron_2007}). This effect is known as short-range \textit{dynamic correlation}, and is always a consideration in the treatment of many-electron systems within the finite basis approximation. Intuitively, the features of the exact wavefunction arising from electron-electron interactions cannot be prepared by a state preparation operator which includes only one-body excitations, and must therefore explicitly include two-body excitations. It is thus not possible in general to account for these features by a rotation of the orbital basis.

A better approximation can be made by choosing an ansatz that is a linear combination of Slater determinants, also known as the \emph{configuration interaction} (CI) ansatz, which converges to the exact wavefunction in the limit where the full set of $n \choose \eta$ determinants are included. Because this method approximates a sharp function with a linear combination of smooth functions, the expansion converges rather slowly, and each decimal place of precision in the energy estimate may require the inclusion of exponentially many more determinants \cite{tew_electron_2007}. Arbitrary accuracy is rarely required in practice, however, and quantum chemistry typically operates on a notion of \textit{chemical accuracy}. This is crudely the level of accuracy obtained by the best experiment: a common benchmark value for energy calculations is a constant additive error of $\sim1.6$ mHa \cite{boys_calculation_1997}. In regimes dominated by dynamic correlation, the wavefunction can often be efficiently approximated to within chemical accuracy using perturbative expansion methods such as coupled-cluster \cite{helgaker_coupled-cluster_2000,cizek_correlation_1966}.
\begin{figure}
\includegraphics[scale=0.88]{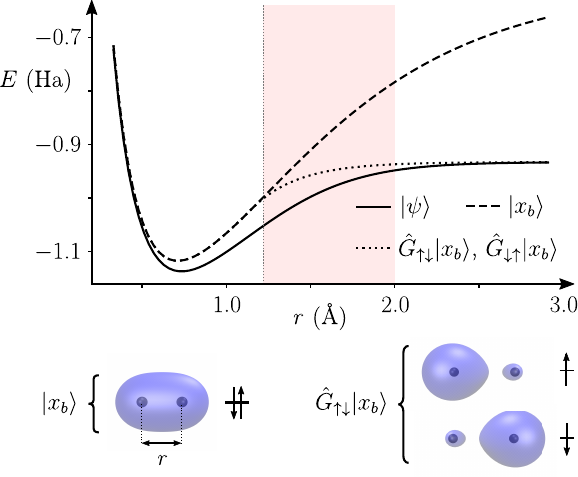} 
\caption{Diatomic hydrogen dissociation in the STO-3G basis set (computed using PySCF \cite{sun_pyscf_2018}). Near the equilibrium geometry ($r_0\approx0.74$\AA), the {FCI ground state} $\ket{\psi}$ is dominated by the RHF determinant, $\ket{x_b}$. The light gray line indicates the Coulson-Fisher point ($r_\text{CF}\approx1.25$\AA), where the unrestricted Hartree-Fock solutions undergo spontaneous symmetry breaking, resulting in two degenerate UHF determinants with up and down spins localized around opposite atomic centers. In the shaded region, the ground state $\ket{\psi}$ is dominated by an equal superposition of these two non-orthogonal bonding determinants, meaning it is not qualitatively close to any single Slater determinant.}
\label{fig:h2_dissoc}
\end{figure}

In addition to the dynamic correlation, in cases where the mean-field approximation computes many near-degenerate orbitals, the full CI wavefunction tends to have significant support (quantified by $|c_x|$) on many determinants, {an effect} known as \textit{static correlation}. For example, consider the covalent dissociation of diatomic H$_2$ (see Figure \ref{fig:h2_dissoc}). In a minimal basis set, each atom contributes two hydrogenic 1s orbitals with opposite spin values, centered at either nuclear coordinate. Near the equilibrium geometry (when the internuclear separation, $r$, is close to $r_0\approx 0.74\text{\AA}$), the mean-field orbitals are linear combinations of the 1s orbitals from each hydrogen atom, which either interfere constructively at the mid-point between the nuclei (\emph{bonding}) or cancel out exactly to zero (\textit{anti-bonding}). In this minimal basis, the full CI ground state works out to be a linear combination of the Slater determinants described by pairing up the electrons with opposite spins in either the bonding orbitals ($\ket{x_b}$) or the anti-bonding orbitals ($\ket{x_a}$). In the region where $r\approx r_0$, these determinants are well-separated in energy, and the full CI wavefunction is dominated by $\ket{x_b}$. In the dissociation limit, $r\rightarrow \infty$, the orbitals belonging to each of the well-separated atomic fragments are degenerate in energy, but there is no physical interaction between them, so the wavefunction can be solved separately within each fragment. 

In the intermediate \textit{dissociation region}, however, the mean-field orbitals are near-degenerate, but the atoms are close enough to observe correlations between the electrons localized around each atomic center. As the H$_2$ bond is stretched, at the Coulson-Fischer point ($r_\text{CF}\approx1.25\text{\AA}$) the spin-unrestricted Hartree-Fock solutions undergo spontaneous symmetry breaking, resulting in two degenerate UHF determinants. These can be prepared by orbital rotation operators relative to the RHF basis, which we will denote $\hat{G}_{\uparrow\downarrow}$ and $\hat{G}_{\downarrow\uparrow}$. For the bonding determinant described by $\hat{G}_{\uparrow\downarrow}\!\ket{x_b}$, the spin-up electron is mostly localized around the left-most atomic center and the spin-down electron around the right-most atom. The degenerate UHF determinant $\hat{G}_{\downarrow\uparrow}\!\ket{x_b}$ is identical but with both of the spins flipped. In this intermediate dissociation region, the exact ground state is dominated by an equal superposition of these two non-orthogonal Slater determinants:
\begin{align}
\ket{\psi} \approx \frac{1}{\sqrt{2+2s}}\left(\hat{G}_{\uparrow\downarrow}\!\ket{x_b}+\hat{G}_{\downarrow\uparrow}\!\ket{x_b}\right),
\label{eq:h2_noci}
\end{align}
where
$s = \braket{x_b|\hat{G}_{\uparrow\downarrow}^\dag\hat{G}_{\downarrow\uparrow}|x_b}\in [0,1]$. The state in Eq. \ref{eq:h2_noci} is an example of a \emph{non-orthogonal configuration interaction} (NOCI) wavefunction \cite{thom_hartreefock_2009}. Because the exact wavefunction is predominantly an equally weighted superposition of these two bonding determinants (with small corrections from the anti-bonding determinants), it is not perturbatively close to any single Slater determinant. 
This effect is compounded in systems with increasing numbers of near-degenerate orbitals, in which case any single Slater determinant may have a vanishingly small overlap with the correlated ground state (this phenomenon has been observed in the ground states of iron-sulfur clusters, for example \cite{lee_evaluating_2023}). {However, as illustrated by the stretched H$_2$ example, the apparent explosion in the number of significant Slater determinants in any \emph{single} orbital basis may, in many cases, be obscuring a much simpler structure in terms of a relatively small number of reference Slater determinants in \emph{different} orbital bases, which would imply a high degree of informational compressibility in the wavefunction description.} Static correlation is generally present in molecules undergoing bond-breaking and dissociation, in addition to the short-range dynamic correlation that is a pervasive feature of many-electron systems. A combined treatment of both types of correlation effects is extremely challenging {for classical methods}, owing to the difficulty of inducing perturbative corrections to a base wavefunction that is not well-described by a single Slater determinant.

\subsection{Application to NOQE (UCCSD or UCJ reference states)}
\label{sec:noqe}
{\subsubsection{Overview of the method}}
{Non-orthogonal quantum eigensolvers} effectively replace the rotated Slater determinants in NOCI with compact, correlated reference states in different orbital bases (see Eq. \ref{eq:multi_ref}). The original NOQE ansatz \cite{baek_say_2023} used UCCD reference states,
\begin{align}
\ket{\phi_i} = \exp(\hat{T}_i-\hat{T}^\dag_i)\ket{x_0},
\label{eq:noqe_def}
\end{align}
where the $\hat{T}_i$ are two-body cluster operators of the form
\begin{align}
\hat{T}_i &= \sum_{pqrs}t^{(i)}_{pqrs}\,\hat{a}_p^\dag\hat{a}_q^\dag\hat{a}_r\hat{a}_s.
\label{eq:twobody_correlator}
\end{align}
The orbital rotations $\hat{G}_i$ in Eq. \ref{eq:multi_ref} transform between a set of $M$ non-orthogonal unrestricted Hartree-Fock solutions (see Appendix \ref{app:el_struct_term}). Within each orbital basis the MOs are ordered by increasing energy, so the Hartree-Fock determinant $\ket{x_0}$ has the same expression in each basis. 
The parameters $t^{(i)}_{pqrs}$ {can be selected via classical methods such as MP2 or projective CCSD}, enabling the NOQE ansatz to capture both static and dynamic electron correlation, without hybrid {quantum-classical} variational optimization \cite{baek_say_2023}. 
The orbital rotation operators $\hat{G}_i$ represent one-body excitations via Eq. \ref{eq:thouless_thm}, so the rotated NOQE reference states $\hat{G}_i\ket{\phi_i}$ are 
{formally equivalent to a specific (not generic) UCCSD ansatz.}

{Ref. \cite{baek_say_2023} employed a low-rank factorization of the UCCD operator \cite{motta_low_2021} {that was selected for easy} hardware implementation. This  {factorization} is structurally equivalent to a unitary cluster Jastrow (UCJ) ansatz \cite{matsuzawa_jastrow-type_2020}, with the generalized form
\begin{align}
\ket{\phi_i} = \prod_{l=1}^{L} \left(\hat{G}^\dag_l \hat{D}_l \hat{G}_l\right)\ket{x_0},
\label{eq:ucj_circuit}
\end{align}
where $\hat{G}_l=\hat{G}_l^{(i)}$ is an orbital rotation operator specific to reference state $i$ and Jastrow layer $l$, and $\hat{D}_l=\hat{D}_l^{(i)}$ is a product of two-body fermionic number excitations of the form
\begin{align}
\hat{D} = \prod_{pq}^{n}\exp(\theta_{pq} \hat{n}_p \hat{n}_q).
\end{align}
{The UCJ circuit parameters in $\hat{G}_l$ and $\hat{D}_l$ can be determined classically, either analytically (for simple cases)~\cite{baek_say_2023} or by numerical optimization for more general cases~
\cite{tkachenko2025beyond}.}
Under the JW transformation, each two-body number excitation maps to a controlled-$Z$ rotation. The total gate count for the UCJ state preparation circuit is $O(n^2L)$, with a circuit depth of $O(nL)$~\cite{baek_say_2023,tkachenko2025beyond}.

We emphasize that there are many forms of UCCSD states that could be utilized here. The unitary cluster Jastrow form is attractive since it does not require Trotterization {and thus has lower circuit depth than a general UCCSD circuit}
\cite{tkachenko2025beyond}. 
Furthermore, since the {reference states} merely constitute a non-orthogonal basis for the Hamiltonian and overlap matrices, the resulting {UCJ states} can be pruned to obtain shallower circuits that can also be made invariant with respect to ordering of individual terms in $\hat{G}_l$ and $\hat{D}_l$ of Eq. \ref{eq:ucj_circuit}.

As explained in Appendix \ref{app:cplex_subs_efficient}, computing the expected energy of the quantum multi-reference ansatz (Eq. \ref{eq:multi_ref}) is quantum-efficient provided that the one-norm of the Hamiltonian $\lambda$, the number of reference states $M$, and the condition number of the subspace $\kappa(\textbf{S})$ all scale polynomially in the system size. For the original NOQE method, the number of reference states $M$ in the quantum multi-reference ansatz (Eq. \ref{eq:multi_ref}) corresponds to the number of dominant spin configurations, which are used as initial states to find different local minima in the UHF training landscape. While the number of distinct spin configurations can grow combinatorially in principle, {for strong electronic correlation this growth is dependent on the nature of the radical centers, i.e., atoms with unpaired electrons, which are usually limited in non-polymeric molecular systems. The required number of reference states $M$ will need to be established on a case-by-case basis.} 

The condition number $\kappa(\textbf{S})$ of the NOCI method can diverge at critical points where the UHF determinants are nearly parallel (see for example the Coulson-Fisher point of the H$_2$ dissociation curve in Figure \ref{fig:h2_dissoc}). In the NOQE method, this can in principle be mitigated by explicitly optimizing the two-body dynamical correlation dressing of each reference state to reduce the overlap, thus capturing a different component of the dynamical correlation within each reference state.
}
{\subsubsection{Barriers to dequantization}}
\label{sec:ucj_barriers}
It is now implied by Corollary \ref{cor:universality_uccsd} that {the generalized UCJ circuit in Eq. \ref{eq:ucj_circuit} can encode arbitrary logical computations with polynomial overhead in the number of qubits and gates, by setting the number of Jastrow layers $L=\text{poly}(n)$.} Therefore, computing a constant additive approximation to the off-diagonal matrix elements is $\mathsf{BQP}$-complete for arbitrary polynomial depth reference states of this general form (Corollary \ref{cor:bqp_complete_formal}). Furthermore, computing the expected energy of the NOQE ansatz with universal reference states is $\mathsf{BQP}$-complete for any well-conditioned subspace of polynomial size in $n$ (see Section \ref{sec:bqp_qexpt} and Appendix \ref{app:cplex_subs_efficient}).

{In the limited depth regime where $L$ is a fixed constant independent of the system size}, the rotated NOQE reference states $\hat{G}_i\ket{\phi_i}$ remain sufficient for encoded universality under postselection by Theorem \ref{thm:main_result}, {since setting $L=1$ is sufficient to prepare the tensor product of magic states (see Appendix \ref{app:twobody_magic_states}).} Corollaries \ref{cor:strong_sim_formal} and \ref{cor:weak_sim_formal} then apply to $\hat{G}_i\ket{\phi_i}$, namely $\mathsf{GapP}$-hardness of strong simulation (Corollary \ref{cor:strong_sim_formal}) and classical hardness of random sampling (Corollary \ref{cor:weak_sim_formal}), under the assumption that the polynomial hierarchy is infinite. {We also argue that the generalized UCJ state preparation circuit is likely to be able to prepare correlated reference states that provide {irreducible} fermionic magic according to Definition \ref{def:itbc}.
\begin{conjecture}
Let $\ket{\phi_i}$ be defined as in Eq. \ref{eq:ucj_circuit}. For some minimal constant Jastrow layer depth $L_\text{min}$, there is at least one instance of $\ket{\phi_i}$ with $L\leq L_\text{min}$ that provides irreducible fermionic magic under Definition \ref{def:itbc}.
\label{conj:jastrow_ifm}
\end{conjecture}
Note that this is a formal conjecture about the {class of reference states} corresponding to this generalized UCJ circuit, and not a statement about typical instances with parameters obtained by any specific choice of classical optimizer for the UCJ parameters. This is a plausible conjecture, because the orbital rotations in the state preparation circuit (Eq. \ref{eq:ucj_circuit}) can generate anticoncentration over the Fock space, while the diagonal number excitations can inject fermionic magic phases that move the quantum state out of the free-fermion manifold. This precludes the existence of an inverse orbital rotation that can perfectly re-concentrate the amplitudes. By setting $L_\text{min}$ to be a constant independent of the system size $n$, this is distinct from the trivial hardness result for $L=\text{poly}(n)$, because the state preparation circuit need not be universal for quantum computation by itself. Instead, computational hardness arises in approximating the off-diagonal matrix elements between orbital-rotated reference states. 

The UCJ circuit with a single Jastrow layer provides a neat example of this potential separation in classical tractability: for any UCJ state $\ket{\phi}$ with $L=1$, the expectation values $\braket{\phi|\hat{H}|\phi}$, where $\hat{H}$ is any local fermionic observable such as the electronic structure Hamiltonian in Eq. \ref{eq:el_ham}, can be efficiently computed to any constant or inverse-polynomial additive error by the mid-circuit $\ell^2$-norm sampling algorithm (see Appendix \ref{app:jastrow_dequantized}). On the other hand, if Conjecture \ref{conj:jastrow_ifm} holds for $L_\text{min}=1$, then approximating the off-diagonal overlap matrix elements $s_{ij}=\braket{\phi_i|\hat{G}_{ij}|\phi_j}$ to inverse-polynomial additive error would be robust against dequantization by the same algorithm due to Propositions \ref{prop:ifmp_sampling} and \ref{prop:ifmp_hardness}, assuming macroscopic overlap between the reference states, and that $\hat{G}_{ij}$ is characteristic of the Haar-random ensemble.} 
{\subsubsection{Importance for molecular electronic structure}}
Molecular systems are typically well described by single-reference coupled-cluster theories around their equilibrium geometry (for example, where $r\approx0.74 \text{\AA}$ in Figure \ref{fig:h2_dissoc}). In this perturbative (dynamic correlation) regime the {full CI} ground state wavefunction is dominated by a single Hartree-Fock Slater determinant, and the two-body amplitudes are small ($|t_{pqrs}|\ll 1$) \cite{helgaker_coupled-cluster_2000}. The wavefunction is then well approximated by a truncated Taylor expansion of the projective CCSD ansatz, the closest tractable classical analogue of the UCCSD ansatz \cite{anand_quantum_2022}, and empirical studies do not suggest a quantum advantage for such systems \cite{lee_evaluating_2023,chan_spiers_2024}. While molecules most commonly occupy their equilibrium configurations under ambient conditions, the \textit{dissociation region} (the shaded region in Figure \ref{fig:h2_dissoc}) is vital for understanding chemical phenomena related to bond breaking and formation, reaction kinetics, and catalysis. Empirically this regime often incurs large $\hat{T}$ amplitudes ($|t_{pqrs}| \sim 1$), and the projective coupled-cluster approximation breaks down \cite{helgaker_coupled-cluster_2000,nielsen_double-substitution-based_1999}. Although there is no single accepted measure of static electron correlation --- a term which connotes the strongly multi-reference character of the wavefunction \cite{tew_electron_2007,ganoe_notion_2024,izsak_measuring_2023} (see Section \ref{sec:correlation}) --- large amplitudes in $\hat{T}$ are considered an indicator \cite{nielsen_double-substitution-based_1999,ganoe_notion_2024}, so this regime is characterized by a combination of static and dynamic correlation effects. 

{Consider that the magic state $\ket{M''}=\frac{1}{\sqrt{2}}(\ket{1100}+\ket{0011})$ can be prepared by 
\begin{align}
\ket{M''} = \exp\left( \frac{\pi}{4}(\hat{a}_1^\dag\hat{a}^\dag_2\hat{a}_3\hat{a}_4 - \text{h.c.})\right)\ket{0011},
\end{align}
and that a two-body excitation with angle $\tfrac{\pi}{2}$ can be used to deterministically transport a pair of electrons. Therefore the tensor product of magic states $\ket{\Phi}=\ket{M''}^{\otimes m}$ can be prepared starting from $\ket{x_0}$ using only two-body excitations with amplitudes $t_{pqrs}\in\{0,\tfrac{\pi}{2},\tfrac{\pi}{4}\}$ and subsequent Givens rotations. The results in this work show that classical simulation of $\hat{G}\ket{\Phi}$, in the strong or weak sense, is not possible by any efficient heuristic unless the polynomial hierarchy collapses. Two-body amplitudes $t_{pqrs}$ of this magnitude are quite common for correlated electronic states within this regime \cite{nielsen_double-substitution-based_1999}, supporting the view that systems exhibiting both static and dynamic electron correlation are prime candidates for achieving exponential quantum advantage in quantum chemistry.}

\subsection{Application to TNQE (orbital-rotated MPS reference states)}
\label{sec:tnqe}
{\subsubsection{Overview of the method}}

The TNQE algorithm \cite{leimkuhler2025quantum} uses the same wavefunction form, that of Eq. \ref{eq:multi_ref}, where the correlated reference states are chosen to be matrix product states,
\begin{align}
\ket{\phi_i} = \sum_{x}\sum_{\{l\}}^\chi A^{x_1}_{l_1}A^{x_2}_{l_1l_2}\cdots A^{x_n}_{l_{n-1}}\ket{x},
\label{eq:mps}
\end{align}
where the $A^{x_p}_{l_{p-1}l_p}$ are three-index tensors with $2\chi^2$ elements corresponding to each fermionic mode, and $\chi$ is the bond dimension. It is implied in Eq. \ref{eq:mps} that the tensor elements can be different for each reference state ($A=A^{(i)}$). The bond dimension  bounds the Schmidt rank over any left-right bipartition of the sites, quantifying both the maximal entanglement of the quantum state and the classical memory requirement \cite{vidal_efficient_2003}. By setting $\chi$ to be independent of $n$, each MPS can be prepared by a quantum circuit of linear depth in $n$ overall, with a constant depth on any one qubit \cite{schon_sequential_2005,ran_encoding_2020,fomichev_initial_2024}. The TNQE algorithm enables DMRG-like variational optimization of a quantum ansatz that does not follow the one-dimensional area law, since each reference state is expressed in molecular orbitals with a unique spatial distribution.

Ref. \citealp{leimkuhler2025quantum} explored an iterative subspace construction scheme in which the MPS reference states are added in stages. At each stage, the tensor parameters are updated by a generalized sweep algorithm based on subspace diagonalization, after which local orbital rotations are inserted at pairs of nearest-neighbor sites on each MPS prior to bond index truncation \cite{leimkuhler2025quantum}. The local orbital update step requires no hybrid optimization or energy gradient calculations, and does not increase the layer depth (this is due to the matchgate structure of the orbital rotation operator, which is re-factorized into a linear depth circuit according to Eq. \ref{eq:givens_factorized}). {Using this optimization method, the condition number of the overlap matrix $\kappa(\textbf{S})$ can be precisely controlled {via the tensor decomposition techniques outlined in Ref. \cite{leimkuhler2025quantum}}. The required number of reference states $M$ to achieve chemical accuracy in the energy estimate is likely to be strongly system-dependent, and determining how this scales for realistic systems is left to future work.}

\begin{figure}
\centering
\includegraphics[scale=0.65]{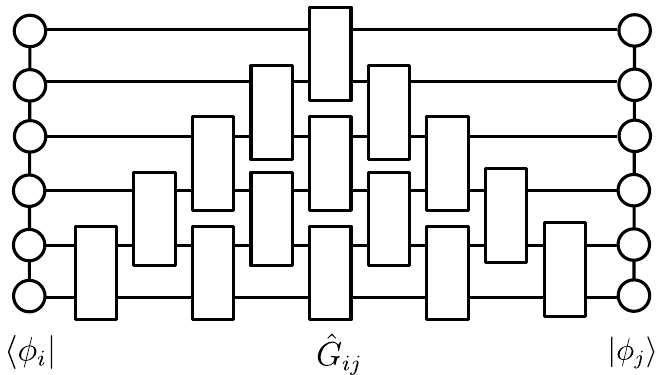}
\caption{A tensor network to compute the overlap matrix element $s_{ij}$ in Eqs. \ref{eq:ovlp_el} between matrix product states $\ket{\phi_i}$ and $\ket{\phi_j}$ expressed in different orbital bases. The orbital rotation operator $\hat{G}_{ij}$ has been factorized into a sequence of Givens rotation gates of depth $O(n)$ following Eq.
\ref{eq:givens_factorized}. $s_{ij}$ can be obtained up to an additive error $\epsilon$ by a linear depth Hadamard test circuit with $O(1/\epsilon^2)$ circuit repetitions \cite{leimkuhler2025quantum}. The Hamiltonian matrix elements $h_{ij}$ can be resolved up to error $\epsilon$ by $O(n^4)$ circuits of the same form with an inserted Pauli string, corresponding to the unique terms in the JW decomposition of $\hat{H}_i$, requiring $O(\lambda^2/\epsilon^2)$ circuit repetitions.}
\label{fig:tnqe_matel}
\end{figure}

The TNQE method relies on the efficient quantum evaluation of the off-diagonal matrix elements in Eqs. \ref{eq:ovlp_el} by a Hadamard test circuit. The overlap matrix element $s_{ij}$ in Eqs. \ref{eq:ovlp_el} is illustrated in Figure \ref{fig:tnqe_matel} in tensor network notation, where the orbital rotation $\hat{G}_{ij}$ has been factorized into a linear depth sequence of Givens rotation gates. The Hamiltonian matrix elements $h_{ij}$ can be computed by $O(n^4)$ tensor networks of this form, each of which has a Pauli string inserted between $\ket{\phi_i}$ and $\hat{G}_{ij}\ket{\phi_j}$ corresponding to a unique term in the JW decomposition of $\hat{H}_i$, which may be contracted into $\ket{\phi_i}$.

{\subsubsection{Barriers to dequantization}}
\label{sec:mps_barriers}

{By Proposition \ref{prop:expectation_equiv}, a proof of worst case hardness for additively approximating the overlaps between orbital-rotated reference states would imply the same worst case hardness for additively approximating the expectation values of local fermionic observables for the multi-reference ansatz, so here we present evidence in support of the conjecture that approximating the overlaps between orbital-rotated MPS to inverse-polynomial additive error in the system size is classically intractable. First, note that} attempting to compute the value of the tensor network in Figure \ref{fig:tnqe_matel} by contracting each Givens rotation into either the MPS on the left or on the right results in a rapid explosion of the bond dimension. For Haar-random orbital rotations $\hat{G}$, the Slater determinant $\hat{G}\ket{x}$ exhibits a linearly scaling von Neumann entropy across any equal bipartition of the orbitals \cite{lydzba_eigenstate_2020,bianchi_page_2021}, implying an exponentially large Schmidt rank (see e.g. Ref. \citealp{vidal_efficient_2003}). It follows that $\hat{G}_{ij}\ket{\phi_i}$ {is not efficiently simulable by an MPS with bond dimension $\chi=\text{poly}(n)$ for Haar-random instances of $\hat{G}_{ij}$. 
}

This observed computational hardness is now {supported by} Theorem \ref{thm:main_result} in the worst case. Because any MPS can be chosen to be the tensor product of fermionic magic states $\ket{\Phi}$ with $\chi=2$ (the magic state $\ket{M}$ as defined in Eq. \ref{eq:magic_state} has a Schmidt rank of $2$), our worst case simulation hardness results, Corollaries \ref{cor:strong_sim_formal} and \ref{cor:weak_sim_formal}, must apply to the rotated reference states. To be clear, while it is possible to efficiently sample from any MPS $\ket{\phi_i}$ in its own \textit{natural} single-particle basis, it is not in general classically efficient to sample from $\hat{G}_i\ket{\phi_i}$ as expressed in any \textit{common} orbital basis. Corollary \ref{cor:strong_sim_formal} now states that {multiplicative} strong simulation of the {output probabilities} by classical tensor network contraction, or by some extension of the matchgate determinant calculation in Eq. \ref{eq:sd_ovlp}, is $\mathsf{GapP}$-hard.

{To rule out additive error dequantization by mid-circuit $\ell^2$-norm sampling under Haar-random orbital rotations, we provide intuition that matrix product states with $\chi=\text{poly}(n)$ are likely to contain instances providing irreducible fermionic magic according to Definition \ref{def:itbc}. 
\begin{conjecture}
Let $\ket{\phi_i}$ be defined as in Eq. \ref{eq:mps}. For some constant power $d>0$, there is at least one instance of $\ket{\phi_i}$ with maximal bond dimension $\chi_\text{max} = O(n^d)$ that provides irreducible fermionic magic under Definition \ref{def:itbc}.
\label{conj:mps_ifm}
\end{conjecture}
We now motivate the conjecture. First, we note that the MPS manifold contains states that provide fermionic magic resources, such as the tensor product of magic states $\ket{M}^{\otimes m}$. The MPS manifold also contains instances which perfectly anticoncentrate, such as the equal positive-valued superposition of all ${n\choose \eta}$ Slater determinants (also known as the Dicke state), which has a maximal bond dimension $\chi_\text{max}=\eta$ when $n=2\eta$. Identifying a specific MPS that simultaneously satisfies both properties in Definition \ref{def:itbc} is left as an open problem. A possible candidate state preparation would be to initialize an MPS in the Dicke state and then apply a sequence of fermionic two-body number excitations between pairs of MPS sites with bounded separation along the chain. These correspond to controlled-$Z$ rotations under the JW transformation, which do not alter the magnitudes of the coefficients in the Slater determinant expansion but can introduce nontrivial phase correlations. If this or some other construction can be shown to satisfy Definition \ref{def:itbc}, this would guarantee a robust barrier to dequantization by Propositions \ref{prop:ifmp_sampling} and \ref{prop:ifmp_hardness}.

We point out that if the absolute overlaps $|s_{ij}|=|\braket{\phi_i|\hat{G}|\phi_j}|$ also vanished exponentially in the large $n$ limit, they could be dequantized to any inverse-polynomial additive error by a trivial algorithm that guesses $s_{ij}=0$. However, this does not follow from the linear scaling von Neumann entropy of $\hat{G}_{ij}\ket{\phi_j}$, because the Schmidt values across any bipartition need not follow a flat distribution to describe volume-law entanglement. While Haar-random orbital rotations serve to illustrate the structural failure of tractable MPS truncation schemes in the limit of free-fermionic scrambling, in practice the orbital rotations are continuously parameterized away from the identity and are optimized to approximate the ground and excited states of molecular systems. We anticipate an intermediate, physically relevant regime where the Schmidt rank of $\hat{G}_{ij}\ket{\phi_j}$ grows super-polynomially in the system size, while the absolute overlap $|s_{ij}|$ decays at worst inverse-polynomially in $n$. A macroscopic overlap with $\ket{\phi_i}$ trivially implies that there exists an efficient MPS representation that has macroscopic fidelity with $\hat{G}_{ij}\ket{\phi_j}$. However, this fact cannot be easily exploited by a classical algorithm, because finding this low Schmidt rank representation would require the classical computer to temporarily store the full volume-law entangled wavefunction representation of $\hat{G}_{ij}\ket{\phi_j}$ in classical memory prior to truncation. If one instead applied local truncations at each bond after contracting each Givens rotation into $\ket{\phi_j}$, enforcing a tractable bond dimension throughout the computation, the errors would compound at each step of the circuit, resulting in a catastrophic cumulative truncation error. Crucially, a Hadamard test quantum circuit can efficiently evaluate the overlaps to \emph{any} inverse-polynomial additive error, including errors that decay much faster than the size of the overlap matrix element itself. We leave the precise characterization of this intermediate nontrivial regime, where $\hat{G}_{ij}$ generates volume law entanglement but the overlap $|s_{ij}|$ remains macroscopically large, as an important topic of future work.}

\subsubsection{Importance for molecular electronic structure}

A single MPS can be efficiently contracted to compute the expectation value of an observable on a classical computer, and variationally optimized to minimize this value by the DMRG sweep algorithm \cite{white_density_1992,baiardi_density_2020}. {In practice, this algorithm can reliably find the ground states of} gapped Hamiltonians satisfying a one-dimensional area law of entanglement \cite{hastings_area_2007}. This property, however, is not typical of chemical Hamiltonians describing Coulomb interactions in three-dimensional space (see Appendix \ref{app:el_struct_term}). The cost of exact {or even approximate} contraction for tensor networks designed to compactly represent ground states with an area law in two or more dimensions can scale exponentially on classical computers \cite{schuch_computational_2007,arad_quantum_2010,malz_computational_2024}. In a related work it has been shown that positive bias in the tensor elements can render approximate tensor network contraction tractable \cite{chen_sign_2025}. However, fermionic systems present a unique challenge in this regard due to their antisymmetry under particle exchange, known in the context of QMC as the fermionic sign problem \cite{troyer_computational_2005}. From this perspective, it is the combination of spatially local interactions in three dimensions, often lacking any translational symmetries, together with {the sign problem arising from} fermionic antisymmetry, which {present the computational roadblocks for scalable} classical simulations of {molecular electronic structure in the hardest cases}.

The TNQE algorithm is designed to address these challenging features of molecular systems with the minimal depth of quantum operations, which can be as low as $O(n)$ in the system size, as discussed above. In the quantum chemical setting, classical DMRG is considered to capture both static and dynamic correlations but to provide a much better treatment of the former \cite{chan_spiers_2024,baiardi_density_2020}. The quantum multi-reference TNQE {is therefore primarily designed to address the static correlation within the most challenging molecular electronic active spaces. The hybrid quantum-classical optimization of this ansatz has been shown to converge reliably in small systems --- a stretched H$_2$O molecule and an octahedral H$_6$ cluster --- with a high tolerance for simulated QPU shot noise (using classical sparse vector emulation for the QPU step), observed by adding a normally distributed noise term to the off-diagonal matrix elements with standard error between $10^{-4}$ and $10^{-5}$ \cite{leimkuhler2025quantum}. Crucially, this was achieved using a smaller total number of tensor parameters than classical DMRG, which supports the utility of the TNQE ansatz for describing molecular electronic ground and excited states that do not obey a one-dimensional area law of entanglement. The superposition of orbital-rotated reference states may also provide} a more flexible treatment of both types of electronic correlation, since each MPS reference state can also capture a different aspect of the dynamic correlation via the inclusion of two-body excitations in different orbital bases.

\subsection{Candidate systems}
\label{sec:candidates}

While the electronic energies of the majority of molecular systems can be adequately represented by {perturbative dynamic} correlation on top of a mean field description, any system with polyradical nature --- i.e., multiple unpaired electrons --- can exhibit {static} correlation. 
Calculation of accurate energies to within a constant additive precision (e.g. $\sim$1.6 mHa, or `chemical accuracy') then requires adding the universal weak (dynamical) correlation to a multi-reference calculation that incorporates the strong (static) correlations. Catalytic systems are a significant class of materials which frequently fall into this category. These often involve transition metals or f-block elements whose valence electrons are strongly correlated, presenting a major challenge for classical computational methods such as projective coupled-cluster or density functional theory.

Canonical examples of catalytic systems are provided by several biological enzyme active sites that operate under ambient conditions, a physically advantageous setting that has spurred considerable effort to understand the electronic changes during the corresponding catalytic cycles. These include the nitrogenase catalysts employed in the fixing of atmospheric nitrogen into ammonia for fertilizers~\cite{hoffman2014mechanism,beinert1997iron,tanifuji2020metal}, 
the cytochrome P450 enzymes enabling drug metabolization in humans~\cite{shaik2010p450}, 
and the oxygen-evolving complex of the photosynthetic apparatus of green plants~\cite{Kern2018,Ibrahim2020,Bhowmick2023a,Bhowmick2023b}. 

Several quantum resource estimation studies have been made for the calculation of the ground state of the iron-molybdenum cofactor (FeMoco), Fe$_7$MoS$_9$C~\cite{reiher2017elucidating,von2021quantum,lee2021even,low2025fast,berry2025rapid}. FeMoco is the primary cofactor of nitrogenase and is responsible for the conversion of dinitrogen, N$_2$ into ammonia, NH$_3$, at room temperature and standard pressure - a process that involves breaking of the strong triple bond in N$_2$. 
While the nuclear structure of the FeMoco is known~\cite{einsle2002nitrogenase,spatzal2011evidence,lancaster2011x}, the large number of possible intermediate charge and spin states within the Fe$_7$MoS$_9$C system, and the difficulty of accurately calculating these given the significant strong and weak electronic correlations, has so far prevented understanding of the mechanism for this catalytic process. We note, however, that recent work has devised a classical approach to rank and filter low-energy spin configurations of broken symmetry, and employed this with coupled cluster and DMRG calculations and extrapolation techniques to estimate upper bounds to the ground state energy that appear to lie within chemical accuracy~\cite{zhai2026classical}, suggesting that the FeMoco system may have lower entanglement than previously believed~\cite{reiher2017elucidating}.

Quantum resource estimation studies for FeMoco have focused primarily on the computational cost of implementing quantum phase estimation (QPE) based on Trotter evolution~\cite{reiher2017elucidating} or qubitization~\cite{von2021quantum,lee2021even,low2025fast}.
One recent study produced an estimate of $7.3\times 10^{10}$ Toffoli gates ~\cite{berry2025rapid},
which is expected to significantly reduce the cost of QPE compared to prior art. Resource estimates have also been made for QPE calculations of energies for the iron-based cytochrome P450 enzyme, with estimates for the total Toffoli count on the order of $10^{9}-10^{10}$~\cite{goings2022reliably}. These calculations require a full fault-tolerant quantum computer to achieve the desired accuracy.

Our third example of a catalytic system requiring accurate treatment of both strong and weak electron correlations to obtain accurate energies, {and thus} to enable mechanistic elucidation, is the oxygen-evolving complex (OEC) of photosystem II (PSII). The OEC catalyzes the oxidation of water to molecular oxygen, protons and electrons in a cycle involving five intermediate states of a manganese-containing core, Mn$_4$O$_5$Ca~\cite{mcevoy2006water,yano2024structure}. Understanding the detailed mechanism of ``water splitting" by the OEC, including the energies and spin characteristics of the low-lying electronic states of the Mn$_4$O$_5$Ca core complex, is crucial for the development of artificial photoelectric systems that harness energy from sunlight and seawater \cite{fujishima1972electrochemical,Limburg1999,Hocking2014,maeda2010photocatalytic,tachibana2012artificial,ye2019artificial,zhang2015synthetic}. 
Despite decades of experimental and theoretical research, including crystallographic determination of the PSII structure revealing the \AA ngstrom scale structure of Mn$_4$O$_5$Ca~\cite{umena2011crystal} and subsequent time-resolved X-ray diffraction studies {to reveal} changes in the electron density along the catalytic cycle~\cite{Kern2018,Pantazis2019,Ibrahim2020,Bhowmick2023a,yehia2025analysis}, a 
quantitative understanding of the electronic states of the OEC and the catalytic cycle they enable has proven elusive. {This is primarily due to} the strong electronic correlations and multiple oxidation states associated with the Mn atoms in its reactive core~\cite{gupta2015high,Kern2018,yamaguchi2024theoretical}. Even though the spin nature of the ground and excited states of all the stable intermediates have been identified experimentally at cryogenic temperature, a prerequisite for understanding the catalytic pathways, the dynamic role  of the intermediates at physiological temperature and the presence of multiple spin coupling states in the  Mn$_4$O$_5$Ca core complex are not well understood~\cite{krewald2015metal,krewald2016spin,askerka2017o2,huo2024dynamic,ablyasova2025high}.

The core complex of the OEC system constitutes the holy grail in the study of natural photosynthesis, since it enables the transformation of energy from the sun into chemical energy that fuels nearly all life on Earth. 
With a valence electron active space of 60 electrons in 56 spatial orbitals, calculation of both the ground and low-lying electronic states of the Mn$_4$O$_5$Ca core complex that are involved in the changes in electronic state during the catalytic cycle of water splitting exceeds the capabilities of classical methods, which are constrained by the strong electronic correlations and multiple oxidation states of the manganese atoms, the oxy-radical character of the high valent Mn-oxo bonds and the complexity of the electron spin states. The NOQE-UCJ method discussed in this work can produce estimates for the ground and low-lying electronic states of Mn$_4$O$_5$Ca in an active space of 112 spin-orbitals using approximately the same number of qubits, with circuits characterized by at most $\sim10^5$ two-qubit gates and $\sim10^6$ T-gates (see Appendix \ref{app:gate_count}). {However, shorter circuits of a few hundred such gates are possible with judicious pruning of the UCJ {reference states} and some reduction of the active space, allowing preliminary computations on current hardware.} Encoding into error-detecting codes such as the Iceberg code~\cite{self2024protecting} or the lowest order hypercube code~\cite{goto2024many} multiplies the gate counts by at most 2--3 orders of magnitude relative to bare-qubit estimates.

{An additional important question regarding the use of non-orthogonal methods to solve for ground and excited states of strongly correlated metallocatalytic systems is the selection of a finite set of {reference} states, as well as the required number of these ($M$) to achieve energies within chemical accuracy. For NOQE, several systematic approaches are suggested by classical quantum chemistry.  One is to carry out efficient polynomial scaling coupled cluster calculations to generate, refine,
and optimize the highest-quality {reference} states, as was very effectively done recently for the ground state of the FeMoco system~\cite{zhai2026classical}, where such filtering identified a single broken symmetry state as providing a good approximation to the ground state. 
A related approach is to distill out multiple {reference} states from an approximate time-dependent wavefunctions.  Obtaining a set of {reference} states that compactly describes the static electronic correlations, with sufficient dynamic correlation to {achieve chemically accurate energy estimates}, will be the key to reducing the value of $M$. 
} 

{An additional factor determining the overall resource cost is} the number of repetitions required for robust and accurate determination of Hamiltonian and overlap matrix elements between reference states in different orbital bases, the essential source of quantum advantage for these methods, which is described above. 
{This is the sampling cost, which depends both on how the measurements are made and how the measurement errors $\epsilon$ in matrix elements propagate to the eigenvalue error $\mathcal{E}$.
Ref.~\cite{baek_say_2023} derived a loose bound estimate for the sampling cost when matrix element measurements are made by the generalized Hadamard test, as $O(M^4/{d_{\text{min}}^2\epsilon^2})$ where $M$ is the number of reference states and $d_{\text{min}}$ is the smallest absolute value of the generalized eigenvalue of $\textbf{H} +i \textbf{S}$, with $\textbf{H}$ the Hamiltonian matrix and $\textbf{S}$ the overlap matrix. 
While estimating $d_{\text{min}}$ is challenging when the eigenvalues are not known, the dependence on $M$ and $\epsilon$ is nevertheless useful.
Ref.~\cite{ren2026error} improved this bound to $O(M^3/{d_{\text{min}}^2\epsilon^2)}$ and also proposed a shadow tomography protocol enhanced by U-statistics that modifies the sample complexity for measurement of a given matrix element from $O(\text{poly}(n)/\epsilon^2)$ to $O(1/\epsilon + 2^n/\epsilon^{1/3})$, where $n$ is the number of qubits encoding the spin-orbitals. 
Detailed analysis showed that the U-statistics-enhanced shadow tomography protocol is beneficial in reducing the circuit depths and hence the overall cost for sizes $n \leq 30$, but not for larger sizes.
{The dependence of the sampling cost on the subspace dimension $M$ has recently been further reduced down to $O(M)$, by careful analysis based on matrix perturbation theory~\cite{kang2026improved}.}}

{A more radical improvement can be achieved by reformulating the matrix element evaluation step as an amplitude amplification task. 
 Ref.~\cite{ren2026towards} showed that employing an iterative quantum amplitude estimation protocol replaces the standard $1/\epsilon^2$ scaling of the sampling cost by near Heisenberg scaling $\tilde{O}(1/\epsilon)$, by coherently amplifying the signal from the state preparation routine before measurement, resulting in a substantial reduction in the overall resource cost. 
 Numerical studies {for} the stretched H$_2$ molecule {have shown} an order-of-magnitude reduction in the overall cost of querying the {reference state} preparation circuit primitives \cite{ren2026towards}. Since these primitives are now queried multiple times during the coherent amplitude amplification, this protocol requires longer-depth circuits than the Hadamard test protocol, placing more demands on the quantum hardware.
}

{The analysis in this work, and the proposed application to systems such as the core complex of the OEC, positions the non-orthogonal quantum eigensolvers NOQE and TNQE favorably within the near-term early fault-tolerant regime, with further applications to the longer-term fully fault-tolerant regime.  
In the near term, the fact that the off-diagonal matrix elements between
orbital-rotated MPS and UCJ states can be realized by relatively low-depth
quantum circuits enables direct energy estimation using the techniques described in this work.
This will become increasingly attractive as implementations with some degree of both active error correction (beyond error detection or postselection) and error mitigation are developed in the coming years. 
For the long-term fully fault-tolerant regime, NOQE and TNQE can be used to generate high-quality initial states with variational guarantees for QPE algorithms, and that have high overlap with the target Hamiltonian eigenstates, a significant bottleneck that dictates the end-to-end runtime of quantum phase estimation for molecular electronic ground and excited state energies \cite{berry2025rapid}. The single-ancilla variants of QPE that also avoid reliance on the quantum Fourier transform, such as QMEGS~\cite{ding2024quantum}, are particularly attractive for eventual deployment in the fully fault-tolerant era.}

Given the {currently accelerating} rate of advancement of quantum computers, we can expect {calculations such as those outlined above for the OEC} to be feasible with logical qubits within the next 5--10 years, enabling the long-sought accurate solution of ground and low-lying electronic states of the Mn$_4$O$_5$Ca complex.

\section{Discussion and Conclusion}
\label{sec:conclusion}

This work has shown that orbital rotation circuits with fermionic magic state inputs are not classically simulable in the worst case, in either the strong or weak sense, under the assumption that the polynomial hierarchy does not collapse. {This analysis provides evidence for the computational hardness of certain molecular electronic structure calculations, focused on obtaining accurate eigenstates and energies of systems characterized by both static and dynamic electronic correlations. This is based on intuition from quantum chemistry about a form of the wavefunction that can accurately describe the electronic ground and low-lying excited states for this important class of systems, namely a linear combination of a relatively small number of compact correlated reference states expressed in different orbital bases.} This insight enables efficient hybrid quantum-classical algorithms for ground and excited-state estimation using extremely low-depth circuits that are suitable for near-term partially fault-tolerant quantum hardware. {In the fully fault-tolerant regime, the quantum states output from these algorithms can be used as high-quality yet efficiently generated trial states for subsequent quantum phase estimation, to mitigate the initial state preparation problem that severely constrains the runtime of QPE~\cite{berry2025rapid}.}

Based on the {complexity} results of this paper, this {choice of quantum non-orthogonal multi-reference ansatz} is now also likely to offer robust quantum speedups in {both} the simulation complexity {of strongly correlated molecular electronic systems and the estimation of} expectation values of local fermionic observables {such as the Hamiltonian}. To summarize this argument {in general terms}, consider that the short-range dynamic correlation arising from the electron-electron cusp is an ever-present feature of many-electron systems, and it is generally impossible to capture this using a rotation of the orbital basis. In the language of quantum computation, it is not possible to represent these correlation features using {a reference state} prepared by a matchgate circuit, requiring {instead the inclusion of either} fermionic magic states or additional gates which have now been shown to enable universal quantum computation, such as the C$Z$ gate (or the double number excitation, see Corollary \ref{cor:universality_uccsd}). 
 
{The analysis in this work was motivated by the recognition that electronic states of molecular systems possessing significant static electronic correlation, such as occur in stretched internuclear bond configurations} (Figure \ref{fig:h2_dissoc}), it is often possible to compactly represent the static correlation in the wavefunction using a linear combination of a small number of reference states expressed in different orbital bases. Furthermore, these orbital bases are often not difficult to find in practice, since the {underlying} matchgate structure enables various fully classical techniques, such as self-consistent field methods, or hybrid quantum-classical techniques which avoid gradient-based optimization over a non-convex parameter landscape, such as local orbital transformations applied to the bonds of a tensor network. These considerations {have motivated the} hybrid quantum-classical non-orthogonal multi-reference methods --- namely TNQE \cite{leimkuhler2025quantum} and NOQE \cite{baek_say_2023} --- which diagonalize within a subspace spanned by compact correlated reference states expressed in different orbital bases. {These methods have been shown empirically to enable an algorithmic quantum advantage relative to the best known classical methods for computing ground and excited states of molecular systems with multiple radical sites, as well as a practical quantum advantage of being both variational and size-consistent} (there is no known tractable classical analogue for systems exhibiting both static and dynamic electron correlation that satisfies both of these requirements). 

The main results of this paper, Theorem \ref{thm:main_result} and its corollaries, now rule out in the worst case any classical simulation algorithm that would require, or otherwise enable, efficient strong simulation of the output probabilities of these {two classes of} quantum ans\"atze up to a multiplicative factor, or efficient sampling from their probability distributions, under the generalized $\mathsf{P}\neq\mathsf{NP}$ conjecture. Furthermore, additively approximating the expectation values of local fermionic observables for the NOQE ansatz, within a well-conditioned subspace of {correlated} reference states prepared by generalized UCCSD circuits of polynomial depth, such as the unitary cluster Jastrow circuits with $L=\text{poly}(n)$, is $\mathsf{BQP}$-complete. This suggests that {molecular electronic ground and excited states} for large complex systems that are characterized by both types of electron correlation will require a quantum computer to be efficiently computed using a scalable wavefunction ansatz.

{We have extended this analysis to compact correlated reference state families which are not individually sufficient for universal quantum computation, but can provide sufficient fermionic magic resources to incur a computational hardness mechanism when evaluating the off-diagonal matrix elements between orbital-rotated reference states. This is based on an average-case multiplicative approximation hardness conjecture for the associated matrix polynomials arising from the overlaps between irreducible fermionic magic states --- which anticoncentrate in any orbital basis according to Definition \ref{def:itbc} --- and Slater determinants that are expressed in a non-orthogonal orbital basis. As a consequence of the assumptions in this definition, we have derived robust barriers to dequantization of the off-diagonal matrix elements, which can be efficiently resolved to any inverse-polyonomial additive error by a quantum Hadamard test. The TNQE and NOQE algorithms use correlated reference state families, prepared by quantum circuits of depth $O(n)$, which are conjectured to provide instances that satisfy Definition \ref{def:itbc}, thus ruling out any known efficient classical algorithm to compute the off-diagonal matrix elements to the same level of approximation (such as approximate tensor network contraction, matchgate simulation, or sample-based dequantization schemes) under Conjectures \ref{conj:jastrow_ifm} and \ref{conj:mps_ifm}.}

{We point out several open problems in the rigorous analysis of these methods. On the complexity theoretic side, identifying specific instances of states that provide irreducible fermionic magic under Definition \ref{def:itbc} (see Conjectures \ref{conj:jastrow_ifm} and \ref{conj:mps_ifm}), and also provide macroscopic overlaps under specific orbital rotations while preserving a robust barrier to sample-based dequantization, will likely require analytic techniques that are beyond the scope of the present work. Furthermore, the results presented in this work assume noiseless quantum hardware. While these results would apply for error-corrected implementations of NOQE or TNQE, incorporating noise models into the analysis to represent existing hardware capabilities is left as an open problem. On the molecular electronic structure side, a rigorous theory describing how the number of {correlated} reference states and the condition number of the overlap matrix scale for realistic systems will be necessary to fully characterize the scalability of non-orthogonal quantum eigensolvers. In addition, there is scope to more rigorously formalize the intuitions about systems exhibiting both static and dynamic correlations. For example, the short-range dynamic correlation intuitively provides fermionic magic in the form of two-body interactions, and may also provide an anticoncentration mechanism in the wavefunction amplitudes, due to the cost of resolving the electron `cusp' (see Section \ref{sec:correlation}), which may be possible to formalize into a more rigorous statement. Finally, on the hardware implementation side, undertaking precise quantum resource estimates for specific cases, including explicit shot counts and gate counts, and the selection of a finite set of correlated reference states to achieve chemical accuracy in the energy estimates, will be a key direction for future work in this area.}

{In summary, we have identified explicit barriers to dequantization for quantum chemistry calculations of molecular electronic ground and excited state energies, in cases where the low-lying electronic eigenstates are well described by linear combinations of correlated reference states expressed in different orbital bases, which can be prepared by circuits of linear depth in the system size, such as orbital-rotated MPS or unitary cluster Jastrow.} We conclude that useful quantum speedups are most likely to be found for systems possessing both static and dynamic electron correlations, {constituting some of the most challenging problems in molecular electronic structure}, exemplified by molecular systems undergoing bond breaking and catalysis, and systems possessing multiple unpaired electrons or multivalent metal atoms.

\section{Acknowledgements}

We thank Jiaqing Jiang, Dominik Hangleiter, Zeph Landau, Junko Yano, and Martin Head-Gordon for helpful discussions. This work was supported by the NSF QLCI program through grant number QMA-2016345, partially by
the U.S. Department of Energy, Office of Science, Office of Advanced Scientific Computing Research under Award Number DE-SC0025526, and as part of a joint development agreement between UC Berkeley and Dow.

\section{AI disclosure}

{The initial version of the manuscript was produced entirely without the use of generative AI. During the peer review process, the authors collaborated with LLMs to refine some aspects of the arguments in Sections \ref{sec:ifm_states}, \ref{sec:ucj_barriers}.2, and \ref{sec:mps_barriers}.2. In particular, this collaboration inspired the proof of Proposition \ref{prop:ifmp_hardness}, the final version of which was proven by the authors. LLMs were not used to write any part of the manuscript, and the authors take full responsibility for all of the results and conclusions presented.}

\section{Note added}

{While preparing this manuscript for publication, the authors were made aware of a recent result \cite{oh2026classicalsimulationfreefermionicdynamics} which proves that the overlaps between orbital-rotated tensor products of fermionic magic states can be efficiently dequantized to a constant or inverse-polynomial additive error approximation by an efficient classical algorithm similar to mid-circuit $\ell^2$-norm sampling. This result holds for arbitrary orbital rotations, including those that would concentrate the probability distribution within a microscopic fraction of the Fock space, and which could yield a potentially macroscopic overlap matrix element. We emphasize that this result does not enable a similar classical algorithm to efficiently compute the overlap matrix elements between states that satisfy Definition \ref{def:itbc} under Haar-random orbital rotations, which is now ruled out by Propositions \ref{prop:ifmp_sampling} and \ref{prop:ifmp_hardness} under {the generalized $\mathsf{P}\neq\mathsf{NP}$ conjecture}.}

\bibliographystyle{apsrev4-1}
\bibliography{refs}

\appendix

\makeatletter 
\@addtoreset{figure}{section} 
\makeatother
\renewcommand{\thefigure}{\thesection\arabic{figure}}

\section{Schematic of theorems, lemmas, corollaries}
\label{app:summary_of_results}

\begin{figure}[h]
\includegraphics[width=0.96\linewidth]{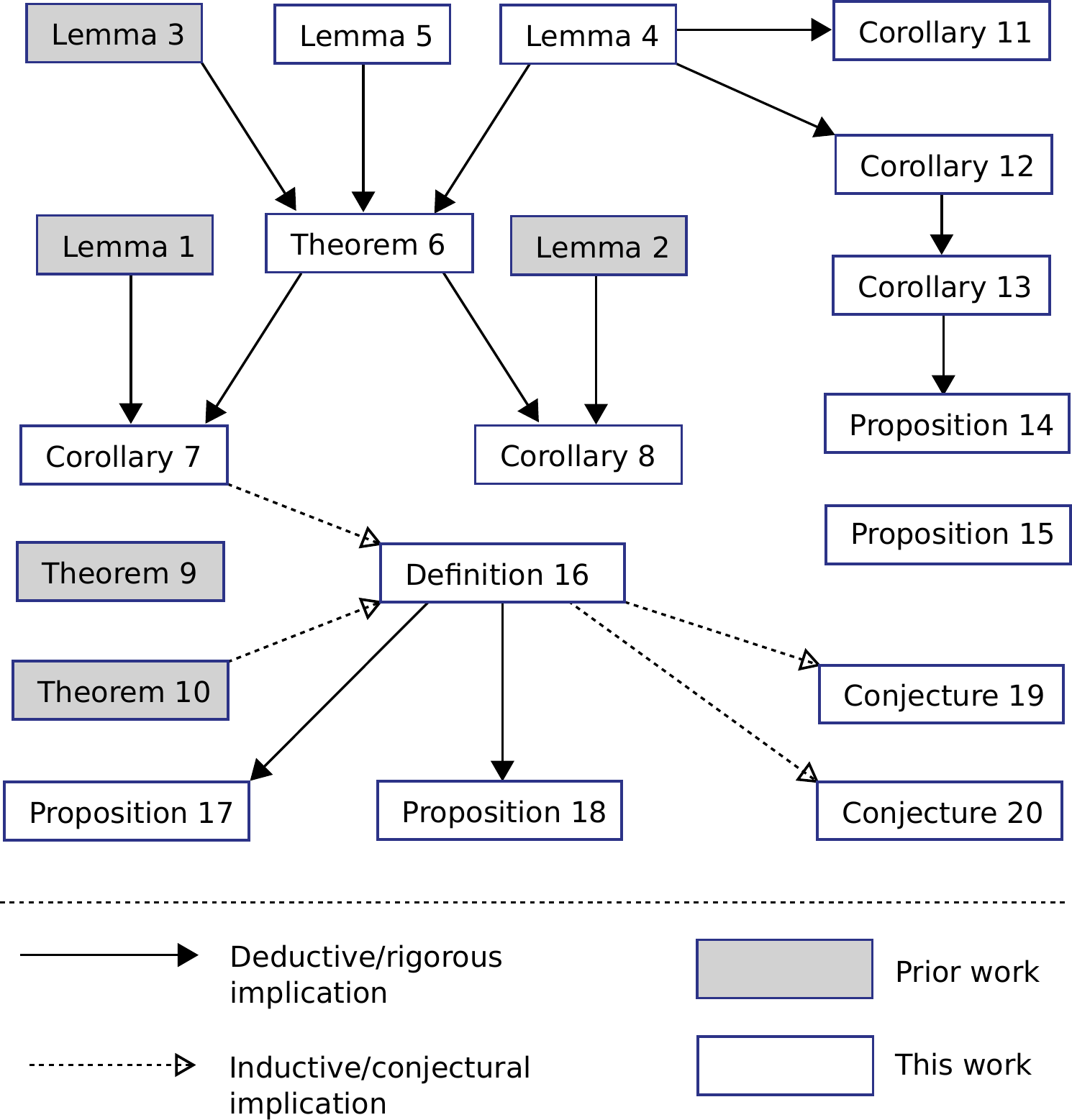}
\caption{{Schematic of the theorems, lemmas, corollaries, definitions, propositions, and conjectures presented in this work{, together with their relationships}. The shaded boxes indicate results of prior work. Lemma \ref{lem:strong_sim} was proven by Fujii and Morimae \cite{fujii_commuting_2017}, Lemma \ref{lem:weak_sim} was proven by Bremner, Jozsa, and Shepherd \cite{bremner_classical_2010}, the proof of Lemma \ref{thm:orbital_matchgates} is contained in works by Terhal and DiVincenzo \cite{terhal_classical_2002} and Kivlichan \emph{et al.} \cite{kivlichan_quantum_2018}, and Theorems \ref{thm:wost_to_avg} and \ref{thm:anticoncentration} were proven by Oszmaniec \emph{et al.} \cite{oszmaniec_fermion_2022}. The solid arrows indicate a deductive rigorous proof (the statement in box $A$ logically implies the statement in box $B$), and the dashed arrows indicate inductive or conjectural implications (Corollary $\ref{cor:strong_sim_formal}$ and Theorem $\ref{thm:wost_to_avg}$ provide supporting evidence for the existence of states satisfying Definition \ref{def:itbc}, and Conjectures \ref{conj:jastrow_ifm} and \ref{conj:mps_ifm} represent specific families of quantum states that are conjectured to provide instances satisfying Definition \ref{def:itbc}).}}
\label{fig:results_schematic}
\end{figure}

See Figure \ref{fig:results_schematic}.

\section{Complexity classes}
\label{app:complexity}

Here we briefly review some complexity classes that are relevant for understanding our results (a more comprehensive presentation can be found in Ref. \citealp{hangleiter_computational_2023}). A standard definition of computational complexity classes is based on \emph{decision problems}, in which one is given a string of bits, $x$, as well as some rule specifying a subset of all the strings of length $|x|$, a.k.a. a \emph{language}, $L$. The decision problem is to determine whether or not $x\in L$. The complexity of $L$ refers to the minimal possible computational resources (typically time or memory) required by any algorithm to solve the decision problem within some model of computation. The class of problems which are efficiently solvable by a deterministic classical algorithm in polynomial time is denoted by $\mathsf{P}$, while $\mathsf{NP}$ is the class of problems for which a solution, once obtained, can be efficiently verified. It is widely believed that $\mathsf{NP}$ contains problems that are not in $\mathsf{P}$, known as the $\mathsf{P}\neq \mathsf{NP}$ conjecture. In other words, it is believed that there are problems in $\mathsf{NP}$ which require super-polynomial time complexity for any classical algorithm.

For the purposes of this paper we will allow for a looser application of these complexity labels, to include not only decision problems, but also other types of computational tasks that could enable the efficient solution of a decision problem. For example, suppose that efficiently computing some function, $f(x)$, would enable one to efficiently decide whether $x\in L$, which is a complete (hardest) decision problem for the class $\mathsf{NP}$. Then we will say that computing $f(x)$ is also $\mathsf{NP}$-hard (i.e., at least as hard as any problem in $\mathsf{NP}$).

An \textit{oracle} is an abstract entity which grants query access to the solution for a complete problem in a given complexity class; the class of problems that can be solved in $\mathsf{A}$ with access to an oracle for a problem in $\mathsf{B}$ is written as $\mathsf{A^B}$. The \textit{polynomial hierarchy} ($\mathsf{PH}$) is the union of a set of nested complexity classes $\mathsf{\Sigma_0}\subseteq\cdots\subseteq\mathsf{\Sigma_\infty}$, defined recursively by
\begin{align}
\mathsf{\Sigma_0}=\mathsf{P}, \qquad \mathsf{\Sigma_{i}}=\mathsf{\Sigma_{i-1}^{NP}}.
\end{align}
For example, $\mathsf{\Sigma_1}=\mathsf{NP}$ and $\mathsf{\Sigma_2}=\mathsf{NP^{NP}}$. A collapse of the polynomial hierarchy to the $i$'th level means that $\mathsf{\Sigma_i}=\mathsf{\Sigma_{i+1}}=\cdots=\mathsf{\Sigma_\infty}$. It is widely conjectured that the polynomial hierarchy does not collapse to \textit{any} finite level, known as the generalized $\mathsf{P}\neq\mathsf{NP}$ conjecture.

The complexity classes $\#\mathsf{P}\subseteq\mathsf{GapP}$ are related to counting the number of solutions to an $\mathsf{NP}$ problem. Broadly speaking, $\#\mathsf{P}$ involves summation over exponentially many non-negative terms $\in\{1,0\}$, each of which is efficiently computed by a classical algorithm, while $\mathsf{GapP}$ involves summation over terms which can be positive or negative ($\in\{1,-1\}$), also known as the closure of $\#\mathsf{P}$ under subtraction. These complexity classes are equivalent under \textit{polynomial-time reductions}, meaning that $\mathsf{P^{\#P}}=\mathsf{P^{GapP}}$. They contain problems that are thought to be well beyond the capabilities of classical algorithms, evidenced by Toda's theorem \cite{toda_pp_1991}:
\begin{align}
\mathsf{PH} \subset \mathsf{P}^{\#\mathsf{P}}.
\label{eq:toda_thm}
\end{align}

The class of problems that can be efficiently solved by a \textit{randomized} (probabilistic) classical algorithm with success probability $>1/2$ is denoted $\mathsf{PP}$. This means there exists an efficient classical algorithm that fails less than half the time. This is a very broad class of problems, as seen by its equivalence to $\#\mathsf{P}$ under polynomial-time reductions ($\mathsf{P}^\mathsf{PP}=\mathsf{P}^{\#\mathsf{P}}$), so by Toda's theorem an oracle for a $\mathsf{PP}$-complete decision problem (i.e., an efficient algorithm which never fails) would enable the efficient solution of any problem in $\mathsf{PH}$. By contrast, $\mathsf{BPP}$ is the far more limited class of problems which can be efficiently solved by a randomized classical algorithm with bounded success probability $\geq2/3$. It is known that $\mathsf{P}\subseteq\mathsf{BPP}\subseteq\mathsf{\Sigma_2}$ \cite{lautemann_bpp_1983}, and it has been conjectured that $\mathsf{P}=\mathsf{BPP}$.

The quantum analogue of $\mathsf{BPP}$ is $\mathsf{BQP}$, which is the class of problems efficiently solvable by a quantum computer with two-thirds success probability \cite{nielsen_quantum_2010}. One definition of $\mathsf{BQP}$ is as follows:
\renewcommand{\thedefinition}{A\arabic{definition}} 
\begin{definition}[$\mathsf{BQP}$]
A language $L$ is in $\mathsf{BQP}$ if and only if there exists a uniform family of quantum circuits, $\hat{U}$, consisting of $\text{poly}(n)$ gates on $n$ qubits, where $n=|x|$, such that the following holds: after preparing the state $\hat{U}\ket{0}^{\otimes n}$, let $P(1)$ be the probability of obtaining the $\ket{1}$ state when the first qubit is measured, then
\begin{align}
x\in L &\implies P(1) \geq 2/3, \\
x\notin L &\implies P(1) \leq 1/3.
\end{align}
\label{def:bqp}
\end{definition}
From this definition, it is easily shown that computing an additive approximation of the output probabilities of an arbitrary polynomial-size quantum circuit is a $\mathsf{BQP}$-complete problem \cite{arad_quantum_2010}. We will say that a quantum circuit family $\hat{U}$ is \textit{universal} for quantum computation if any logical quantum circuit $\hat{W}$ defined on $\nu$ qubits, with $\mu=\text{poly}(\nu)$ two-qubit gates, can be encoded within an instance of $\hat{U}$ defined on a larger register of $n=\text{poly}(\nu,\mu)$ qubits using $\text{poly}(n)$ gates. The output probabilities of $\hat{U}$ then cannot be efficiently approximated up to an additive error by a probabilistic classical algorithm unless $\mathsf{BQP}=\mathsf{BPP}$. Furthermore, exactly computing the output probabilities of $\hat{U}$ is $\mathsf{GapP}$-hard \cite{hangleiter_computational_2023}. While $\#\mathsf{P}$ and $\mathsf{GapP}$ are equivalent under polynomial-time reductions, the approximation of a $\mathsf{\#P}$ sum up to a multiplicative factor is enabled by Stockmeyer's algorithm in $\mathsf{BPP}^\mathsf{NP}\subseteq\mathsf{\Sigma_3}$ \cite{stockmeyer_complexity_1983}, whereas approximating a $\mathsf{GapP}$ sum up to a multiplicative factor is also a $\mathsf{GapP}$-hard problem \cite{hangleiter_computational_2023}. This is the basis for quantum speedups in the weak simulation of universal quantum circuits, as an efficient classical sampler would enable a multiplicative approximation of the output probabilities by Stockmeyer's algorithm in $\mathsf{\Sigma_3}$ \cite{aaronson_computational_2011,hangleiter_computational_2023}. Then, by Toda's theorem, we would have that $\mathsf{PH}\subseteq\mathsf{P^{GapP}}\subseteq\mathsf{\Sigma_3}$, so the polynomial hierarchy would collapse to the third level.

\begin{figure}
\centering
\includegraphics[scale=0.78]{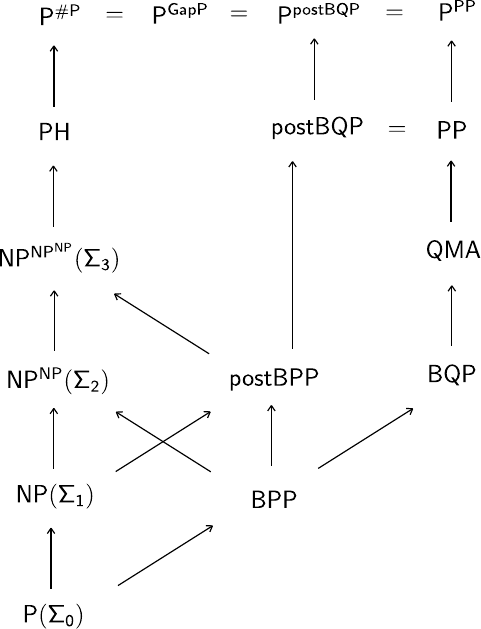}
\caption{A summary of inclusion relations between complexity classes. An arrow drawn from $\mathsf{A}$ to $\mathsf{B}$ implies $\mathsf{A}\subseteq\mathsf{B}$. Harder complexity classes generally appear higher up in the diagram, but this is not necessarily the case. For example, $\mathsf{BQP}$ appears above $\mathsf{NP}$ for convenience, but this is not meant to imply that $\mathsf{BQP}$ is more powerful than $\mathsf{NP}$.}
\label{fig:detailed_inclusion}
\end{figure}

In summary, the complexity classes referred to in this work come in two varieties. The first of these are the `easy' or tractable classes,
\begin{align}
\mathsf{P}\subseteq\mathsf{BPP}\subseteq\mathsf{BQP},
\end{align}
describing the problems that can be solved in polynomial time by a classical or quantum computer. The second variety are the `hard' or intractable classes that are not believed to be efficiently solvable using either a classical or a quantum computer, which includes all of the remaining complexity classes we have mentioned. The prototypical `hard' class is $\mathsf{NP}$, which is extended by $\mathsf{NP}$ oracles to construct a set of nested complexity classes known as the polynomial hierarchy ($\mathsf{PH}$). The hardest complexity classes we have covered include the counting complexity classes $\mathsf{\#P}$ and $\mathsf{GapP}$, as well as the probabilistic class $\mathsf{PP}$ and the postselected universal quantum class, $\mathsf{postBQP}$, which are all equivalent under polynomial-time reductions:
\begin{align}
\mathsf{P^{\#P}}=\mathsf{P^{GapP}}=\mathsf{P^{PP}}=\mathsf{P^{postBQP}}.
\end{align}
In addition to these, we have seen the postselected probabilistic class $\mathsf{postBPP}$, which sits between the first and third levels of the polynomial hierarchy. Another frequently encountered quantum complexity class is $\mathsf{QMA}$, which in some sense can be thought of as a `quantum analogue' for $\mathsf{NP}$, and which lies between $\mathsf{BQP}$ and $\mathsf{PP}$. All of these inclusion relations are summarized in Figure \ref{fig:detailed_inclusion}. Note that due to the constraints of the diagram, the known inclusion relation $\mathsf{NP}\subseteq\mathsf{QMA}$ has not been indicated.

\section{Mid-circuit $\ell^2$-norm sampling}
\label{app:sampling_dequantization}

Consider a unitary quantum circuit $\hat{U}$ that can be factorized into a pair of unitaries $\hat{U}=\hat{U}_a\hat{U}_b$. It was demonstrated in Ref. \citealp{tang_quantum-inspired_2019} that if one can efficiently sample computational basis states $\ket{z}\sim\hat{U}_b\ket{x}$, and efficiently compute the overlaps $\braket{y|\hat{U}_a|z}$ and $\braket{z|\hat{U}_b|x}$ {up to an exponentially small additive error}, then one can efficiently obtain a {constant additive error} approximation to $\braket{y|\hat{U}|x}$. We will refer to this dequantization scheme as \textit{mid-circuit} $\ell^2$-\textit{norm sampling}

To see how this works, let $\hat{U}=\hat{U}_a\hat{U}_b$, and define 
\begin{align}
\ket{\alpha}\equiv\hat{U}_a^\dag\ket{y}, \quad \ket{\beta}\equiv\hat{U}_b\ket{x}
\end{align} 
to be real-valued normalized wavefunctions, so that $\braket{y|\hat{U}|x}=\braket{\alpha|\beta}$, which is also real-valued. Let this decomposition be chosen such that $\ket{\alpha}$, $\ket{\beta}$ are both strongly simulable, i.e., the overlaps $\braket{z|\beta}$ and $\braket{z|\alpha}$ are efficiently computable for any computational basis vector $\ket{z}$. Furthermore, let $\ket{\beta}$ be weakly simulable, i.e., we may efficiently sample $\ket{z}\sim P(z)=|\braket{z|\beta}|^2$. Now let $\mathcal{Z}$ be the random variable $\braket{z|\alpha}/\braket{z|\beta}$, which is efficiently computable for any $\ket{z}$, sampled with probability $P(z)=|\braket{z|\beta}|^2$. Then
\begin{align}
\text{E}[\mathcal{Z}] &= \sum_z P(z)\frac{\braket{z|\alpha}}{\braket{z|\beta}} = \sum_z\braket{\beta|z}\braket{z|\alpha} = \braket{\alpha|\beta}, \\
\text{Var}[\mathcal{Z}] &\leq \sum_zP(z)\left(\frac{\braket{z|\alpha}}{\braket{z|\beta}}\right)^2 = \sum_z|\braket{z|\alpha}|^2 = 1.
\end{align}
Because $\text{Var}[\mathcal{Z}]$ is bounded by a constant, we may efficiently compute an additive error approximation to $\braket{\alpha|\beta}$ with $O(1/\epsilon^2)$ samples, completely independent of the number of qubits $n$. In general, the availability of a factorization of $\hat{U}$ into a pair of efficiently simulable circuits is not typical of unitary transformations, so this particular scheme only applies to circuit families with special structure. 

{Now suppose that when $\braket{z|\alpha}$ is queried by some randomized classical algorithm, for any bitstring state $\ket{z}$, one instead obtains $\braket{z|\alpha}+\mathcal{E}$, where $\mathcal{E}$ is a random variable with mean zero and variance $\sigma^2$. Let $\mathcal{Z}'$ be the random variable $(\braket{z|\alpha}+\mathcal{E})/\braket{z|\beta}$ thus obtained when $\ket{z}$ is sampled according to $P(z)=|\braket{z|\beta}|^2$. Let us choose $\ket{\beta}$ to have uniform positive support over all $N$ computational basis vectors. Then we may write $\mathcal{Z}' = \mathcal{Z}+N^{\frac{1}{2}}\mathcal{E}$, which is the sum of two independent random variables (the distribution of $\mathcal{E}$ is assumed not to depend on the value of $\braket{z|\alpha}$). We then have that
\begin{align}
\text{E}[\mathcal{Z}'] &= \text{E}[\mathcal{Z}] + \text{E}[N^{\frac{1}{2}}\mathcal{E}] = \braket{\alpha|\beta}, \\
\text{Var}[\mathcal{Z}'] &= \text{Var}[\mathcal{Z}] + \text{Var}[N^{\frac{1}{2}}\mathcal{E}] = 1 + N\sigma^2.
\end{align}
Now, to bound the variance of $\mathcal{Z}'$ by a constant, it is necesary to set $\sigma^2\propto N^{-1} \implies \sigma \propto N^{-\frac{1}{2}}$. Therefore, computing $\braket{\alpha|\beta}$ up to a constant additive error $\epsilon$ by mid-circuit $\ell^2$-norm sampling requires one to resolve each partial overlap $\braket{x|\alpha}$ up to an exponentially small additive error $O(\epsilon/N^{\frac{1}{2}})$. If $\hat{U}$ conserves particle number, and $\ket{x}$ and $\ket{y}$ have Hamming weight $\eta$, then $N={n\choose \eta}$, so the partial overlaps must be resolved up to an additive error $O(\epsilon/{n\choose \eta}^{\frac{1}{2}})$. If this is ruled out under complexity theoretic assumptions for even a constant fraction of $\eta$-particle computational basis states, then the additive error dequantization scheme fails.}

\section{Electronic structure concepts}
\label{app:el_struct_term}

Here we provide a brief overview of some important concepts in electronic structure for the benefit of non-specialists (these are covered comprehensively in Ref. \citealp{helgaker_second_2000}). A \textit{molecular orbital}, a.k.a. a \textit{spin-orbital}, is a single-electron function $\omega_p(\vec{r},\sigma)$, where $\vec{r}$ is a spatial coordinate vector $(x,y,z)\in\mathbb{R}^3$, and $\sigma$ is a discrete spin coordinate $\in\{\tfrac{1}{2},-\tfrac{1}{2}\}$. \textit{Restricted} and \textit{unrestricted} spin-orbitals are delta-functions in the spin coordinate: restricted spin-orbitals come in pairs consisting of a spin-up and spin-down orbital with the same spatial distribution, while unrestricted orbitals allow for these to have different spatial distributions. \textit{General} spin-orbitals, on the other hand, can have linear combinations of spin-up and spin-down character within a single orbital. In the following we will assume restricted or unrestricted orbitals, and we will drop the spin coordinate notation, noting that whenever orbitals $p$ and $q$ with opposite spins are integrated over the same set of coordinates the result is zero (i.e., we neglect any spin coupling or fine structure corrections in the Hamiltonian). Molecular orbitals are constructed from linear combinations of primitive basis functions (the \textit{basis set}, e.g., atom-centered Gaussian functions or plane waves) to form an orthonormal set, indexed by $p=1,\ldots,n$:
\begin{align}
\int_{\vec{r}}\omega^*_p(\vec{r}\,)\omega_q(\vec{r}\,) d\vec{r}=\delta_{pq}.
\end{align}
The electronic structure Hamiltonian (Eq. \ref{eq:el_ham}) then has one- and two-body coefficients given by
\begin{align}
h_{pq} &= \int_{\vec{r}} \omega^*_p(\vec{r}\,) \left(-\frac{1}{2}\nabla^2 - \sum_a \frac{Z_a}{\|\vec{r}-\vec{R}_a\|} \right) \omega_q(\vec{r}\,) d\vec{r}, \label{eq:h_pq} \\
h_{pqrs} &= \int_{\vec{r},\vec{r}\,'} \omega^*_p(\vec{r}\,)\omega^*_q(\vec{r}\,') \frac{1}{\|\vec{r}-\vec{r}\,'\|} \omega_r(\vec{r}\,)\omega_s(\vec{r}\,') d\vec{r}d\vec{r}\,', \label{eq:h_pqrs}
\end{align}
where $a$ indexes the atomic nuclei with fixed coordinates $\vec{R}_a$ (under the \textit{Born-Oppenheimer} approximation) and atomic charge numbers $Z_a$, with all quantities expressed in atomic units. 

Any sufficiently expressive primitive basis set is \textit{localizable}, meaning that there is an orbital rotation $\hat{G}$, as defined in the main text, which rotates the basis functions according to 
\begin{align}
\tilde{\omega}_p(\vec{r}\,) = \sum_qg_{pq}\omega_q(\vec{r}\,),
\end{align}
such that the rotated functions $\tilde{\omega}_p$ are largely confined to certain regions of space and decay rapidly outside of these regions. In the case of atom-centered Gaussian primitives, $\tilde{\omega}_p$ might correspond to a linear combination of atomic orbitals around a single atomic center, while in the case of plane wave primitives, $\tilde{\omega}_p$ may correspond to a wavepacket with a compact envelope. When the orbitals are spatially localized, the rotated Hamiltonian coefficients $\tilde{h}_{pq}$ and $\tilde{h}_{pqrs}$ (obtained by Eqs. \ref{eq:h_pq}, \ref{eq:h_pqrs} substituting $\omega_p$ with $\tilde{\omega}_p$, etc.) are largest for interactions between MOs confined to neighboring spatial regions, and decay with increasing separation following the $1/r$ dependence of the Coulomb potential. This illustrates the importance of the molecular orbital basis in quantum chemistry, which can significantly affect the computational hardness of the electronic structure problem.

A \textit{Slater determinant} is a multi-electron wavefunction describing uncorrelated indistinguishable fermions, constructed so as to respect wavefunction antisymmetry under exchange of any pair of particle coordinates. Let $\omega_1,\ldots,\omega_\eta$ be a set of occupied MOs. Then the corresponding $\eta$-particle Slater determinant $\Omega$ is given by
\begin{align}
\Omega(\vec{r}_1,\ldots,\vec{r}_\eta) = \frac{1}{\sqrt{\eta!}}\begin{vmatrix}
\omega_1(\vec{r}_1) & \cdots & \omega_\eta(\vec{r}_1) \\
\vdots & \ddots & \vdots \\
\omega_1(\vec{r}_\eta) & \cdots & \omega_\eta(\vec{r}_\eta)
\end{vmatrix},
\end{align}
where $|\cdot|$ indicates the matrix determinant operation, and the `matrix elements' are single-particle wavefunctions. The result is a sum over all products of the occupied orbitals with permuted particle coordinates, multiplied by a sign of $\pm1$ according to the parity of the permutation. In the Fock space representation, each Fock space vector $\ket{x}=\ket{x_1\ldots x_n}$, where $x_p=1$ if $\omega_p$ is occupied or $0$ otherwise, corresponds to a unique Slater determinant $\Omega_x$. The \textit{Hartree-Fock} (HF) self-consistent field method is used to find the set of molecular orbitals that give rise to the lowest energy $\eta$-electron Slater determinant. We denote this by $\ket{x_0}$, to indicate that the orbitals are filled in order of increasing energy, so the HF method finds the optimal set of occupied molecular orbitals $\{\omega_p\}_{p=1}^\eta$, as linear combinations of the primitive basis set, in order to minimize $\braket{x_0|\hat{H}|x_0}$. This algorithm can be constrained to optimize over spin-restricted orbitals (RHF) or unrestricted orbitals (UHF).

Configuration interaction (CI) refers to diagonalization in a basis of Slater determinants. For example, the lowest energy vector in the complete basis of $\eta$-electron Slater determinants provides the full configuration interaction (FCI) wavefunction,
\begin{align}
\ket{\Psi_\text{FCI}} = \sum_xc_x\ket{x},
\end{align}
where $x$ runs over all computational basis vectors with the correct particle number (and total $z$-spin component). Within the non-relativistic quantum theory, and under the Born-Oppenheimer and finite basis approximations, this provides the exact ground state energy,
\begin{align}
\braket{\Psi_\text{FCI}|\hat{H}|\Psi_\text{FCI}} = E_0. 
\end{align}
Because diagonalizing over all $n \choose \eta$ basis vectors is computationally intractable, an efficiently representable \textit{ansatz} state $\ket{\psi}$ is chosen to approximate this wavefunction. Any normalized vector in the Fock space satisfies the \textit{variational principle}. this means that if $\ket{\psi}$ can be reliably and efficiently normalized, then $\braket{\psi|\hat{H}|\psi}\geq E_0$ can be guaranteed. This is an important property for the design of reliable quantum chemistry methods to minimize the expected energy, as the obtained estimate then never goes below the true value. 

Another desired property of $\ket{\psi}$ is \textit{size-consistency}. Consider two spatially separated molecular fragments $A$ and $B$. Using sets of localized MOs confined to each fragment, in the limit of $r_{AB}\rightarrow\infty$ we may write 
\begin{align}
\hat{H}_{AB} = \hat{H}_A\otimes \hat{\mathds{1}}_B + \hat{\mathds{1}}_A\otimes \hat{H}_B.
\label{eq:composite_ham}
\end{align}
Suppose that a classical or quantum algorithm, when applied separately to subsystems $A$ and $B$, produces ansatz states $\ket{\psi_A}$ with energy estimate $E_A$ and $\ket{\psi_B}$ with estimate $E_B$. If the method is size-consistent, then when applied to the composite system described by Eq. \ref{eq:composite_ham} it should produce the energy estimate $E_{AB}=E_A+E_B$, corresponding to the product wavefunction $\ket{\psi_A}\otimes\ket{\psi_B}$. Simply put, increasing the size of the system should not reduce the accuracy of the ansatz within each fragment.

Correlated electronic wavefunctions are categorized using loose descriptors for different types of electronic correlation. Broadly speaking, \textit{dynamic correlation} (also known as \textit{weak correlation}, not to be confused with dynamical evolution) is characterized by many small contributions from highly excited determinants. For example, consider a coupled-cluster doubles (CCD) wavefunction,
\begin{align}
\ket{\psi_\text{CCD}} = \exp(\hat{T})\ket{x_0} = \sum_{k=0}^\infty \frac{1}{k!}\hat{T}^k\ket{x_0},
\label{eq:ccd}
\end{align}
where the two-body excitation generator $\hat{T}$ is defined as in Eq. \ref{eq:twobody_correlator}, with $|t_{pqrs}|\ll 1$ (by which we mean that the largest double excitation amplitude is smaller than unity in absolute value by at least roughly an order of magnitude). This wavefunction is dominated by a single determinant, $\ket{x_0}$, with small contributions from many excited determinants which decay with increasing excitation order (here parameterized by $k$). This can be thought of as a perturbative correction to the \textit{reference state} $\ket{x_0}$. 

\textit{Static correlation}, on the other hand, generally refers to the non-perturbative features of the electronic wavefunction, which must be described using multiple reference states of large amplitude (also known as \textit{strong correlation}). For example, a wavefunction exhibiting a moderate amount of static correlation, but with little dynamic correlation, might be well described by a configuration interaction ansatz obtained by diagonalizing in the subspace of $O(n^4)$ doubly excited determiants (CID),
\begin{align}
\ket{\psi_\text{CID}} = \sum_{pqrs}^nc_{pqrs}\hat{a}^\dag_p\hat{a}^\dag_q\hat{a}_r\hat{a}_s\ket{x_0}.
\label{eq:sci}
\end{align}
Alternatively, the non-orthogonal configuration interaction (NOCI) ansatz may be applicable to model such a system, which takes the form of Eq. \ref{eq:multi_ref}, where the reference states are single Slater determinants in rotated orbital bases (for instance, these could be chosen as degenerate UHF solutions \cite{thom_hartreefock_2009}). In the main text we have expanded this notion of a reference state from a single Slater determinant to include compact correlated states in different orbital bases, in particular MPS or UCCD states, enabling far more flexible treatment of both types of electronic correlation. We stress that the line between static and dynamic correlation is not precise, and their features are emphasized differently depending on the context \cite{tew_electron_2007,ganoe_notion_2024,izsak_measuring_2023}.

Note that the CCD wavefunction in Eq. \ref{eq:ccd} is size-consistent but is not variational (the same is true for CCSD and CCSD(T), regarded as the ``gold standard'' of quantum chemistry). On the other hand, the CID wavefunction in Eq. \ref{eq:sci} is variational but is not size-consistent. Developing a computationally tractable wavefunction ansatz which is both variational and size-consistent, and which faithfully represents ground state wavefunctions exhibiting both static and dynamic electronic correlation, is the essential challenge of ground state quantum chemistry \cite{tew_electron_2007}. The quantum subspace framework of TNQE and NOQE ensures strict variationality in the obtained energy estimate, and in both cases the reference states are size-consistent, which enables the multi-reference ansatz to recover size-consistency in the limit of increasing subspace dimension (i.e., a sufficient number of reference states, $M$). The scaling of $M$ with the system size in order to achieve a constant additive error in the energy estimate will be system dependent, and the overall cost in terms of classical and quantum comptuational resources may be benchmarked against other hybrid quantum-classical approaches. Empirical studies have so far suggested highly favorable performance and resource estimates versus comparable quantum and classical methods \cite{leimkuhler2025quantum,baek_say_2023}.

\section{Dual-rail encodings for one- and two-qubit gates}
\label{app:dualrail}

A logical single qubit rotation can be decomposed into a sequence of Euler angle rotations \cite{nielsen_quantum_2010} as
\begin{align}
U=R_z(\varphi_1)R_y(\theta)R_z(\varphi_2),
\end{align}
up to a global phase, where 
\begin{align}
R_z(\varphi) &= \begin{pmatrix}
e^{-i\tfrac{\varphi}{2}} & 0 \\
0 & e^{i\tfrac{\varphi}{2}}
\end{pmatrix}, \\ R_y(\theta) &= \begin{pmatrix}
\cos(\tfrac{\theta}{2}) & -\sin(\tfrac{\theta}{2}) \\
\sin(\tfrac{\theta}{2}) & \cos(\tfrac{\theta}{2})
\end{pmatrix}.
\end{align}
Under the dual-rail encoding, the $\ket{0}$, $\ket{1}$ states are mapped to the $\ket{01}$, $\ket{10}$ states respectively (the $\ket{00}$ and $\ket{11}$ states of the dual-rail are not accessed throughout the computation). Now the PN conserving matchgates
\begin{align} \label{eq:Givens}
R(\varphi)\otimes\mathds{1} &= \begin{pmatrix}
1 & 0 & 0 & 0 \\
0 & 1 & 0 & 0 \\
0 & 0 & e^{i\varphi} & 0 \\
0 & 0 & 0 & e^{i\varphi}
\end{pmatrix}, \\
G(\theta/2) &= \begin{pmatrix}
1 & 0 & 0 & 0 \\
0 & \cos(\tfrac{\theta}{2}) & -\sin(\tfrac{\theta}{2}) & 0 \\
0 & \sin(\tfrac{\theta}{2}) & \cos(\tfrac{\theta}{2}) & 0 \\
0 & 0 & 0 & 1
\end{pmatrix},
\end{align}
where $G(\theta/2)$ is a Givens rotation by $\theta/2$ in the single excitation subspace, encode the exact same computations as the logical single-qubit $R_z$ and $R_y$ gates within the single-particle block, up to an unimportant global phase. Hence the circuit decomposition in Figure \ref{fig:singlequbit} implements an arbitrary logical single-qubit rotation in the dual-rail encoding.

In the logical space, an arbitrary two-qubit unitary decomposes into three CNOT gates interleaved with single-qubit rotations \cite{vidal_universal_2004}. A CNOT gate can be further decomposed into a C$Z$ gate conjugated by Hadamard gates on the target qubit. Hence the same two-qubit gate decomposition can be achieved with three C$Z$ gates interleaved with single-qubit rotations. In the dual-rail encoding, a logical C$Z$ operation induces a phase flip on the $\ket{1010}$ basis vector, leaving the other dual-rail computational basis vectors ($\ket{1001}$, $\ket{0110}$, $\ket{0101}$) unchanged. This can be achieved using only nearest-neighbor gates by a C$Z$ gate acting on the middle two qubits followed by a $Z$ gate on the third qubit, inducing a resultant phase flip only when the middle two qubits are in the $\ket{01}$ state. Then by decomposing each logical single-qubit rotation into Euler angles and encoding within the dual-rail space according to Figure \ref{fig:singlequbit}, followed by some rearrangement of the phase gates, we arrive at the dual-rail circuit decomposition of an arbitrary logical two-qubit gate shown in Figure \ref{fig:twoqubit}.

Note that the C$Z$ gate acting between neighboring dual-rail qubits generates entanglement through a controlled phase flip, thus requiring no transport of particles between the dual-rails. The transmission of logical quantum information is thus entirely by means of entangling the phases of the dual-rail qubit states. Note also that encoded universality, and the implementation of arbitrary logical two-qubit gates, is only possible because the C$Z$ gate is not a matchgate. Attempting to draw an equivalence between the C$Z$ gate,
\begin{align}
\text{C}Z = \begin{pmatrix}
1 & 0 & 0 & 0 \\
0 & 1 & 0 & 0 \\
0 & 0 & 1 & 0 \\
0 & 0 & 0 & -1
\end{pmatrix},
\end{align}
and the matchgate definition (Eq. \ref{eq:matchgate}), we would have that
\begin{align}
\textbf{u} = \begin{pmatrix}
1 & 0 \\
0 & -1
\end{pmatrix}, \quad \textbf{v} = \begin{pmatrix}
1 & 0 \\
0 & 1
\end{pmatrix},
\end{align}
so that $\text{det}(\textbf{u})=-\text{det}(\textbf{v})$. It is worth remarking that this single phase difference between the submatrix determinants is sufficient to elevate the {classically} simulable family of PN conserving matchgate circuits to a universal model of quantum computation.
\section{Universality under postselection with real-valued orbital rotations}
\label{app:real_valued}
Under the Jordan-Wigner mapping, the single-qubit phase gate $R(\varphi)$ is equivalent to a complex phase change on one of the orbitals (see Eq. \ref{eq:phase_fermop}). We now show that universal quantum computation can be achieved under postselection using only real-valued PN-conserving matchgates. This follows from the well known fact that a complex-valued unitary computation can be encoded within a real-valued one by the use of ancilla qubits (see e.g. Ref. \citealp{aharonov_simple_2003}). For example, consider a single qubit in the state
\begin{align}
(a+ib)\ket{0} + (c+id)\ket{1},
\end{align}
where 
\begin{align}
|a+ib|^2 + |c+id|^2 = |a|^2 + |b|^2+|c|^2+|d|^2 = 1.
\end{align}
We may equivalently encode this quantum state with an additional ancilla qubit that represents the complex phase as 
\begin{align}
a\ket{00} + b\ket{01} + c\ket{10} + d\ket{11}.
\end{align}
Now consider the implementation of the complex phase gate $R(\varphi)$ which shifts the phase of the $\ket{1}$ state by $e^{i\varphi}=\cos\varphi + i\sin\varphi$. The same transformation of the coefficients can be achieved in the real-valued encoding by a controlled-$R_y$ gate on the second qubit with control from the first qubit. By adding a separate phase qubit for each system qubit, it is then straightforward to see that an arbitrary quantum circuit can be implemented within this encoding using only orthogonal nearest-neighbor two-qubit gates (including SWAP gates). Then this logical real-valued circuit may be encoded under postselection within the PN conserving dual-rail representation with entirely real-valued matchgates and magic state inputs, where each logical qubit is now represented by four physical qubits instead of two.

\section{Conditions for efficient ground state estimation by {non-orthogonal quantum eigensolvers}}
\label{app:cplex_subs_efficient}

\begin{figure}
    \centering
    \includegraphics[scale=0.72]{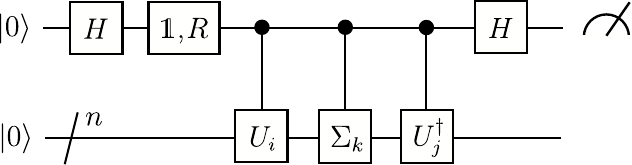}
    \caption{A Hadamard test circuit used to evaluate the off-diagonal Hamiltonian matrix elements in quantum subspace methods. By setting the second gate applied to the ancilla qubit to be the identity, the real part of the expectation value $\braket{0|\hat{U}_j^\dag \hat{\Sigma}_k \hat{U}_i|0}$ is computed up to an additive error $\epsilon$ from the probabilities $P(0)-P(1)$ for obtaining each outcome upon measuring the ancilla qubit, requiring $O(1/\epsilon^2)$ measurements. The imaginary component of the expectation value is obtained by setting the second gate on the ancilla register to be a phase gate, $R$, with a rotation angle of $\pi/2$. The off-diagonal Hamiltonian matrix element $\braket{u_i|\hat{H}|u_j}$ can therefore be efficiently computed with $O(\lambda^2/\epsilon^2)$ measurements.}
    \label{fig:generic_htest}
\end{figure}

The goal of quantum subspace methods is to approximate the low-energy spectrum of a Hamiltonian with an efficient Pauli string decomposition,
\begin{align}
\hat{H}=\sum_{k=1}^{K} h_{k} \hat{\Sigma}_k,
\end{align}
where $\hat{\Sigma}_k$ is a tensor product of Pauli matrices, $\hat{\Sigma}_k=\hat{\sigma}^k_1\otimes\hat{\sigma}^k_2\otimes \cdots\otimes \hat{\sigma}^k_n$, with each $\hat{\sigma}^k_p\in\{{X},{Y},{Z},{\mathds{1}}\}$, and $K=\text{poly}(n)$. If $\hat{H}$ is the second quantized electronic structure Hamiltonian (Equation \ref{eq:el_ham}) under the Jordan-Wigner transformation, then $K=O(n^4)$. We will also assume that the $|h_k|$ are bounded such that $\lambda\equiv\sum_{k}|h_k|$ also scales as $\text{poly}(n)$. For simplicity, let us take as our goal to compute the \textit{exact} lowest energy eigenvalue within the finite basis approximation, $E_1^{(\text{ex})}$, to within a chemical accuracy threshold, $\Delta=1.6$ mHa, which has no dependence on $n$. Subspace methods consist of three parts:

\begin{enumerate}
\item Compute a subspace of \textit{reference state vectors}, $\{\ket{u_i}\}_{i=1}^M$, where $M=\text{poly}(n)$, each of which can be efficiently prepared by a quantum circuit of $\text{poly}(n)$ depth, $\ket{u_i}=\hat{U}_i\ket{0}$. Thus, the entire subspace can be uniquely characterized using $\text{poly}(n)$ parameters.
\item Evaluate the Hamiltonian and overlap matrices $\textbf{H}$ and $\textbf{S}$, with matrix elements given by 
\begin{align}
h_{ij} = \braket{u_i|\hat{H}|u_j}, \qquad s_{ij} = \braket{u_i|u_j},
\end{align}
which can be efficiently evaluated up to an additive error $\epsilon$, with $N_\text{m}=O(1/\epsilon^2)$ measurements for the overlap matrix elements and $N_\text{m}=O(\lambda^2/\epsilon^2)$ measurements for the Hamiltonian matrix elements, using a Hadamard test circuit (see Figure \ref{fig:generic_htest}). Because these matrices are Hermitian, each has $(M^2+M)/2$ unique matrix elements that must be evaluated.
\item Solve the generalized eigenvalue problem, 
\begin{align}
\textbf{H}\textbf{C}=\textbf{S}\textbf{C}\textbf{E},
\label{eq:gen_eig}
\end{align}
on a classical computer. Exact diagonalization requires $O(M^3)$ flops. The columns of the solution matrix $\textbf{C}$ describe the \textit{Ritz vectors}, and $\textbf{E}$ is a diagonal matrix of \textit{Ritz values} $E_1\leq \cdots\leq E_M$ that approximates the low energy spectrum.
\end{enumerate}

With exact matrix elements, the Ritz vector approximating the ground state is
\begin{align}
\ket{\psi} = \sum_{i=1}^M c_i\ket{u_i},
\end{align}
where the coefficients $c_i$ are given by the first column of $\textbf{C}$, which is subject to the normalization constraint $\textbf{C}^\dag\textbf{S}\textbf{C}=\mathds{1}$. The exact Ritz vector is therefore bounded by the Rayleigh-Ritz variational principle,
\begin{align}
\braket{\psi|\hat{H}|\psi} = E_1 \geq E^{(\text{ex})}_1.
\end{align}

The quantum computer provides the matrix elements up to an additive error $\epsilon$, so in practice an approximate Ritz value $\tilde{E}_1$ is obtained. The total error in the energy estimate is therefore
\begin{align}
|\tilde{E}_1-E^{(\text{ex})}_1| \leq |\tilde{E}_1-E_1| + |E_1-E^{(\text{ex})}_1|.
\end{align}
If we want to guarantee that we can resolve $|\tilde{E}_1-E^{(\text{ex})}_1|\leq \Delta$ with a reasonable number of shots, then we require $|E_1-E^{(\text{ex})}_1|$ to be not only smaller than $\Delta$, but to be smaller by a sufficient margin. The specific value we choose for this is somewhat arbitrary. We can state this by splitting the window of chemical accuracy into two parts, $\Delta=\Delta_0+\Delta_1$, where $\Delta_0,\Delta_1<\Delta$ (for example, we could choose $\Delta_0=\Delta_1=\Delta/2$). Then assuming $|E_1-E^{(\text{ex})}_1|\leq \Delta_0$, we are guaranteed to resolve our energy estimate to within chemical accuracy provided that $|\tilde{E}_1-E_1|\leq\Delta_1$.

Now we will conduct a simple error analysis. Let the perturbed matrices be denoted $\tilde{\textbf{H}}=\textbf{H}+\delta\textbf{H}$ and $\tilde{\textbf{S}}=\textbf{S}+\delta\textbf{S}$. From the evaluation of the matrix elements to within additive error $\epsilon$, we have that $\|\delta\textbf{H}\|,\|\delta\textbf{S}\|\leq\sqrt{M}\epsilon$ in spectral norm. First, we can write
\begin{align}
\textbf{U} \equiv \textbf{S}^{1/2}\textbf{C}, \qquad \textbf{H}'\equiv\textbf{S}^{-1/2} \textbf{H}\textbf{S}^{-1/2}.
\end{align}
Provided that $\textbf{U}$ is invertible, it follows from the normalization condition on the Ritz vectors that $\textbf{U}^{-1}=\textbf{U}^\dag$, which is the condition for unitarity. Then we can recast Equation \ref{eq:gen_eig} as
\begin{align}
\textbf{H}'=\textbf{U}\textbf{E}\textbf{U}^\dag,
\end{align}
which is a standard eigenvalue problem. Now writing $\tilde{\textbf{H}}'=\tilde{\textbf{S}}^{-1/2} \tilde{\textbf{H}}\tilde{\textbf{S}}^{-1/2}$, from Weyl's theorem it follows that the approximate eigenvalues are perturbed as
\begin{align}
|\tilde{E}_i-E_i|&\leq\|\tilde{\textbf{H}}'-\textbf{H}'\| \\
&\leq \frac{\sqrt{m}\epsilon \kappa(\textbf{S})}{\|\textbf{S}\|}\left(1 + \frac{\|\textbf{H}\|\kappa(\textbf{S})}{\|\textbf{S}\|}\right) + O(\epsilon^2),
\end{align}
where we have expanded $(\textbf{S}+\delta\textbf{S})^{-1/2}$ to first order in $\epsilon$. The upshot is that the error in the eigenvalues remains controlled provided that the condition number of the overlap matrix, $\kappa(\textbf{S})=\|\textbf{S}\|\cdot\|\textbf{S}^{-1}\|$, does not blow up. Because $\|\textbf{H}\|\leq\lambda$, which we assumed grows at most as $\text{poly}(n)$, and provided we can guarantee that $\kappa(\textbf{S})$ also scales at most as $\text{poly}(n)$, then we can guarantee $|\tilde{E}_i-E_i|\leq\Delta_1$ by resolving the matrix elements up to some $\epsilon=1/\text{poly}(n)$, requiring $N_\text{m}=\text{poly}(n)$ measurements. Note that the higher order terms in the expansion are in powers of $\kappa(\textbf{S})\epsilon$, so by choosing $\epsilon^{-1}$ to grow by a higher polynomial than $\kappa(\textbf{S})$, all of the higher order terms can be polynomially suppressed in $n$. 

Therefore, with access to an ideal quantum computer, one can efficiently resolve the ground state energy $E_1^{(\text{ex})}$ up to chemical accuracy, $\Delta$, with steps 2 and 3 of the quantum subspace routine being strictly polynomially scaling in the system size, \emph{provided} there is an efficient algorithm to compute a subspace in step 1 that satisfies the following properties:
\begin{enumerate}
\item The subspace consists of $M=\text{poly}(n)$ reference states.
\item Each reference state can be prepared by a quantum circuit of layer depth $D=\text{poly}(n)$.
\item The lowest Ritz value in the subspace satisfies $|E_1-E_1^{\text{(ex)}}|\leq\Delta_0<\Delta$.
\item The condition number of the subspace grows as $\kappa(\textbf{S})=\text{poly}(n)$. 
\end{enumerate}
Note that the third and fourth requirements imply that the off-diagonal Hamiltonian matrix elements $h_{ij}$ should be relatively large, while the off-diagonal overlap elements $s_{ij}$ should be small.

\section{Proof of Proposition \ref{prop:expectation_equiv}}
\label{app:offd_matel_from_qexpt}
Here we prove Proposition \ref{prop:expectation_equiv} in the main text.

\emph{Proof.} For any pair of reference states $\ket{u_1}$ and $\ket{u_2}$ prepared by a given circuit family on $n$ qubits, we can add a single dual-rail qubit to define a pair of reference states from the same family on $n+2$ qubits:
\begin{align}
\ket{v_1}&=\ket{01}\otimes\ket{u_1}, \\
\ket{v_2}&=\ket{10}\otimes\ket{u_2},
\end{align}
so that trivially $\braket{v_1|v_2}=0$. Let
\begin{align}
\ket{\psi}=\frac{1}{\sqrt{2}}(\ket{v_1}+\ket{v_2}),
\label{eq:example_psi_def}
\end{align}
and let $\hat{A}$ be any quantum operator. We then have
\begin{align}
\braket{\psi|\hat{A}|\psi} = \frac{1}{2}\left(\braket{v_1|\hat{A}|v_1} + \braket{v_2|\hat{A}|v_2}\right) + \text{Re}\braket{v_1|\hat{A}|v_2}.
\label{eq:qexpt_A}
\end{align}
Now let $\hat{H}$ be any two-body Hermitian fermionic observable with the form of Eq. \ref{eq:el_ham}, and let us define 
\begin{align}
\hat{H}'=\hat{H}+\hat{a}_1^\dag\hat{a}_2+\hat{a}_2^\dag\hat{a}_1,
\end{align}
meaning that $\hat{H}'-\hat{H}$ is an operator which acts only on the first two qubits. Then, under the JW transformation,
\begin{align}
\braket{v_1|(\hat{H}'-\hat{H})|v_2} = \braket{u_1|u_2}.
\label{eq:op_diff_ovlp}
\end{align}
Then substituting $\hat{H}'-\hat{H}$ for $\hat{A}$ in Equation \ref{eq:qexpt_A} and rearranging yields
\begin{align}
\text{Re}\braket{u_1|u_2} = &\braket{\psi|\hat{H}'|\psi} - \braket{\psi|\hat{H}|\psi} \nonumber\\
& -\frac{1}{2}\Big(\braket{v_1|\hat{H}'|v_1}-\braket{v_1|\hat{H}|v_1}\\
&+\braket{v_2|\hat{H}'|v_2}-\braket{v_2|\hat{H}|v_2}\Big). \nonumber
\end{align}
By computing each of the expectation values on the RHS up to an additive error $\epsilon$, one computes the LHS up to an additive error $4\epsilon$. One can similarly compute the imaginary part of $\braket{u_1|u_2}$ by inserting a complex phase between $\ket{u_1}$ and $\ket{u_2}$ in Eq. \ref{eq:example_psi_def}. Having thus obtained the real and imaginary components of $\braket{u_1|u_2}$ to within an additive error $4\epsilon$, one thereby obtains the overlap matrix element up to an additive error with norm bounded by at most $ 4\sqrt{2}\epsilon$, which is a constant independent of $n$. It is thereby shown that six queries to an oracle for $\textsc{Expv}(\hat{H},\psi,\epsilon)$ on $n+2$ qubits are sufficient to solve any instance of $\textsc{Ovlp}(u_i,u_j,4\sqrt{2}\epsilon)$ on $n$ qubits. $\Box$

\section{Proof of Proposition \ref{prop:ifmp_hardness}}
\label{app:proof_prop18}
{Here we prove Proposition \ref{prop:ifmp_hardness} in the main text.

\emph{Proof.} The irreducible anticoncentration property in Definition \ref{def:itbc} states that for any orbital rotation $\hat{G}$, and for some $\delta>0$, we have $P(x)=|\braket{x|\hat{G}|\phi}|^2 > \delta/{n\choose\eta}$ for a constant fraction $f$ of Slater determinants. The average-case hardness property in Definition \ref{def:itbc} states that computing $|\braket{x|\hat{G}|\phi}|^2$ up to a constant multiplicative factor $c$ is $\mathsf{\# P}$-hard in the overwhelming majority of Haar-random instances of $\hat{G}$ in the asymptotic limit, and for at least half of the `heavy' fraction $f$ of Slater determinants. By Markov's inequality, for any constant $C$, the fraction of computational basis states with a probability greater than $C/{n \choose \eta}$ is upper bounded by $1/C$, therefore by choosing $C>2/f$ we ensure that there is a constant positive fraction of computational basis states
\begin{align}
f' > \frac{f}{2}-\frac{1}{C}
\end{align}
for which the probabilities are bounded by 
\begin{align}
\frac{\delta}{{n \choose \eta}} < P(x) \leq \frac{C}{{n \choose \eta}},
\label{eq:probability_bound}
\end{align} 
and for which computing an estimate $\tilde{P}(x)$ to within an additive approximation window
\begin{align}
|\tilde{P}(x) - P(x)| \leq \delta\left(1-\frac{1}{c}\right)\cdot {n \choose \eta}^{-1}
\label{eq:probability_bound2}
\end{align}
would allow one to solve a $\mathsf{\# P}$-hard decision problem. Now consider a black-box algorithm that computes $\braket{x|\hat{G}|\phi}$ up to an additive error $\epsilon / {n\choose\eta}^{\frac{1}{2}}$, which would allow one to compute a probability estimate satisfying
\begin{align}
|\tilde{P}(x)-P(x)| \leq 2\epsilon|\braket{x|\hat{G}|\phi}|\cdot{n\choose \eta}^{-\frac{1}{2}} + \epsilon^2\cdot{n\choose\eta}^{-1}.
\end{align}
If $P(x)$ satisfies the bound in Eq. \ref{eq:probability_bound}, then this estimate is bounded by
\begin{align}
|\tilde{P}(x)-P(x)| \leq 2\sqrt{C}\epsilon \cdot{n\choose\eta}^{-1} +  \epsilon^2\cdot{n\choose\eta}^{-1}.
\end{align}
By comparison with Eq. \ref{eq:probability_bound2}, we see that this is within the $\mathsf{\# P}$-hard additive approximation window for $P(x)$ whenever
\begin{align}
2\sqrt{C}\epsilon + \epsilon^2 &< \delta\left(1-\frac{1}{c}\right) \\
\implies \epsilon &< \sqrt{C + \delta\left(1-\frac{1}{c}\right)} - \sqrt{C}.
\end{align}
Because $C$, $\delta$, and $c$ are constants independent of $n$, then there is a constant $\epsilon>0$ such that computing $\braket{x|\hat{G}|\phi}$ up to an exponentially small additive error $\epsilon / {n\choose\eta}^{\frac{1}{2}}$ is $\mathsf{\# P}$-hard for a constant fraction $f'>f/2-1/C>0$ of Slater determinants, and for a fraction $1-o(1)$ of Haar-random instances of $\hat{G}$.
$\Box$}

\section{Universality and fermionic magic state preparation with UCJ circuits}
\label{app:twobody_magic_states}

{Let $\ket{\phi}$ be a state prepared by a UCJ circuit,
\begin{align}
\ket{\phi} = \prod_{l=1}^{L} \left(\hat{G}^\dag_l \hat{D}_l \hat{G}_l\right)\ket{x_0},
\label{eq:ucj_circuit_app}
\end{align}
where $\ket{x_0}=\ket{0}^{\otimes{n-\eta}}\otimes\ket{1}^{\otimes \eta}$, consisting of orbital rotations $\hat{G}_l$ and diagonal operators corresponding to products of two-body number excitations,
\begin{align}
\hat{D}_l = \prod_{pq} \exp(i\theta^{(l)}_{pq} \hat{n}_p\hat{n}_q).
\end{align}
First, we show how to encode arbitrary logical quantum computations into $\ket{\phi}$ with a number of Jastrow layers $L=\text{poly}(n)$ and in $n=2\eta$ orbitals. Take any logical quantum circuit $\hat{U}$ consisting of $\mu$ two-qubit gates on $\nu$ qubits, and then encode this in the dual-rail representation as a PN conserving quantum circuit $\hat{\mathcal{U}}$ on $n=2\nu$ qubits, with at most $3\mu$ two-body fermionic number excitations of the form $\exp(i\pi\hat{n}_p\hat{n}_q)$ under the JW transformation, and $\text{poly}(\nu,\mu)$ PN conserving matchgates (Lemma \ref{lem:dualrail_encoding} and Corollary \ref{cor:universality_uccsd}). We can write this circuit in terms of interleaved layers of products of two-body number excitations and products of Givens and phase rotation gates. Let the $l$-th layer of Givens and phase rotation gates be denoted $\hat{G}^{(0)}_l$, then we can write 
\begin{align}
\hat{\mathcal{U}} = \prod_{l=1}^{\text{poly}(n)}\hat{D}_l\hat{G}_l^{(0)}.
\end{align}
This circuit encodes any amplitude output from the logical quantum {circuit} $\hat{U}$ in the dual-rail encoding, given the input state $\ket{01}^{\otimes \nu}$. There is trivially a fermionic SWAP network $\hat{F}$ that prepares
\begin{align}
\ket{01}^{\otimes \nu}=\hat{F}\ket{x_0}.
\end{align}
By setting $\hat{G}_{L}=\hat{F}$ and $\hat{G}_{l-1}=\hat{G}_l^{(0)}\hat{G}_l$ we obtain $\ket{\phi}$ as defined in Eq. \ref{eq:ucj_circuit_app}, such that
\begin{align}
\ket{\phi} = \hat{\mathcal{U}}\ket{01}^{\otimes \nu},
\end{align}
with $L=\text{poly}(n)$.
}

{We now show that a single Jastrow layer ($L=1$) is already sufficient to prepare the tensor product of magic states $\ket{M}^{\otimes m}$. In this case we have
\begin{align}
\ket{\phi}=\hat{G}^\dag\hat{D}\hat{G}\ket{x_0}.
\end{align}
Again we use the facts that the C$Z$ gate between qubits $p$ and $q$ is equivalent to $\text{exp}(i\pi\hat{n}_p\hat{n}_q)$, and that all diagonal operators commute with each other. One can then prepare the four-qubit state $\tfrac{1}{2}(\ket{10}+\ket{01})^{\otimes 2}$ starting from the $\ket{1010}$ state by applying two Givens rotations. 
Following this, application of three C$Z$ gates in any order yields the magic state $\ket{M}$ defined in Eq. \ref{eq:magic_state}. 
This analysis shows that we can therefore write 
\begin{align}
\ket{M}^{\otimes m}=\hat{D}\hat{G}\ket{x_0} 
\end{align}
for appropriately chosen $\hat{G}$ and $\hat{D}$. 
The inverse orbital rotation $\hat{G}^\dag$ of the UCJ 
{reference} may then be absorbed into any subsequent orbital rotation $\hat{G}'$ that transforms between single-particle bases. 
This implies that efficiently sampling from $\hat{G}'\ket{\phi}$ by a classical algorithm is ruled out in the worst case by Corollary \ref{cor:weak_sim_formal} unless there is a polynomial hierarchy collapse, even for $L=1$.
}

\section{Dequantizing the single-reference UCJ ansatz with a single Jastrow layer}
\label{app:jastrow_dequantized}

Here an example is presented which illustrates why expectation values of ansatz states in the non-orthogonal multi-reference framework can be significantly more difficult to dequantize than for ansatz states prepared by a single reference circuit. The single-layer UCJ ansatz has the form
\begin{align}
\ket{\phi} = \hat{G}\hat{D}\hat{G}^\dag\ket{x_0},
\end{align}
where $\ket{x_0}$ is a Hartree-Fock Slater determinant, $\hat{G}$ is an orbital rotation, and $\hat{D}$ is a diagonal operator,
\begin{align}
\hat{D} = \exp\Big(i\sum_{pq}J_{pq}\hat{n}_p\hat{n}_q\Big)
\end{align}
with $\hat{n}_p=\hat{a}^\dag_p\hat{a}_p$. Now write
\begin{align}
\braket{\phi|\hat{H}|\phi} = \braket{x_0|(\hat{G}\hat{D}^\dag\hat{G}^\dag)\hat{H}(\hat{G}\hat{D}\hat{G}^\dag)|x_0}.
\label{eq:ham_expt}
\end{align}
We can define 
\begin{align}
\hat{H}'\equiv\hat{G}^\dag\hat{H}\hat{G}
\end{align}
then decompose
\begin{align}
\hat{H}'=\sum_kh_k\hat{\Sigma}_k
\end{align}
which is a sum over $O(n^4)$ tensor products of Pauli matrices, $\hat{\Sigma}_k$, where the coefficients $h_k$ are efficiently computable, and $\lambda\equiv\sum_k|h_k|$ is polynomial in $n$. {Substituting} into Equation \ref{eq:ham_expt} we get
\begin{align}
\braket{\phi|\hat{H}|\phi} = \sum_kh_k\braket{x_0|\hat{G}\hat{D}^\dag\hat{\Sigma}_k\hat{D}\hat{G}^\dag|x_0}.
\end{align}
We can then use the mid-circuit $\ell^2$-norm sampling technique to compute each of the braket products on the RHS up to an additive error $\epsilon$ using $O(1/\epsilon^2)$ samples (see Section \ref{sec:simulation}). To see this, consider the circuit partitioning
\begin{align}
\ket{\alpha} = \hat{D}^\dag\hat{\Sigma}_k\hat{D}\hat{G}^\dag\ket{x_0}, \quad \ket{\beta} = \hat{G}^\dag\ket{x_0}.
\end{align}
Then sampling bitstring states $\ket{z}\sim\ket{\beta}$ and computing the partial overlaps $\braket{z|\beta}$ are both classically efficient, because $\ket{\beta}$ is prepared by a non-interacting fermion circuit ($\hat{G}^\dag$) \cite{terhal_classical_2002}. It remains to show that computing $\braket{z|\alpha}$ is also classically efficient. This follows because
\begin{align}
\braket{z|\hat{D}^\dag\hat{\Sigma}_k\hat{D}\hat{G}^\dag|x_0} = e^{i\varphi}\braket{z'|\hat{G}^\dag|x_0},
\end{align}
where the transformed bitstring $\ket{z'}$ and the phase $\varphi$ are efficiently computable from the operator $\hat{D}^\dag\hat{\Sigma}_k\hat{D}$ in $\text{poly}(n)$ time.

Putting this all together, the expectation value $\braket{\phi|\hat{H}|\phi}$ is {classically} computable up to an additive error $\epsilon$ in  time $\text{poly}(n)\text{poly}(1/\epsilon)$.
{This theoretical result has recently been confirmed with an experimental demonstration on quantum hardware~\cite{belagali2026efficient}.}

\section{NOQE-UCJ gate count estimates for the Mn$_4$O$_5$Ca complex}

\label{app:gate_count}

Here we estimate {upper bounds on} the per-circuit CNOT and T-gate counts to compute the off-diagonal Hamiltonian matrix elements between single-layer UCJ reference states for the Mn$_4$O$_5$Ca complex in an active space of 112 spin-orbitals. The CNOT and $Z$-rotation counts are obtained directly from the circuit resource estimates presented in Section V of Ref. \citealp{baek_say_2023}. For 112 spin-orbitals and a single Jastrow layer ($k=L=m=1$), we obtain
\begin{align}
N^{h_{ij}}_\text{CNOT} &\approx 7.8 \times 10^4, \\
N^{h_{ij}}_{R_z} &\approx 7.2 \times 10^4.
\end{align}
The T-gate count is estimated using the repeat-until-success (RUS) synthesis routine \cite{bocharov2015efficient}, which implements an $R_z$ gate to within error $\epsilon$ using $\approx 1.15 \log_2(1/\epsilon)+9.2$ T-gates. Assuming an acceptable error threshold of $\epsilon=10^{-10}$ thus requires $\approx 47$ T-gates per $Z$-rotation. We therefore arrive at a total per-circuit T-gate count of $\approx 3.4\times 10^6$.
{We note that these are upper bounds on the gate counts and can be considerably reduced by pruning the UCJ 
{reference states} and/or reducing the size of the active space.}

\end{document}